\documentstyle[12pt]{article}
\makeatletter \oddsidemargin  -.1in \evensidemargin -.1in
\newcommand{\singlespacing}{\let\CS=\@currsize\renewcommand{\baselinestretch}{1}\tiny\CS}
\newcommand{\oneandahalfspacing}{\let\CS=\@currsize\renewcommand{\baselinestretch}{1.25}\tiny\CS}
\newcommand{\doublespacing}{\let\CS=\@currsize\renewcommand{\baselinestretch}{1.35}\tiny\CS}

\newtheorem{rule-def}[theorem]{Rule}

\RequirePackage[dvips]{graphicx} \textheight 22.5cm

\begin{document}

\title{\bf Non-Newtonian characteristics of peristaltic flow of blood in micro-vessels}
\author{\small S. Maiti$^{1}$\thanks{Presently at Department of Applied Mathematics, IIT (BHU), Varanasi 221005, India. Email address: {\it maiti0000000somnath@gmail.com/somnathm@cts.iitkgp.ernet.in (S. Maiti)}},
  ~~~J.C.Misra$^{2}$\thanks{Email address: {\it misrajc@gmail.com/misrajc@rediffmail.com (J. C. Misra)}}  \\
\it$^{1}$School of Medical Science and Technology $\&$ Center for Theoretical Studies, \\Indian Institute of Technology, Kharagpur-721302, India\\
$^{2}$\it Professor of Mathematics, Institute of Technical Education
  $\&$ Research,\\Siksha 'O' Anusandhan University, Bhubaneswar-751030, India}
\date{}
\maketitle \noindent \doublespacing

\begin{abstract}
Of concern in the paper is a generalized theoretical study of the
non-Newtonian characteristics of peristaltic flow of blood through
micro-vessels, e.g. arterioles. The vessel is considered to be of
variable cross-section and blood to be a Herschel-Bulkley type of
fluid. The progressive wave front of the peristaltic flow is supposed
sinusoidal/straight section dominated (SSD) (expansion/contraction
type); Reynolds number is considered to be small with reference to
blood flow in the micro-circulatory system. The equations that govern
the non-Newtonian peristaltic flow of blood are considered to be
non-linear. The objective of the study has been to examine the effect
of amplitude ratio, mean pressure gradient, yield stress and the power
law index on the velocity distribution, wall shear stress, streamline
pattern and trapping. It is observed that the numerical estimates for
the aforesaid quantities in the case of peristaltic transport of the
blood in a channel are much different from those for flow in an
axisymmetric vessel of circular cross-section. The study further shows
that peristaltic pumping, flow velocity and wall shear stress are
significantly altered due to the non-uniformity of the cross-sectional
radius of blood vessels of the micro-circulatory system. Moreover, the
magnitude of the amplitude ratio and the value of the fluid index are
important parameters that affect the flow behaviour. Novel features of
SSD wave propagation that affect the flow behaviour of blood have also
been discussed.\\ \it Keywords: {\small Non-Newtonian Fluid; Flow
  Reversal; Wall Shear Stress; Trapping; SSD Wave.}
\end{abstract}

\section{Introduction}
Peristaltic transport of fluids through different vessels of human
physiological systems is known to physiologists as a natural mechanism
of pumping materials. The phenomenon of peristalsis has important
applications \cite{Chen,Jaggy,Hansbro,Nisar} in the design and
construction of many useful devices of biomedical engineering and
technology, artificial blood devices such as heart-lung machine, blood
pump machine, dialysis machine etc. Apart from physiological studies,
many of the essential fluid mechanical characteristics of peristalsis
have also been elucidated in analyses of different engineering
problems carried out by several researchers. These characteristics
become more prominent when the flow is induced by progressive waves of
area contraction/expansion along the length of the boundary of a
fluid-filled distensible tube. Benefits that can be derived from
studies on peristaltic movement have been elaborately discussed in our
earlier communications
\cite{Misra1,Misra2,Misra3,Misra4,Misra5,Misra6,Misra7,Misra8,Maiti1,Maiti2}.
Further discussion on peristalsis was made by several other researchers
(cf. Guyton and Hall \cite{Guyton}, Jaffrin and Shapiro
\cite{Jaffrin}, Srivastava and Srivastava \cite{Srivastava2},
Vajravelu et al. \cite{Vajravelu}, Hayat et al. \cite{Hayat}).
\begin{center}
\begin{tabular}{|l l|}\hline
{~\bf Nomenclature} &~ \\
~~$R,\theta,Z $ & Cylindrical co-ordinates\\
~~$c$ & Speed of the travelling wave\\
~~$d$ & Wave amplitude\\
~~$a_0 $ & Radius of the micro-vessel at the inlet\\
~~$R_0 $ & Radius of the plug flow region\\
~~$H $ & Displacement of the wall in the radial direction\\
~~$I$ & Identity matrix\\
~~$n$ & Flow index number\\
~~$k $ & Reciprocal of n \\
~~$k_1 $ & A parametric constant\\
~~$P $ & Fluid pressure\\
~~$Q $ & Flux at any location\\
~~~~$t $  & Time\\
~~$Re $ & Reynolds number\\
~~$U,V,W$ & Velocity components in Z-, R- and $\theta$- directions respectively\\
~~$\delta $ & Wave number\\
~~$\Delta P$ & Pressure difference between the ends of the vessel segment\\
~~$\dot{\gamma}$ & Strain rate of the fluid\\
~~$\lambda $ & Wave length of the travelling wave motion at the wall\\
~~$\lambda_{c} $ & A portion of length at which SSD expansion/contraction waves are confined\\
~~$\mu $ & Blood viscosity\\
~~$\mu_0 $ & A constant denoting the limiting viscosity of blood\\
~~$\nu $ & Kinematic viscosity of blood\\
~~$\phi$ & Amplitude ratio \\
~~$\Pi$ & The second invariant of the strain-rate tensor\\
~~$\Pi_0$ & Limiting value of $\Pi$ defined by (\ref{cnsns_manuscript_mu_0})\\
~~$\rho$ & Density of blood \\
~~$\tau_0$ & Yield stress of blood\\
~~$\tau_h$ & Wall shear stress\\
\hline
\end{tabular}
\end{center}
 A theoretical foundation of peristaltic transport in inertia-free
 Newtonian flows driven by sinusoidal transverse waves of small
 amplitude was suggested by Fung and Yih \cite{Fung1}. Shapiro et
 al. \cite{Shapiro} presented the closed form solution for an infinite
 train of peristaltic waves for small Reynolds number flow, where the
 wave length is large and wave amplitude is arbitrary. The said
 investigations witnessed a variety of important applications that are
 very useful in explaining the functioning of various physiological
 systems, such as flows in ureter, gastro-intestinal tract, small
 blood vessels of the micro-circulatory system and glandular
 ducts. Conditions for the the existence of physiologically
 significant phenomena of trapping and reflux in peristaltic transport
 were also presented in the aforesaid communication by Shapiro et
 al. \cite{Shapiro}. References to some of the earlier literatures on
 peristaltic flow of physiological fluids are available in
 communications of Jaffrin and Shapiro \cite{Jaffrin} as well as
 Srivastava and Srivatava \cite{Srivastava1}, while brief reviews of
 some recent literatures have been made by Tsiklauri and Beresnev
 \cite{Tsiklauri}, Mishra and Rao \cite{Mishra}, Yaniv et
 al. \cite{Yaniv}, Jimenez-Lozano et al.  \cite{ Jimenez-Lozano1},
 Nadeem and Akbar \cite{Nadeem}, Hayat et al. \cite{Hayat} as well as
 by Pandey and Chaube \cite{Pandey}. Several other studies dealing
 with analysis of different problems of peristaltic transport of
 various physiological fluids were reported by different
 investigators, e.g. Takabatake and Ayukawa \cite{Takabatake1},
 Jimenez-Lozano and Sen \cite{Jimenez-Lozano2}, Bohme and Friedrich
 \cite{Bohme}, Srivastava and Srivastava \cite{Srivastava2}, Provost
 and Schwarz \cite{Provost} and Chakraborty \cite{Chakraborty}.

Past experimental observations indicate that the non-Newtonian
behaviour of whole blood mainly owes its origin to the presence of
erythrocytes. As early as in the sixth decade of the last century,
different groups of scientists, viz. Rand et al. \cite{Rand},
Bugliarello et al. \cite{Bugliarello} and Chien et al. \cite{Chien}
made an observation that the non-Newtonian character of blood is
prominent as soon as the hematocrit rises above 20$\%$. However, it
plays a dominant role, when hematocrit level lies between 40$\%$ and
70$\%$.  It is known that the nature of blood flow in small vessels
(radius $<$0.1 mm) at low shear rate ($<20 sec^{-1}$) can be
represented by a power law fluid (cf. Charm and Kurland
\cite{Charm1,Charm2}). However, Merrill et al. \cite{Merrill}
observed that Casson model is somewhat satisfactory for blood
flowing in tubes of 130-1000$\mu$ diameter. Later on Scott-Blair and
Spanner \cite{Scott-Blair} also reported that blood obeys Casson
model for moderate shear rate flows. Further, they observed that
Herschel-Bulkley model is more appropriate than Casson model, more
particularly for cow's blood. However, they did not report any
difference between Casson and Herschel-Bulkley plots over the range
where Casson plot is valid for blood.

It is known that most vessels of physiological system are of varying
cross-sectional diameter (cf. Wiedman \cite{Wiedman}, Wiederhielm
\cite{Wiederhielm}, Lee and Fung \cite{Lee}). Some initial attempts to
perform theoretical studies pertaining to peristaltic transport of
physiological fluids in vessels of varying cross section were made by
several researchers (cf. Srivastava and Srivastava \cite{Srivastava1},
Wiederhielm \cite{Wiederhielm}, Lee and Fung \cite{Lee}, Gupta and
Seshadri \cite{Gupta}, Srivastava and Srivastava
\cite{Srivastava3}). In all these studies, the physiological fluids
were considered either as Newtonian fluids or non-Newtonian fluids
described by Casson/power-law models. However, different flow
characteristics have not been adequately explained in these
studies. Herschel-Bulkley models are, of course, more suitable to
represent some physiological fluids like blood, because the fluids
represented by this model describe very well material flows with a
non-linear constitutive relationship depicting the behaviour of
shear-thinning/shear-thickening fluids that are of much relevance in
the field of biomedical engineering (\cite{Malek}). Moreover, among
the various types of non-Newtonian models used to represent blood,
Herschel-Bulkley fluid model is more general. Use of this model has
the advantage that the corresponding results for fluids represented by
Bingham plastic model, power law model and Newtonian fluid model can
be derived from those of the Herschel-Bulkley fluid model, as
different particular cases. Also, Herschel-Bulkley fluid model yields
more accurate results than many other non-Newtonian models.

In view of all the above, a study concerning the analysis of a
non-linear problem of peristaltic flow of blood in a vessel of the
micro-circulatory system has been undertaken here, by treating blood
as a non-Newtonian fluid of Herschel-Bulkley type and the vessel to be
of varying cross-section. It is worthwhile to mention that although
flow through axisymmetric tubes is qualitatively almost similar to the
case of flow in channels, magnitude of different physiological
quantities associated with flow and pressure differ appreciably. Since
for blood flow in the micro-circulatory system, the Reynolds number is
low and since the ratio between the radius of the tube and the wave
length is small, the theoretical analysis has been performed by using
the lubrication theory \cite{Shapiro}. Based upon the analytical
study, extensive numerical calculations have been made. Keeping a
specific situation of micro-circulation in view, the pumping
performance has been investigated numerically. The computational
results are presented for the velocity distribution of blood, the wall
shear stress distribution as well as the streamline pattern and
trapping. Influence of SSD wave front on various flow variables
concerned with peristaltic transport has been discussed. The plots
give a clear view of the qualitative variation of various fluid
mechanical parameters. The results presented for shear thinning and
shear thickening fluids are quite relevant in the context of various
studies of blood rheology. It is important to mention here that normal
blood usually behaves like a shear thinning fluid for which with the
increase in shear rate, the viscosity decreases. As mentioned in
\cite{Fung2,White,Xue}, in the case of hardened red blood cell
suspensions, the fluid behaviour is similar to that of a shear
thickening fluid for which the fluid viscosity is enhanced due to
increase in shear rate.

The study provides some novel information by which we can have a
better insight of blood flows taking place in the micro-circulatory
system. More particularly, it has an important bearing on the clinical
procedure of extra-corporeal circulation of blood through the use of
the heart-lung machine that may result in damage of erythrocytes owing
to high variation of the wall shear stress. In addition, the results
should find useful application in the development of roller pumps and
arthro-pumps by which fluids are transported in living organs.

\section{Formulation}
A non-linear problem concerning the peristaltic motion of blood in a
micro-vessel will be studied here, by considering blood as an
incompressible viscous non-Newtonian fluid. The non-Newtonian viscous
behaviour will be represented by Herschel-Bulkley fluid model. The
vessel will be considered to be of varying cross-sectional
radius. However, it will be considered as an axisymmetric vessel and
the flow that takes place through it as also axisymmetric.

\begin{figure}
\begin{center}
\includegraphics[width=3.in,height=2.0in]{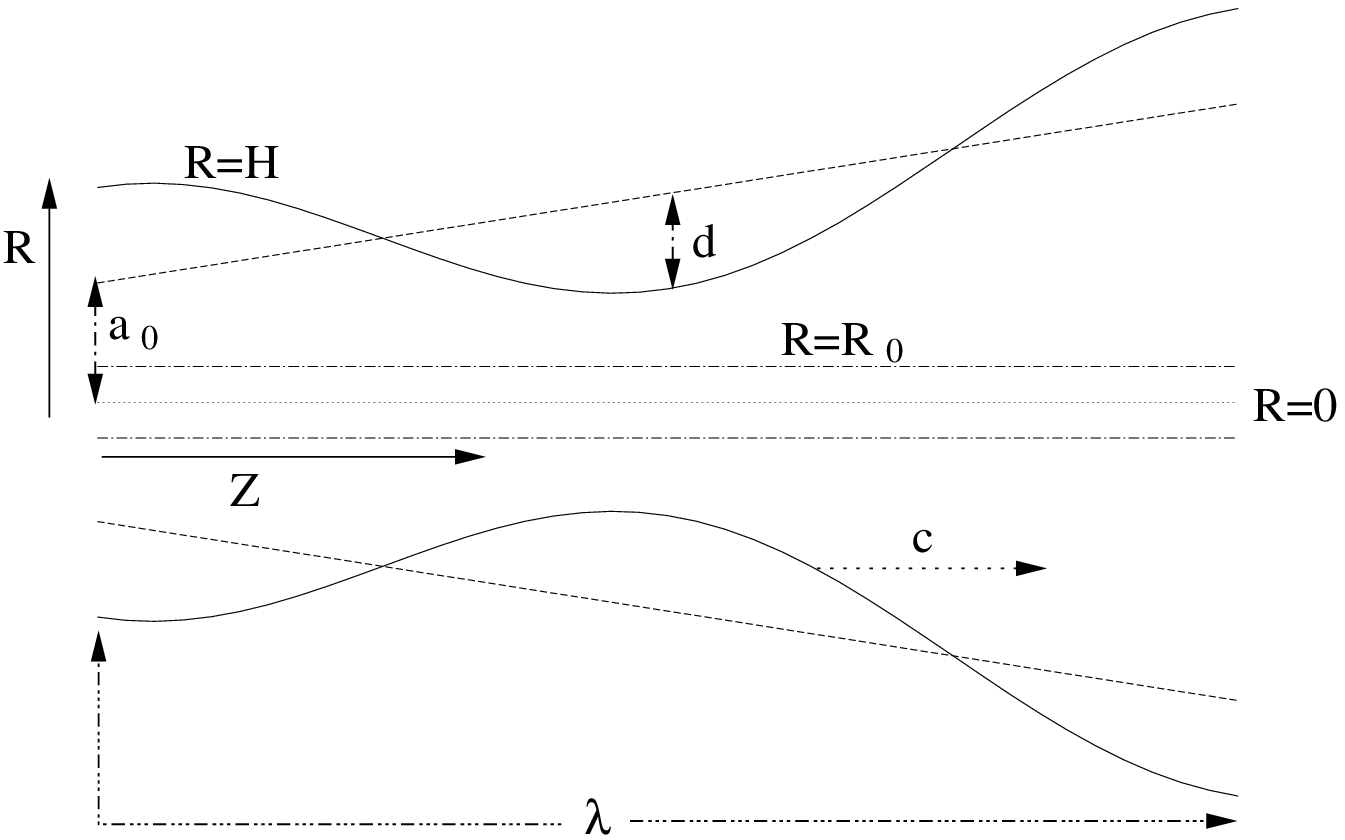}\includegraphics[width=3.in,height=2.0in]{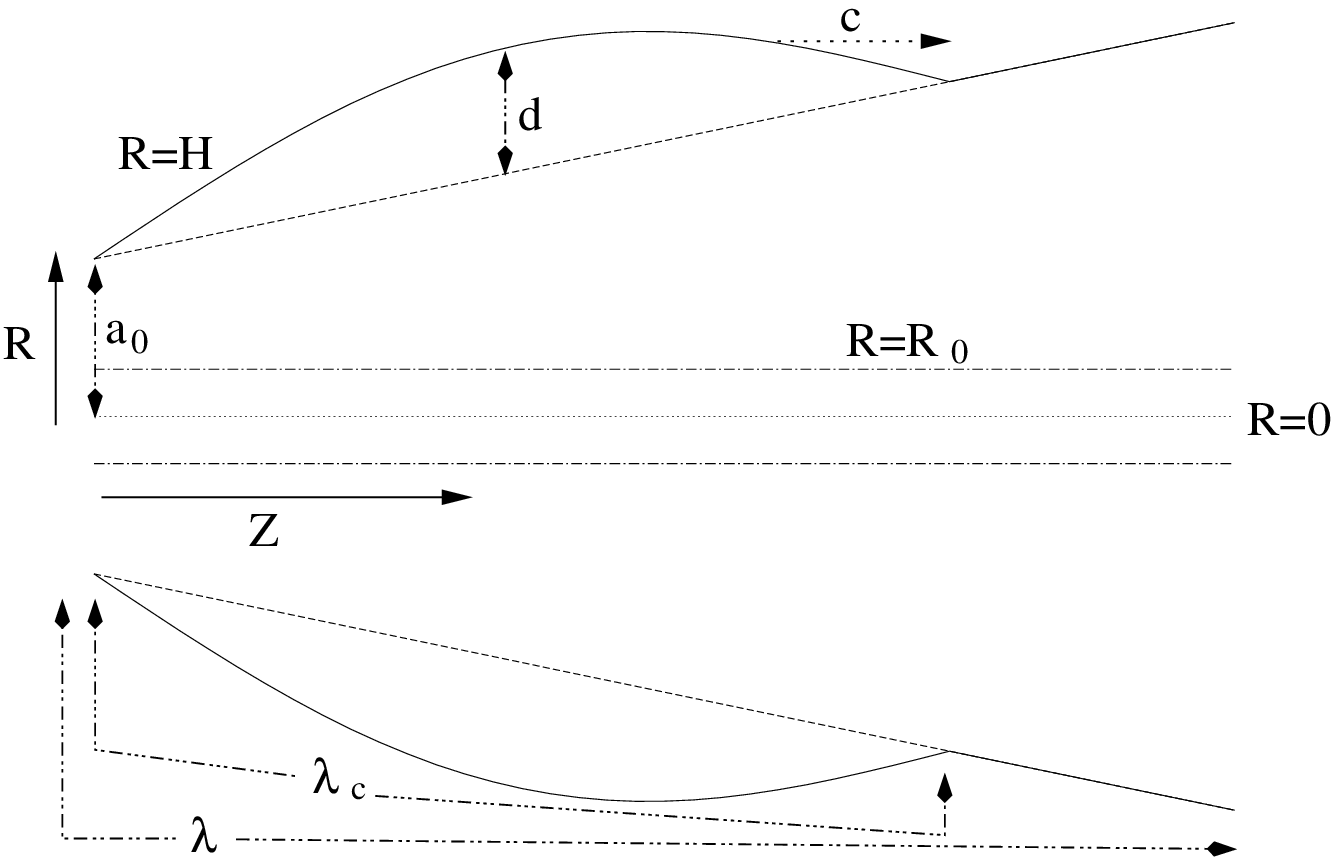}
\\$~~~~~~~~~~~~~~~~~(a)~~~~~~~~~~~~~~~~~~~~~~~~~~~~~~~~~~~~~~~~~~~~~~~~~~~~~~~~~~~~~~~~~(b)~~~~~~~~~~~~~~~~~~~~~~~~$
\includegraphics[width=3.in,height=2.0in]{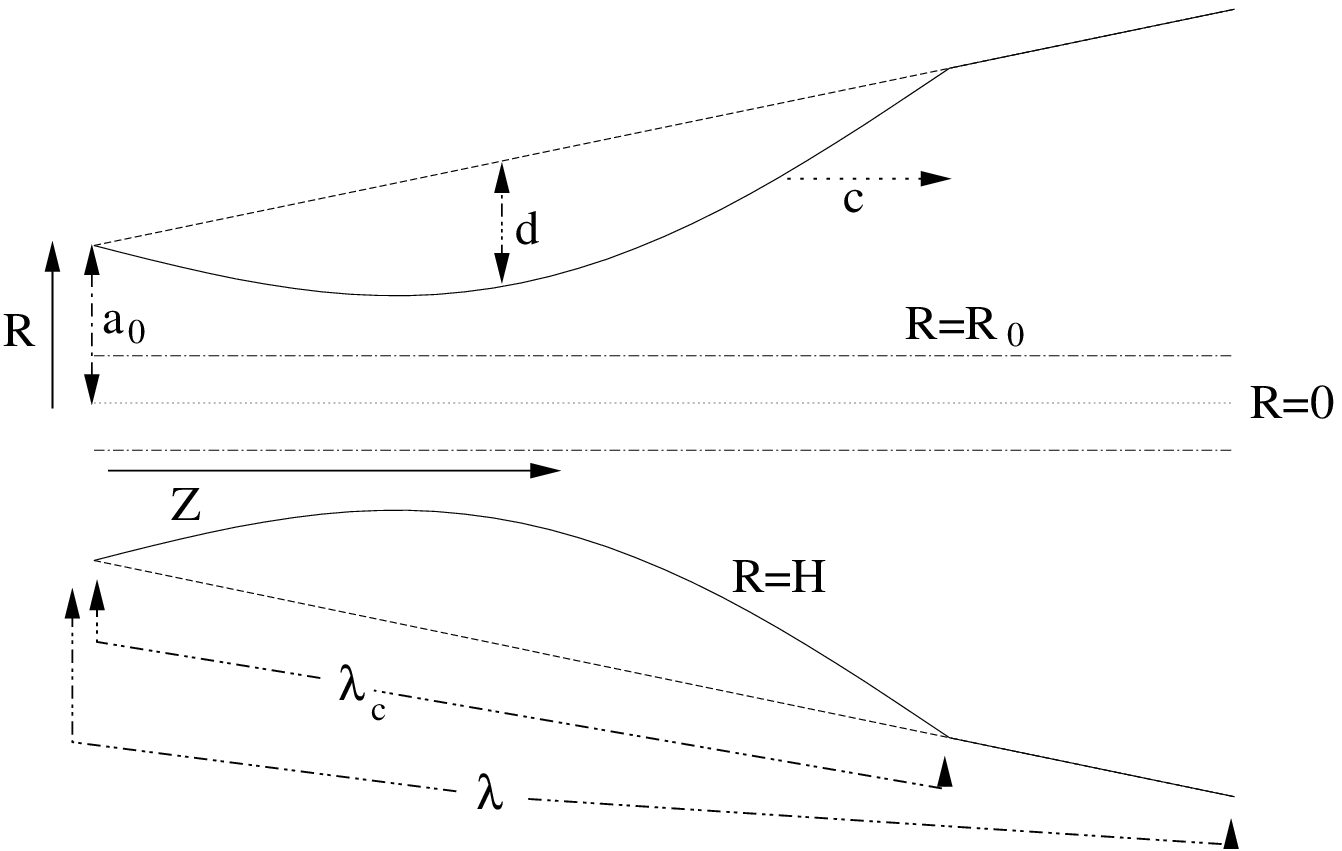}
\\(c)
\caption{A physical sketch of the problem for a tapered vessel in the case of (a)
  Sinusoidal wave, (b) SSD expansion wave and (c) SSD contraction wave}
\label{cnsns_manuscript_geo5.1}
\end{center}
\end{figure}

We take (R,$\theta$,Z) as the cylindrical coordinates of the
location of any fluid particle, Z being measured in the direction of
wave propagation. $ R=H$ (cf. Fig. \ref{cnsns_manuscript_geo5.1})
represents the wall of the micro-vessel and the flow is supposed to be
induced by either a progressive sinusoidal wave or an SSD expansion/contraction wave
train propagating with a constant speed $c$ travelling
down the wall, such that
\begin{eqnarray}
H=a(Z)+d~\sin\left(\frac{2\pi}{\lambda}(Z-ct)\right),
\label{cnsns_manuscript_sinusoidal_wave}\\
H=\left\{\begin{array}{r@{\quad : \quad}l}a(Z)+
d~\sin\left(\frac{\pi}{\lambda_c}(Z-ct)\right)& if~~ 0\le Z-ct\le
\lambda_c\\ a(Z) & if~~ \lambda_c\le Z-ct\le \lambda, \\
\end{array} \right.
\label{cnsns_manuscript_SSD_expansion_wave}\\
H=\left\{\begin{array}{r@{\quad : \quad}l}a(Z)-
d~\sin\left(\frac{\pi}{\lambda_c}(Z-ct)\right)& if~~ 0\le Z-ct\le
\lambda_c
\\ a(Z) & if~~\lambda_c\le Z-ct\le \lambda; \\
\end{array} \right.
\label{cnsns_manuscript_SSD_contraction_wave}
\end{eqnarray}
The SSD expansion/contraction waves defined by equations
(\ref{cnsns_manuscript_SSD_expansion_wave}) and
(\ref{cnsns_manuscript_SSD_contraction_wave}) are confined to a
portion of length $\lambda_c$. Let us consider a(Z)=$a_0+k_1$Z, where
$a(Z)$ represents the radius of the vessel at any axial distance Z
from the inlet, $a_0$ is the radius at the inlet and $k_1(<$1) is a
constant whose magnitude depends on the length of the vessel as well
as on the dimensions of the inlet and the exit; $b$ is the wave
amplitude, $t$ is the time variable and $\lambda$ denotes the wave
length.

\section{Analysis}
Using the fixed frame of reference we shall perform the analysis of
the non-linear problem. For the model formulated in the preceding
section, flow of blood in the micro-vessel can be assumed to be
governed by the partial differential equations
\begin{eqnarray}
\nabla \cdot{\bf V}=0
\end{eqnarray}
\begin{eqnarray}
\rm{and~~~~}\rho \frac{d{\bf V}}{dt}=\nabla\sigma+ \rho f,
\end{eqnarray}
where ${\bf V}$ is the velocity, f the body force per unit mass,
$\rho$ the fluid density and $\frac{d}{dt}$ the material
time-derivative. $\sigma$ represents the Cauchy stress tensor defined
by\\ $~~~~~~~~~~~~~~~~~~~~~~~~~~~~~~~~~~~~~~~~~~~~~~~~~~~~~\sigma=-PI+T$,\\
 $~~~~~~~~~~~~~~~~~~~~~~~~~~~~~~~~~~~~~~~~~~~\rm{in~which}~~T=2\mu
E_{ij}+\eta I S$,\\
$~~~~~~~~~~~~~~~~~~~~~~~~~~~~~~~~~~~~~~~~~~~~~~~\rm{and}~~S=\nabla \cdot
{\bf V}$,\\ $E_{ij}$ being the symmetric part of the velocity
gradient, defined
by\\$~~~~~~~~~~~~~~~~~~~~~~~~~~~~~~~~~~~~~~~~~~~~~~~~~~~~E_{ij}
=\frac{1}{2}[L+L^T]$,\\$~~~~~~~~~~~~~~~~~~~~~~~~~~~~~~~~~~~~~~~~~~~~~~~~\rm{where}~~L=\nabla{\bf
V}$. \\ $-PI$ represents the
indeterminate part of the stress due to the constraint of
incompressibility, while $\mu$ and $\eta$ denote viscosity
parameters.

As mentioned in Sec. 1, blood in micro-vessels is considered in the
present study as a Herschel-Bulkley fluid \cite{Herschel,Sahu}. It
may be mentioned that Herschel-Bulkley model is the representative
of the combined effect of Bingham plastic and power-law behavior of
the fluid. When strain-rate $\dot{\gamma}$ is low such that
$\dot{\gamma}<\frac{\tau_0}{\mu_0}$, the fluid behaves like a
viscous fluid with constant viscosity $\mu_0$. But as the strain
rate increases and the yield stress threshold ($\tau_0$) is
reached, the fluid behavior is better described by a power law of the form
\begin{eqnarray*}
\mu=\frac{\tau_0+\alpha\left\{\dot{\gamma}^n-\left(\frac{\tau_0}{\mu_0}\right)^n\right\}}{\dot{\gamma}},
\end{eqnarray*}
in which $\alpha$ and n denote respectively the consistency factor
and the power law index. $n<1$ corresponds to a shear thinning
fluid, while for a shear thickening fluid $n>1$.

In the case of a uniformly circular vessel, if the length of the
vessel is an integral multiple of the wavelength and the pressure
difference between the ends of the vessel is a constant, the flow is
steady in the wave frame. Since in the present study, the geometry
of wall surface is considered non-uniform, the flow is inherently
unsteady in the laboratory frame as well as in the wave frame of
reference. Disregarding the body forces (i.e. taking f=0) the
Herschel-Bulkley equations that are being considered here as the
governing equations of the incompressible fluid motion in the
micro-vessel, in the fixed frame of reference may be put in the form

\begin{equation}
\rho \left (\frac{\partial U}{\partial t}+U\frac{\partial
U}{\partial Z}+V\frac{\partial U}{\partial R}\right
)=-\frac{\partial P}{\partial Z}+\frac{1}{R}\frac{\partial
  (R\tau_{RZ})}{\partial R}+\frac{\partial \tau_{ZZ}}{\partial Z}
\end{equation}
\begin{equation}
\rho\left (\frac{\partial V}{\partial t}+U\frac{\partial V}{\partial
Z}+V\frac{\partial V}{\partial R}\right )=-\frac{\partial
P}{\partial R}+\frac{1}{R}\frac{\partial
  (R\tau_{RR})}{\partial R}+\frac{\partial \tau_{RZ}}{\partial Z}
\end{equation}
\begin{eqnarray}
with ~\tau_{ij}=2\mu E_{ij}=\mu \left(\frac{\partial U_i}{\partial
  X_j}+\frac{\partial U_j}{\partial X_i}\right)~,\\
\mu=\left\{\begin{array}{r@{\quad : \quad}l}\mu_0~ & for~\Pi\le\Pi_0,
\\ \alpha\Pi^{n-1}+\tau_0\Pi^{-1} & for~\Pi\ge\Pi_0
\end{array} \right.\\
\Pi=\sqrt{2E_{ij}E_{ij}}
\end{eqnarray}
The limiting viscosity $\mu_0$ is considered such that
\begin{equation}
\mu_0=\alpha\Pi_0^{n-1}+\tau_0\Pi_0^{-1}.
\label{cnsns_manuscript_mu_0}
\end{equation}
The following non-dimensional variables will be introduced in the
analysis that follows:
\begin{eqnarray}
\bar{Z}=\frac{Z}{\lambda},~~\bar{R}=\frac{R}{a_0},~~\bar{U}=\frac{U}{c},~\bar{V}=\frac{V}{c\delta},~
\delta=\frac{a_0}{\lambda},~\bar{P}=\frac{a_0^{n+1}P}{\alpha
  c^n\lambda},~\bar{t}=\frac{ct}{\lambda},
~h=\frac{H}{a_0},~\phi=\frac{d}{a_0},\nonumber\\ Re=\frac{\rho
  a_0^n}{\alpha
  c^{n-2}},~\bar{\lambda}_c=\frac{\lambda_c}{\lambda},~\bar{\tau}_0=\frac{\tau_0}{\alpha\left(\frac{c}{a_0}\right)^n},~\bar{\tau}_{RZ}=\frac{\tau_{RZ}}{\alpha\left(\frac{c}{a_0}\right)^n},~\bar{\Psi}=\frac{\Psi}{a_0
  c}.~~~~~~~~
\label{cnsns_manuscript_non-dimensionalize}
\end{eqnarray}
The equation governing the  flow of the fluid  can now be rewritten in the form (dropping the bars over the symbols)
\begin{equation}
 Re\delta \left (\frac{\partial U}{\partial t}+U\frac{\partial
U}{\partial Z}+V\frac{\partial U}{\partial R}\right
)=-\frac{\partial P}{\partial Z}+\frac{1}{R}\frac{\partial
  \left(\Phi\left(R\frac{\partial U}{\partial R}+R\delta^2\frac{\partial V}{\partial
Z}\right)\right)}{\partial R}+2\delta^2\frac{\partial \left(\Phi\frac{\partial
U}{\partial Z}\right)}{\partial Z}
\end{equation}
\begin{equation}
Re\delta^3\left (\frac{\partial V}{\partial t}+U\frac{\partial V}{\partial
Z}+V\frac{\partial V}{\partial R}\right )=-\frac{\partial
P}{\partial R}+\delta^2\frac{1}{R}\frac{\partial
  \left(R\Phi\frac{\partial V}{\partial R}\right)}{\partial R}+\delta^2\frac{\partial\left(\Phi\left(\frac{\partial U}{\partial R}+\delta^2\frac{\partial V}{\partial
Z}\right)\right)}{\partial Z}
\end{equation}
\begin{eqnarray}
where~~~\Phi=\left|\sqrt{2\delta^2\left\{\left(\frac{\partial V}{\partial R}\right)^2+\left(\frac{V}{R}\right)^2+\left(\frac{\partial
U}{\partial Z}\right)^2\right\}+\left(\frac{\partial U}{\partial R}+\delta^2\frac{\partial V}{\partial
Z}\right)^2}\right|^{n-1}\nonumber\\+\tau_0\left|\sqrt{2\delta^2\left\{\left(\frac{\partial V}{\partial R}\right)^2+\left(\frac{V}{R}\right)^2+\left(\frac{\partial
U}{\partial Z}\right)^2\right\}+\left(\frac{\partial U}{\partial R}+\delta^2\frac{\partial V}{\partial
Z}\right)^2}\right|^{-1}
\end{eqnarray}
We now use the long wavelength approximation ($\delta\ll 1$) and the
lubrication approach \cite{Shapiro,Mishra}. Then the governing
equations stated earlier describing the flow in the laboratory frame
of reference in terms of the dimensionless variables
(\ref{cnsns_manuscript_non-dimensionalize}), assume the form
\begin{equation}
0=-\frac{\partial P}{\partial Z}+ \frac{1}{R}\frac{\partial
  (R\frac{\partial U}{\partial R}|\frac{\partial U}{\partial
    R}|^{n-1}+\tau_0)}{\partial R}
\label{cnsns_manuscript_Z_momentum_lubrication}
\end{equation}
\begin{equation}
\rm{and~~~}0=-\frac{\partial P}{\partial R}.
\end{equation}
Also the form of the boundary conditions will now be
 \begin{eqnarray}
 \Psi=0,~\frac{\partial U}{\partial R}=\frac{\partial
   \left(\frac{1}{R}\frac{\partial \Psi}{\partial R}\right)}{\partial
   R}=0,~ \tau_{RZ}=0~ at~ R=0;~~
\label{cnsns_manuscript_non-dimensional_boundary-condition_central}\\\rm{and~~~}U=\frac{1}{R}\frac{\partial
  \Psi}{\partial R}=0~ at~ R=h.~~~~~~~~~~~~~~~~~~~~~~~~~~~~~~~
\label{cnsns_manuscript_non-dimensional_boundary-condition_wall}
\end{eqnarray}
The solution of equation (\ref{cnsns_manuscript_Z_momentum_lubrication}) subject to the conditions (\ref{cnsns_manuscript_non-dimensional_boundary-condition_central}) and (\ref{cnsns_manuscript_non-dimensional_boundary-condition_wall}) is
found to be given by
\begin{equation}
U(R,Z,t)=\frac{1}{(k+1)P_1}\left[(P_1h-\tau_0)^{k+1}-(P_1R-\tau_0)^{k+1}\right],
0\le R\le h
\label{cnsns_manuscript_axial_velocity}
\end{equation}
where~$P_1=-\frac{1}{2}\frac{\partial P}{\partial Z} \rm{~~and~}~k=\frac{1}{n}$.
\\If $R_0$ be the radius of plug flow region (where $\mu_0=\infty$),
\begin{eqnarray*}
\frac{\partial U}{\partial R} =0~~at~R=R_0.
\end{eqnarray*}
It follows from (\ref{cnsns_manuscript_axial_velocity}) that
\begin{eqnarray*}
R_0=\tau_0/P_1.
\end{eqnarray*}
If $\tau_{RZ}=\tau_h$ at R=h, we find $h=\tau_h/P$.
\begin{eqnarray}
Thus~~~\frac{R_0}{h}=\frac{\tau_0}{\tau_h}=\tau~(say),~~0<\tau<1.~~~~~~~~~~~~
\end{eqnarray}
Then the expression of the plug velocity turns out to be
\begin{eqnarray}
U_P=\frac{(P_1h-\tau_0)^{k+1}}{(k+1)P_1}.~~~~~~~~~~~~~~~~~~~~~~~~~~~~~
\label{cnsns_manuscript_axial_plug_velocity}
\end{eqnarray}
In order to determine the stream function $\Psi$ we use the boundary
conditions\\ $~~~~~~~~~~~~~~~~~~~~~~~~~~~~~~~~~~\Psi_P=0~~at~R=0~\\~~~~~~~~~~~~~~~~~~~~~~~~~~~~and~\Psi=\Psi_P~at~R=R_0$.
\\Integrating (\ref{cnsns_manuscript_axial_velocity}) and (\ref{cnsns_manuscript_axial_plug_velocity}), the stream function is found in the form
 \begin{eqnarray}
\Psi=\frac{P_1^k}{k+1}\left[\frac{R^2}{2}(h-R_0)^{k+1}-\frac{(R-R_0)^{k+2}((k+2)R+R_0)}{(k+2)(k+3)}\right],~~~~
R_0\le R \le h
\label{cnsns_manuscript_stream_function}
\end{eqnarray}
\begin{eqnarray}
\rm{and}~~\Psi_P=\frac{P_1^k(h-R_0)^{k+1}R^2}{2(k+1)},~~~~~~0\le R \le R_0
\end{eqnarray}
The instantaneous rate of volume flow through the micro-vessel, $\bar{Q}$ is given by
\begin{eqnarray}
\bar{Q}(Z,t)=2\int_{0}^{R_0}RUdR+2\int_{R_0}^{h}RUdR~~~~~~~~~~~~~\nonumber~~~\\=\frac{P_1^k(h-R_0)^{k+1}(h^2(k^2+3k+2)+2R_0h(k+2)+2R_0^2)}{(k+1)(k+2)(k+3)},~~k=\frac{1}{n}.
\label{cnsns_manuscript_volume_flow_rate}
\end{eqnarray}
From (\ref{cnsns_manuscript_volume_flow_rate}), we have
\begin{eqnarray}
P_1=-\frac{1}{2}\frac{\partial P}{\partial
  Z}=\left[\frac{\bar{Q}(Z,t)(k+1)(k+2)(k+3)}{(h-R_0)^{k+1}(h^2(k^2+3k+2)+2R_0h(k+2)+2R_0^2)}\right]^n~~~~~~~~~~~\nonumber~~~~\\=\left[\frac{\bar{Q}(Z,t)(k+1)(k+2)(k+3)}{h^{k+3}(1-\tau)^{k+1}((k^2+3k+2)+2\tau
    (k+1)+2\tau^2)}\right]^n.~~~~~~~~~~~~~~~
\label{cnsns_manuscript_pressure_gradient}
\end{eqnarray}
The average pressure difference per wave length can now be
calculated by using the formula
\begin{eqnarray}
 \Delta P=\int_{0}^{1}\int_{0}^{1}\left(\frac{\partial P}{\partial
   Z}\right)dZdt~.
\label{cnsns_manuscript_pressure_difference_wave_length}
\end{eqnarray}
We observe that when $n=1,~\tau=0~and~k_1=0$, the equations
(\ref{cnsns_manuscript_axial_velocity}),
(\ref{cnsns_manuscript_stream_function}),
(\ref{cnsns_manuscript_volume_flow_rate}) and
(\ref{cnsns_manuscript_pressure_gradient}) reduce to those obtained by
Shapiro et al. \cite{Shapiro} for a Newtonian fluid. Our results also match with those of
Lardner and Shack \cite{Lardner}, when the eccentricity of the
elliptical motion of cilia tips is set equal to zero in their
analysis for a Newtonian fluid flowing on a uniform channel. Also,
for $n=1~and~R_0=0$, equation
(\ref{cnsns_manuscript_pressure_gradient}) tallies with that obtained
by Gupta and Seshadri \cite{Gupta} for a Newtonian fluid of constant
viscosity.

To obtain the quantitative results, the instantaneous rate of volume
flow $\bar{Q}(Z,t)$ has been considered to be periodic in (Z-t)
\cite{Gupta,Srivastava3}, so that $\bar{Q}$ appearing in equation
(\ref{cnsns_manuscript_pressure_gradient}) can be represented as
\begin{eqnarray}
\bar{Q}^n(Z,t)=\left\{\begin{array}{r@{\quad :
\quad}l}Q^n+\phi^2\sin^2 2\pi(Z-t)+\frac{2\phi(1+k_1\lambda Z)\sin 2\pi(Z-t)}{a_0}-\frac{\phi^2}{2}~ & for~sinusoidal~wave,~
\\ Q^n+h^2-1-\frac{\lambda_c\phi^2}{2}-\frac{4\lambda_c\phi}{\pi} & for~SSD~expansion~wave,~ \\
Q^n+h^2-1-\frac{\lambda_c\phi^2}{2}+\frac{4\lambda_c\phi}{\pi} ~&
for~SSD~contraction~wave.~
\end{array} \right.
\label{cnsns_manuscript_time_average_flow_flux_relation}
\end{eqnarray}
In (\ref{cnsns_manuscript_time_average_flow_flux_relation}), $Q$ represents the time-averaged flow flux. Since the right hand side of
equation (\ref{cnsns_manuscript_pressure_difference_wave_length})
cannot be integrated in closed form, for non-uniform/uniform
geometry, for further investigation of the problem under
consideration, we had to resort to the use of software Mathematica.
This helped us calculate the numerical estimate of the pressure
difference, $\Delta P$ given by
(\ref{cnsns_manuscript_pressure_difference_wave_length}).

\section{Quantitative Investigation}

\begin{figure}
\begin{center}
\includegraphics[width=3.5in,height=2.0in]{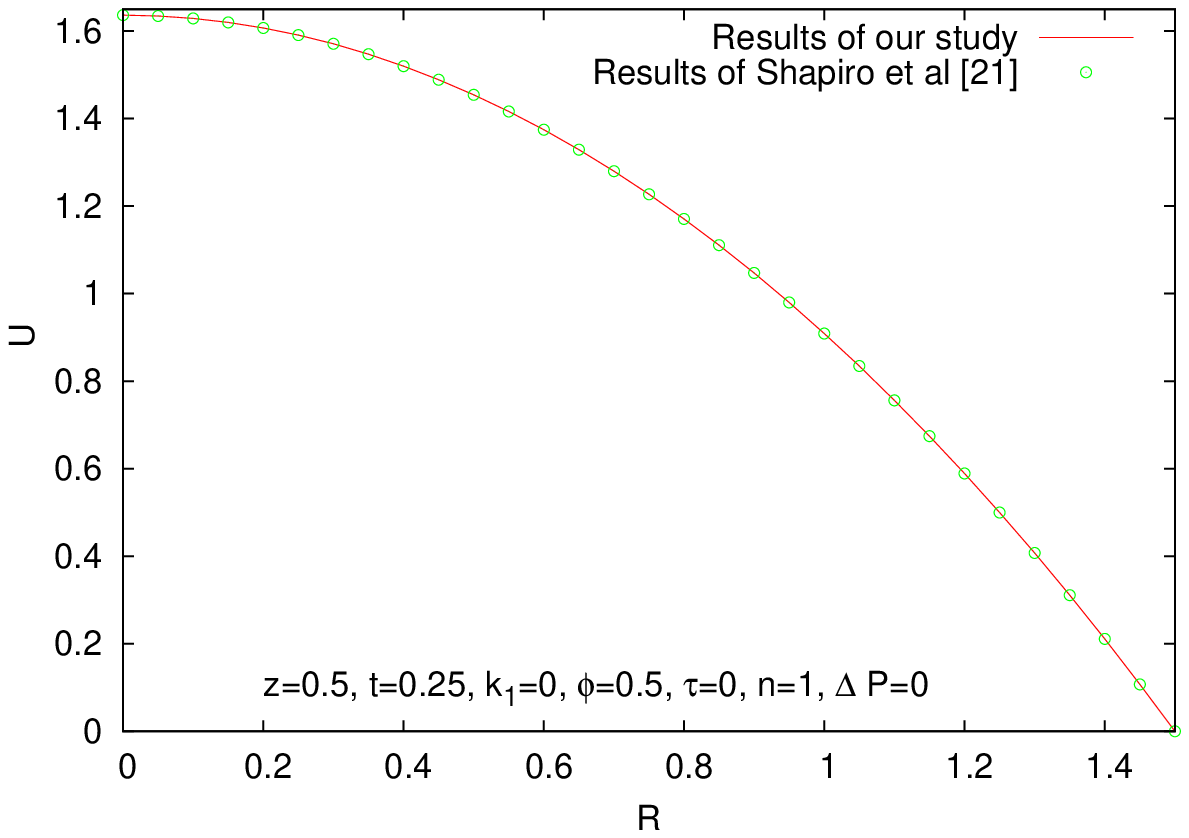}\includegraphics[width=3.5in,height=2.0in]{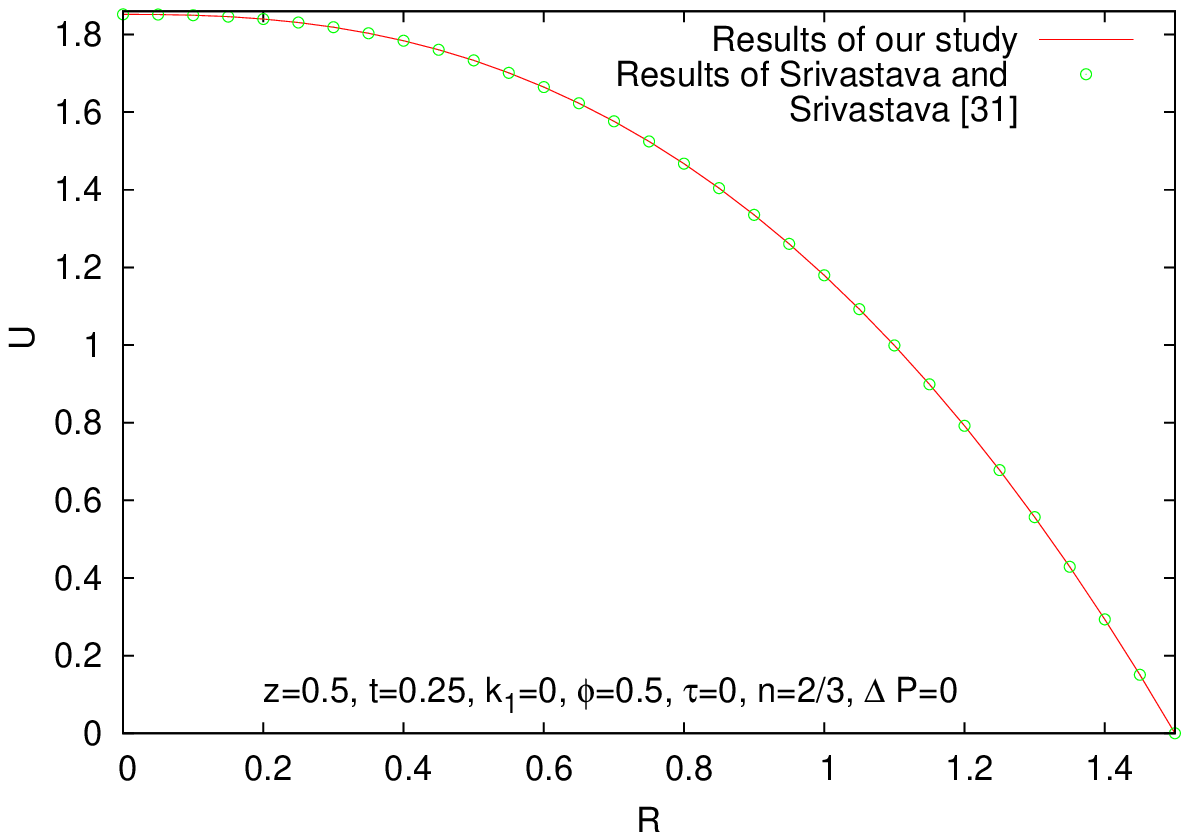}
\\$~~~~~~~~~~~~~~~~~(a)~~~~~~~~~~~~~~~~~~~~~~~~~~~~~~~~~~~~~~~~~~~~~~~~~~~~~~~~~~~~~~~~~~~(b)~~~~~~~~~$
\caption{Variation of axial velocity
in the radial direction at Z=0.5, (a) Newtonian fluid, (b) Shear-thinning fluid}
\label{cnsns_manuscript_velocompare5.1.1}
\end{center}
\end{figure}

\begin{figure}
\includegraphics[width=3.5in,height=2.0in]{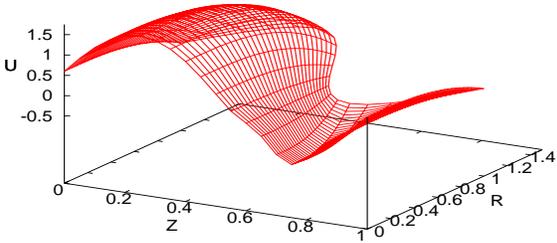}\includegraphics[width=3.5in,height=2.0in]{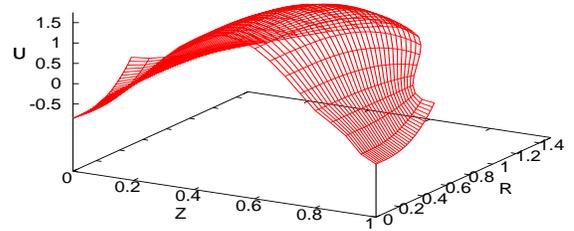}\\$~~~~~~~~~~~~~~~~~(a)~~~~~~~~~~~~~~~~~~~~~~~~~~~~~~~~~~~~~~~~~~~~~~~~~~~~~~~~~~~~~~~~~~~(b)~~~~~~~~~$
\includegraphics[width=3.5in,height=2.0in]{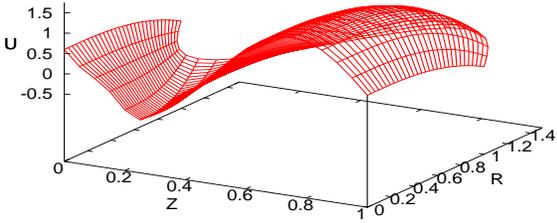}\includegraphics[width=3.5in,height=2.0in]{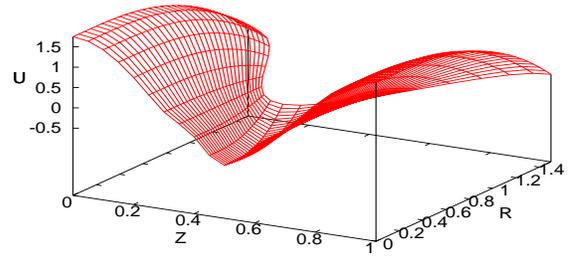}\\$~~~~~~~~~~~~~~~~~(c)~~~~~~~~~~~~~~~~~~~~~~~~~~~~~~~~~~~~~~~~~~~~~~~~~~~~~~~~~~~~~~~~~~~(d)~~~~~~~~~$
\caption{Aerial view of the velocity distribution at different
instants of time, when $n=2/3,~k_1=0,~\Delta P=0,~\tau=0.1,~\phi=0.5$ for a non-Newtonian
 fluid of shear thinning type (a) t=0.0 (b) t=0.25 (c) t=0.5 (d)
 t=0.75}
\label{cnsns_manuscript_velo3D5.1.5-5.1.8}
\end{figure}

Theoretical estimates of different physical quantities that are of
relevance to the physiological problem of blood flow in the
micro-circulatory system have been obtained on the basis of the
present study. For this purpose, the following data that are
valid in the physiological range
\cite{Guyton,Srivastava1,Fung2,Barbee} have been used:\\ $a_0= 10~
to~ 60\mu m$, $\phi=$ 0.1~ to~ 0.9,
$\frac{a_0}{\lambda}=0.01~to~0.02$, $\Delta P =-300~to~50$;~
$\tau=0.0$~ to ~0.2; Q=0~ to~2, n= $\frac{1}{3}$ to 2. The value of
$k_1$ for the non-uniform geometry of the micro-vessel, has been
chosen to match with physiological system. Thus for arterioles,
which are of converging type, the width of the outlet of one wave
length has been taken to be $25\%$ less than that of the inlet; in the case of
diverging vessels (e.g. venules), the width of the outlet of one
wave length has been taken $25\%$ more than that of the inlet.

\begin{figure}
\includegraphics[width=4.5in,height=3.0in]{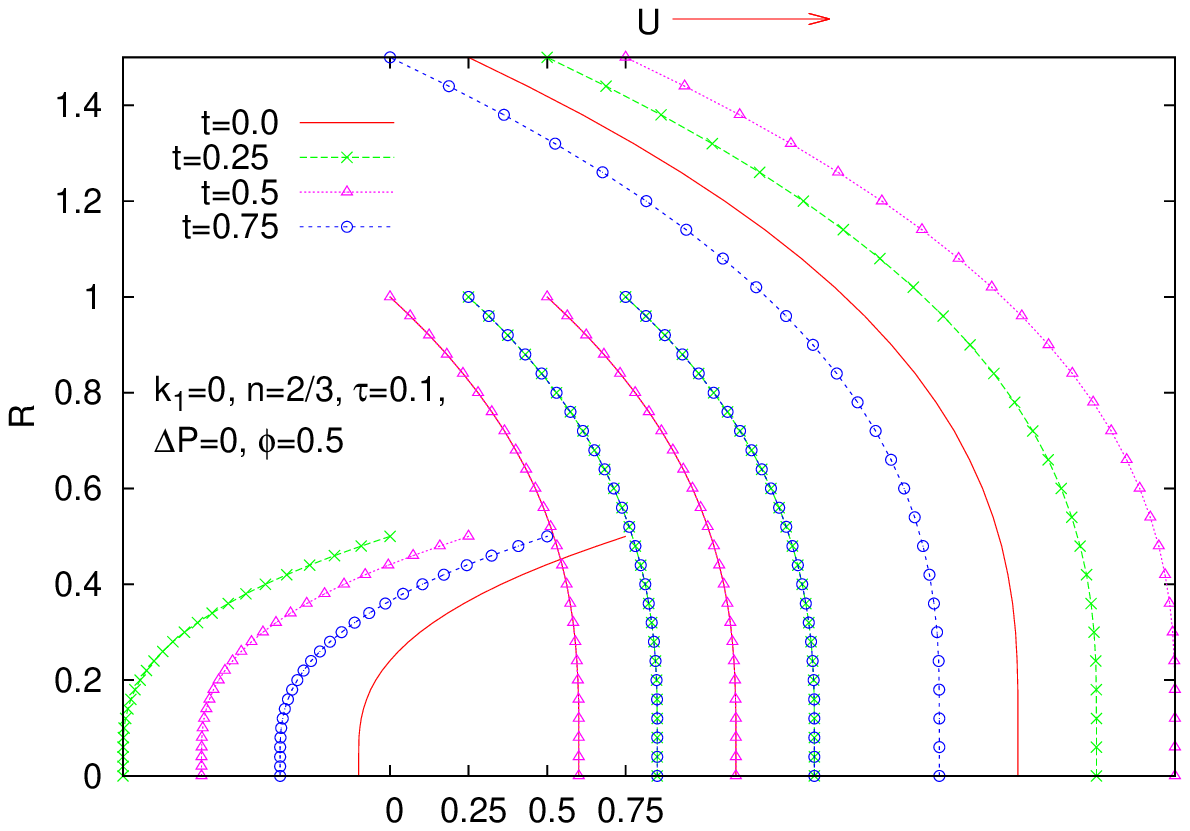}
\caption{Velocity distribution at different instants of time.}
\label{cnsns_manuscript_velo5.1.5}
\end{figure}

\subsection{Distribution of Velocity}
For different values of the amplitude ratio $\phi $, flow index number
n, $k_1$ and $\tau$, Figs. 2-6 present the distribution of axial
velocity of blood in the cases of free pumping, pumping and co-pumping
zones. Fig.  \ref{cnsns_manuscript_velocompare5.1.1} shows that the
results computed on the basis of our study for the particular case of
a Newtonian fluid and shear-thinning fluid ($n=2/3$) tally well with
the results reported by Shapiro et al. \cite{Shapiro} and Srivastava
and Srivastava \cite{Srivastava2}, respectively when the amplitude
ratio $\phi=0.5$. Since the velocity profiles along with the radius of
the blood vessels change with time, we have investigated the
distribution of velocity at a time interval of a quarter of a wave
period. Fig.  \ref{cnsns_manuscript_velo3D5.1.5-5.1.8} gives us the
aerial view of a few typical axial velocity distributions for a
Newtonian fluid and a shear-thinning fluid (n=2/3) flowing in uniform
micro-vessels, while Fig. \ref{cnsns_manuscript_velo5.1.5} presents
the velocity distribution in a vessel at different instants of
time. Fig.  \ref{cnsns_manuscript_velo5.3.1-5.14.1} reveals that at
any instant of time, there exists a retrograde flow region. However,
the forward flow region is predominant in this case, since the
time-averaged flow rate is positive. For a shear-thinning fluid
(n=2/3), the present study indicates that there exist two stagnation
points on the axis. For example, at time t=0.25, one of the stagnation
points lies between Z=0.0 and Z=0.25, while the other lies between
Z=0.75 and Z=1.0. Similar observations were made numerically by
Takabatake and Ayukawa \cite{Takabatake1} for a Newtonian fluid. Figs.
\ref{cnsns_manuscript_velo5.3.1-5.14.1}(a)-(b) depict that in both the
regions, as $\tau$ increases, the magnitude of velocity decreases for
both types of fluids mentioned above. It can be observed from Fig.
\ref{cnsns_manuscript_velo5.3.1-5.14.1}(c-d) that for a
Herschel-Bulkley fluid (with n=2/3) when $\Delta P=0$, the velocity in
both the regions of backward and forward flows is enhanced as the
value of $\phi$ increases in the interval $0<\phi<0.6$, but beyond
$\phi=0.6$, the velocity in the backward flow region decreases with
$\phi$ increasing. These observations
(cf. Figs. \ref{cnsns_manuscript_velo5.3.1-5.14.1}(c,d)) are in
contrast to the case of two dimensional channel flow.

\begin{figure}
 \includegraphics[width=3.3in,height=2.0in]{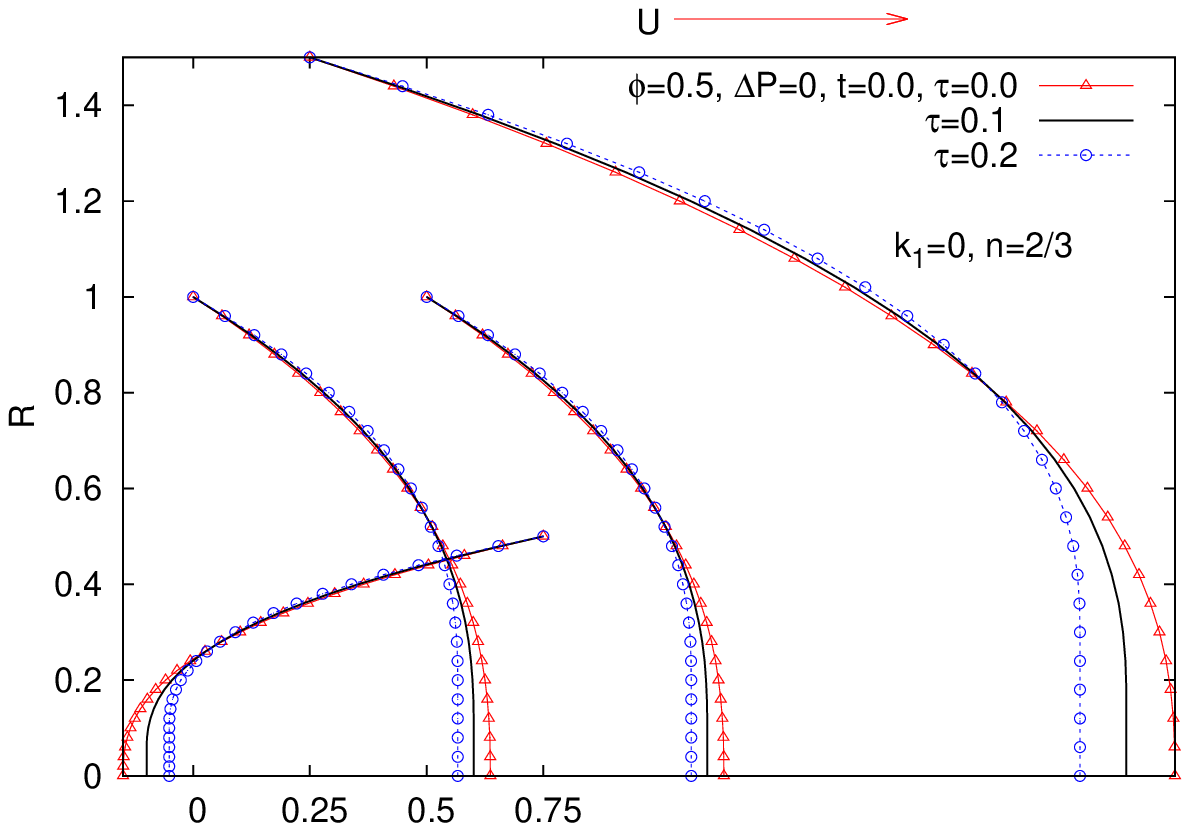}\includegraphics[width=3.3in,height=2.0in]{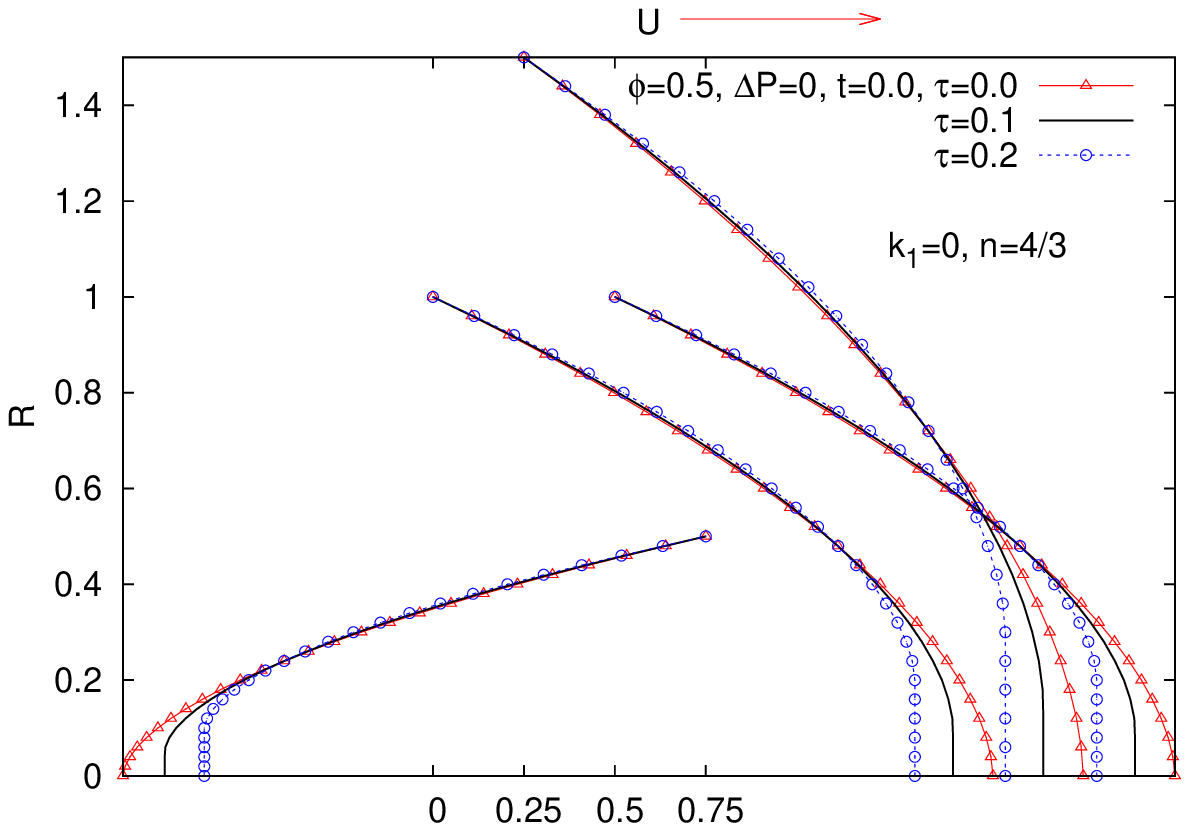}\\$~~~~~~~~~~~~~~~~~~~~~~~~(a)~~~~~~~~~~~~~~~~~~~~~~~~~~~~~~~~~~~~~~~~~~~~~~~~~~~~~(b)~~~~~~~~~~~~~~~$
 \includegraphics[width=3.3in,height=2.0in]{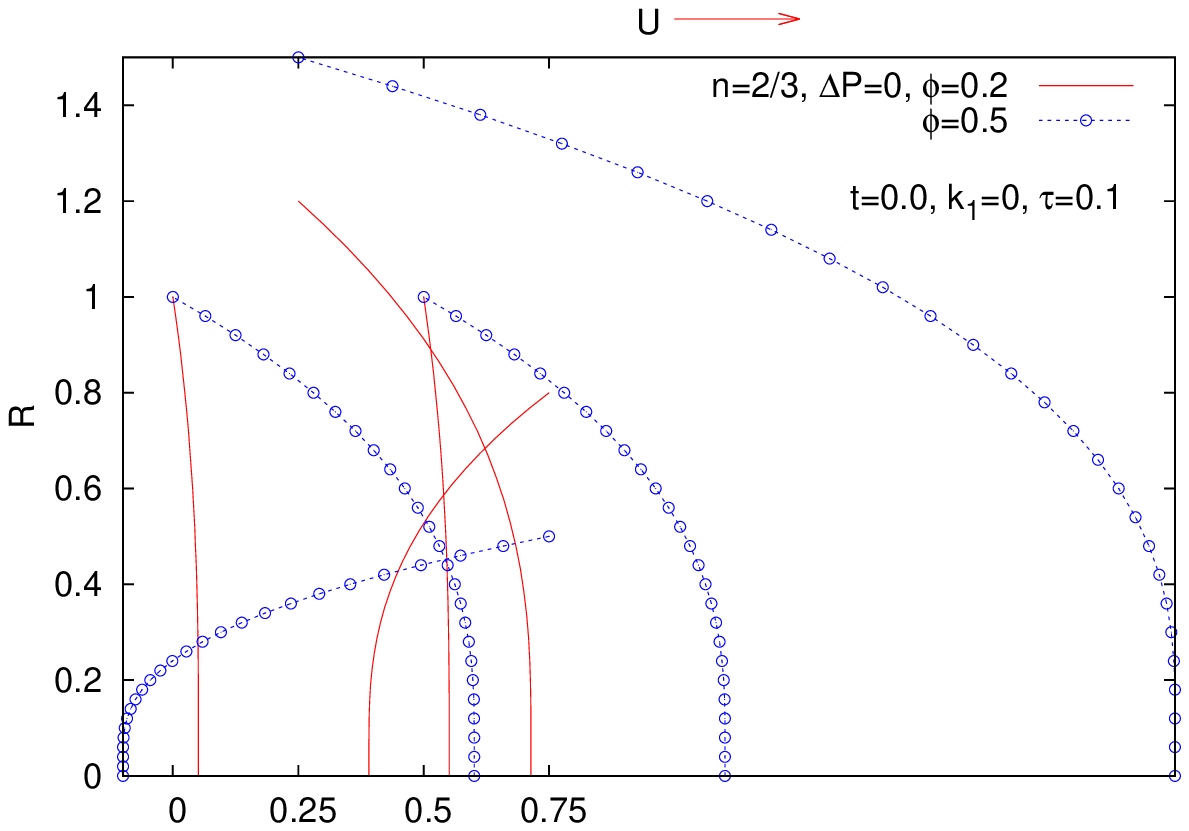}\includegraphics[width=3.3in,height=2.0in]{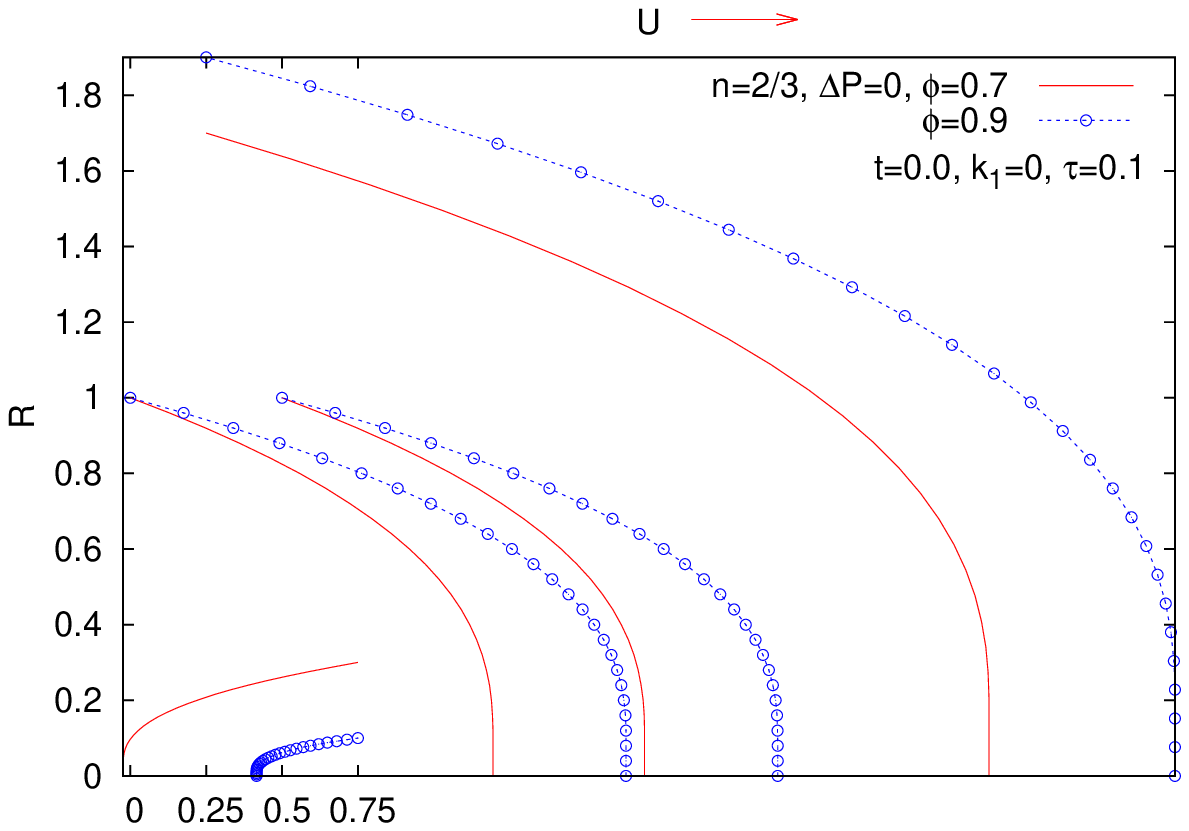}\\$~~~~~~~~~~~~~~~~~~~~~~~~(c)~~~~~~~~~~~~~~~~~~~~~~~~~~~~~~~~~~~~~~~~~~~~~~~~~~~~~(d)~~~~~~~~~~~~~~~$
\includegraphics[width=3.3in,height=2.0in]{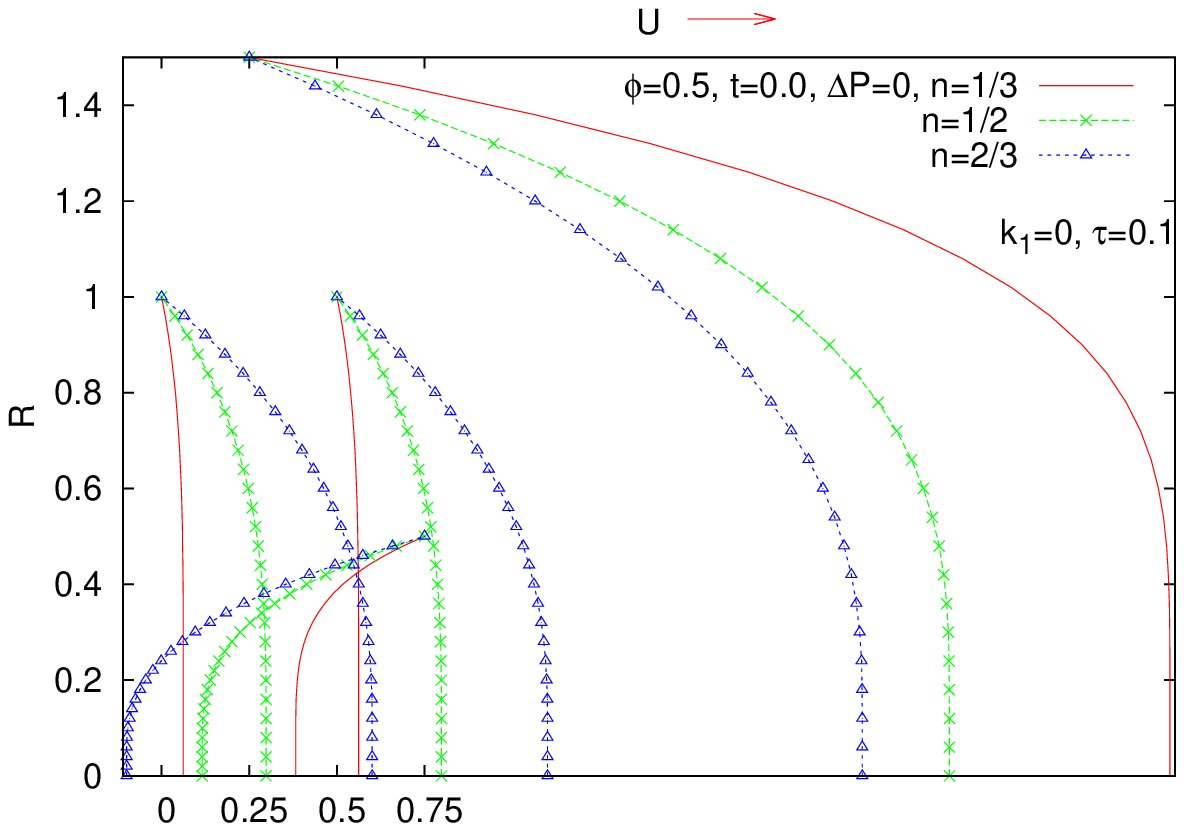}\includegraphics[width=3.3in,height=2.0in]{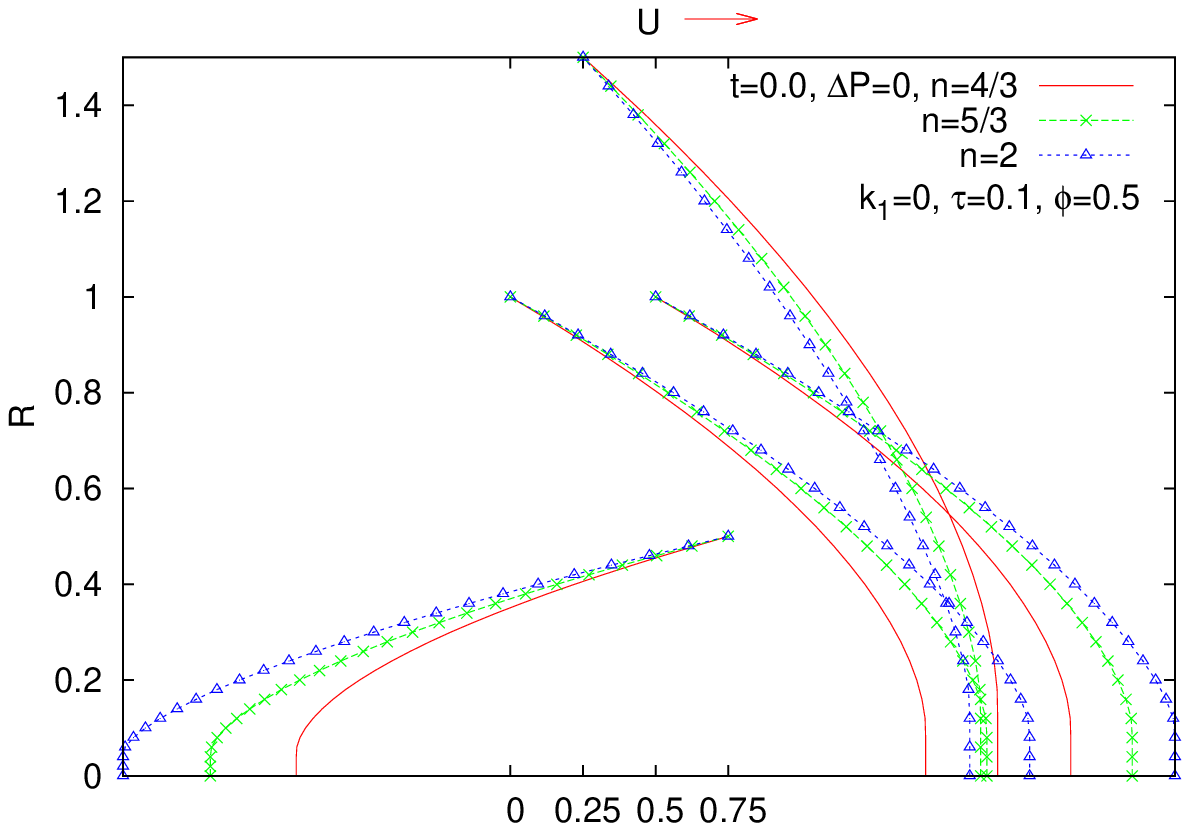}\\$~~~~~~~~~~~~~~~~~~~~~~~~~~~~~~~(e)~~~~~~~~~~~~~~~~~~~~~~~~~~~~~~~~~~~~~~~~~~~~~~~~~~~~~~~~~~~~~~~~~~~(f)~~~~~~~~~~~~~~~$
\end{figure}

\begin{figure}
\includegraphics[width=3.3in,height=2.0in]{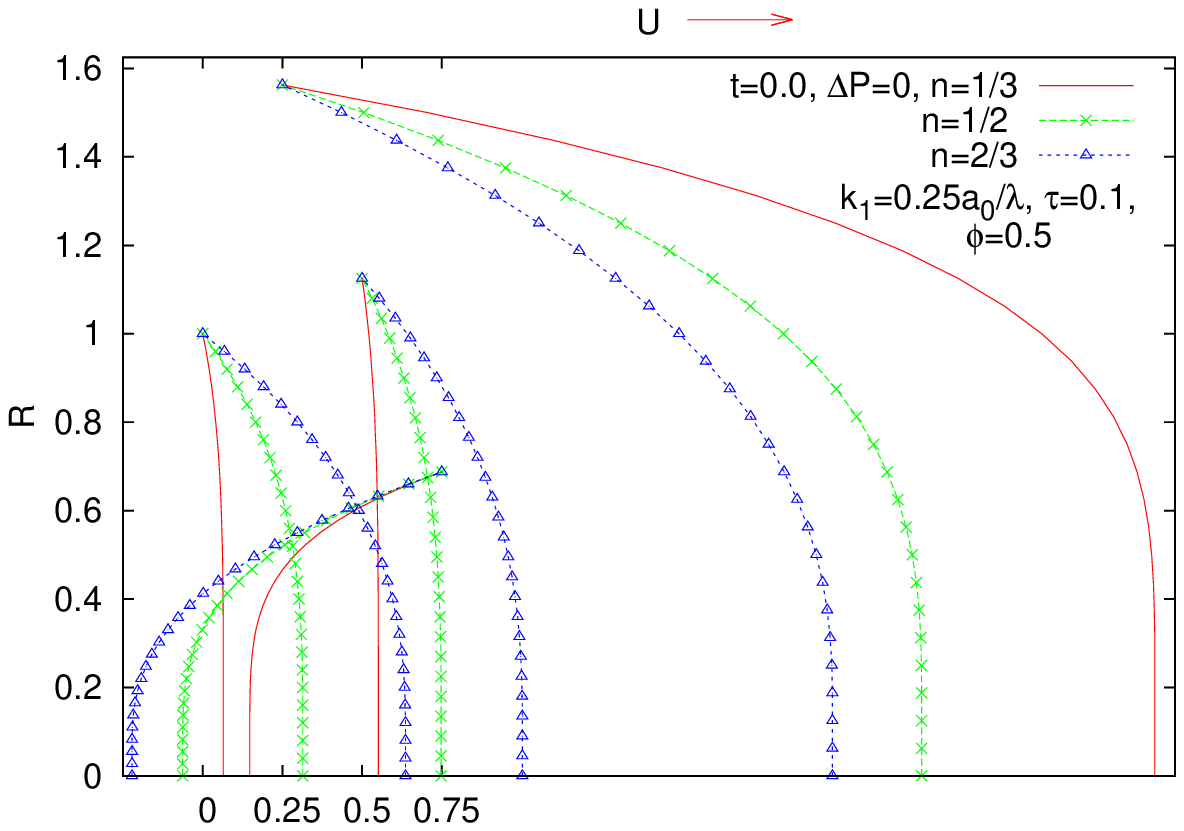}\includegraphics[width=3.3in,height=2.0in]{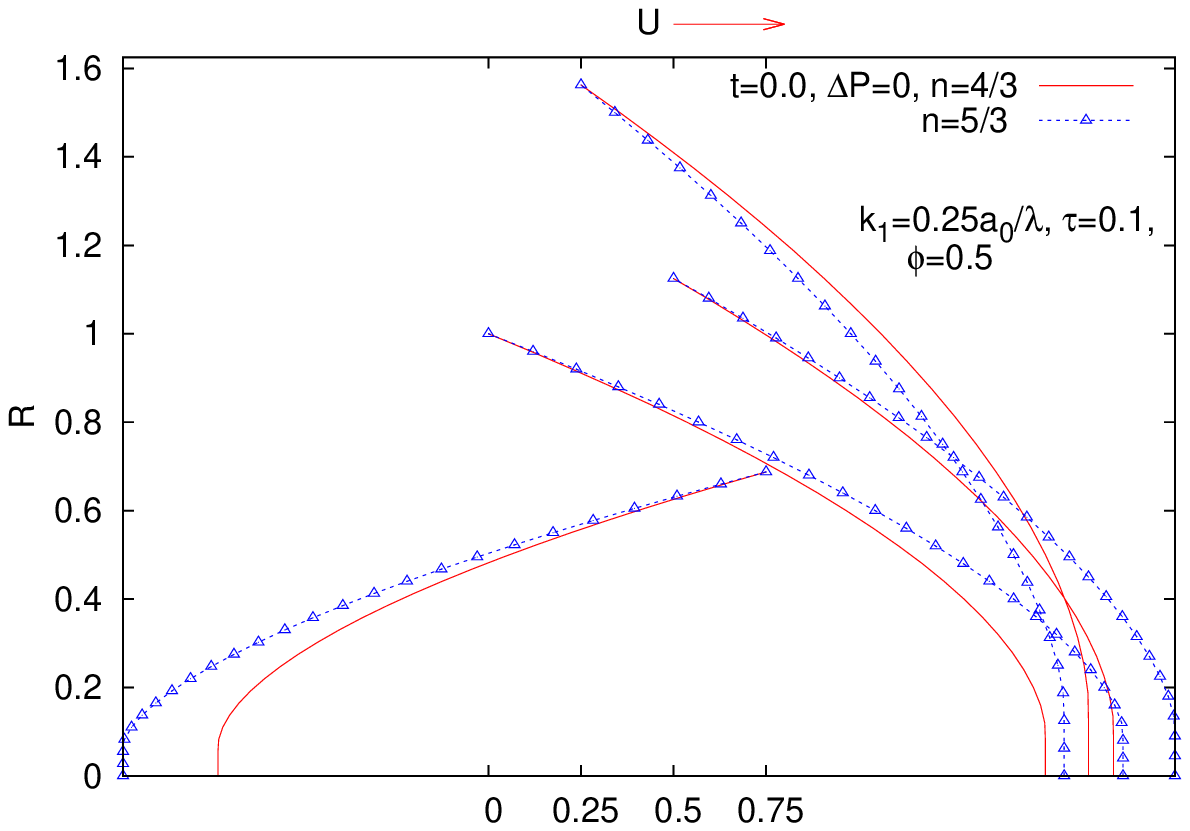}\\$~~~~~~~~~~~~~~~~~~~~~~~~~~~(g)~~~~~~~~~~~~~~~~~~~~~~~~~~~~~~~~~~~~~~~~~~~~~~~~~~~~~~~~~~~~~~~~~~~~~(h)~~~~~~~~~~~~~~~$
\includegraphics[width=3.3in,height=2.0in]{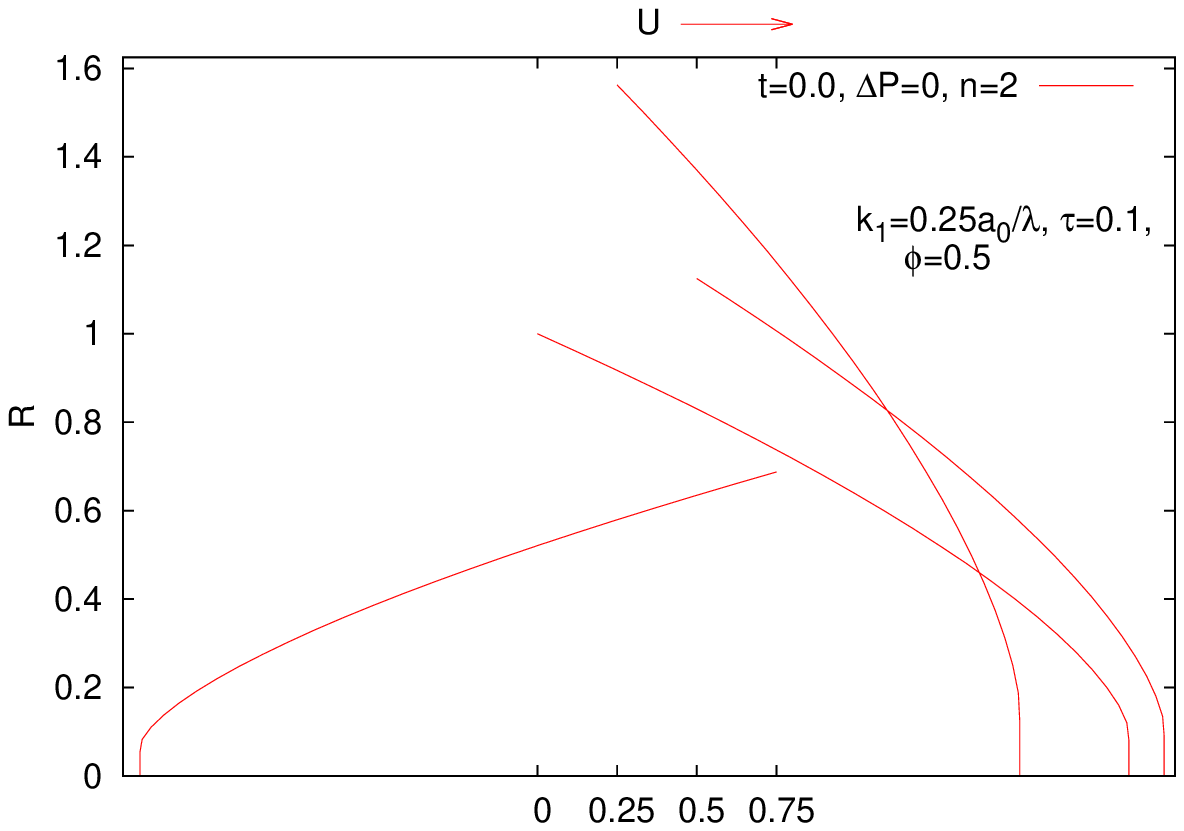}\includegraphics[width=3.3in,height=2.0in]{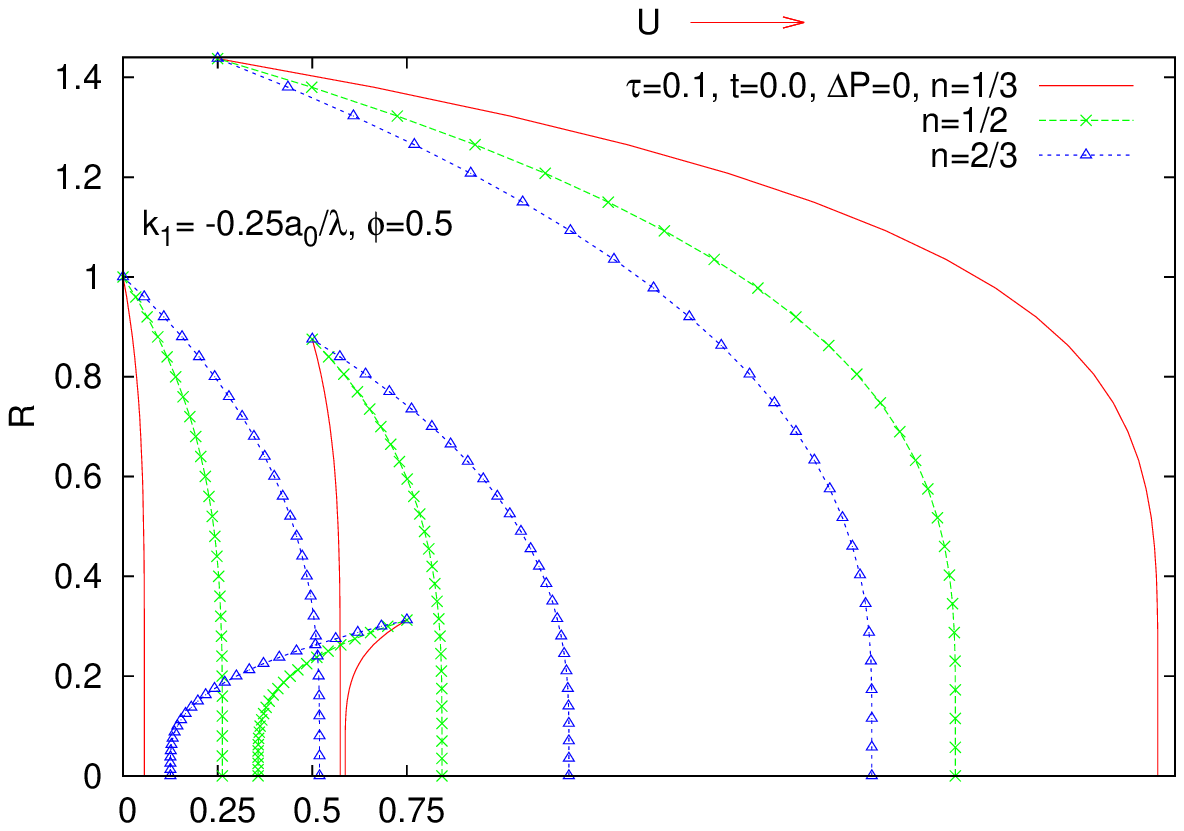}\\$~~~~~~~~~~~~~~~~~~~~~~~~~~~(i)~~~~~~~~~~~~~~~~~~~~~~~~~~~~~~~~~~~~~~~~~~~~~~~~~~~~~~~~~~~~~~~~~~~~~(j)~~~~~~~~~~~~~~~$
\includegraphics[width=3.3in,height=2.0in]{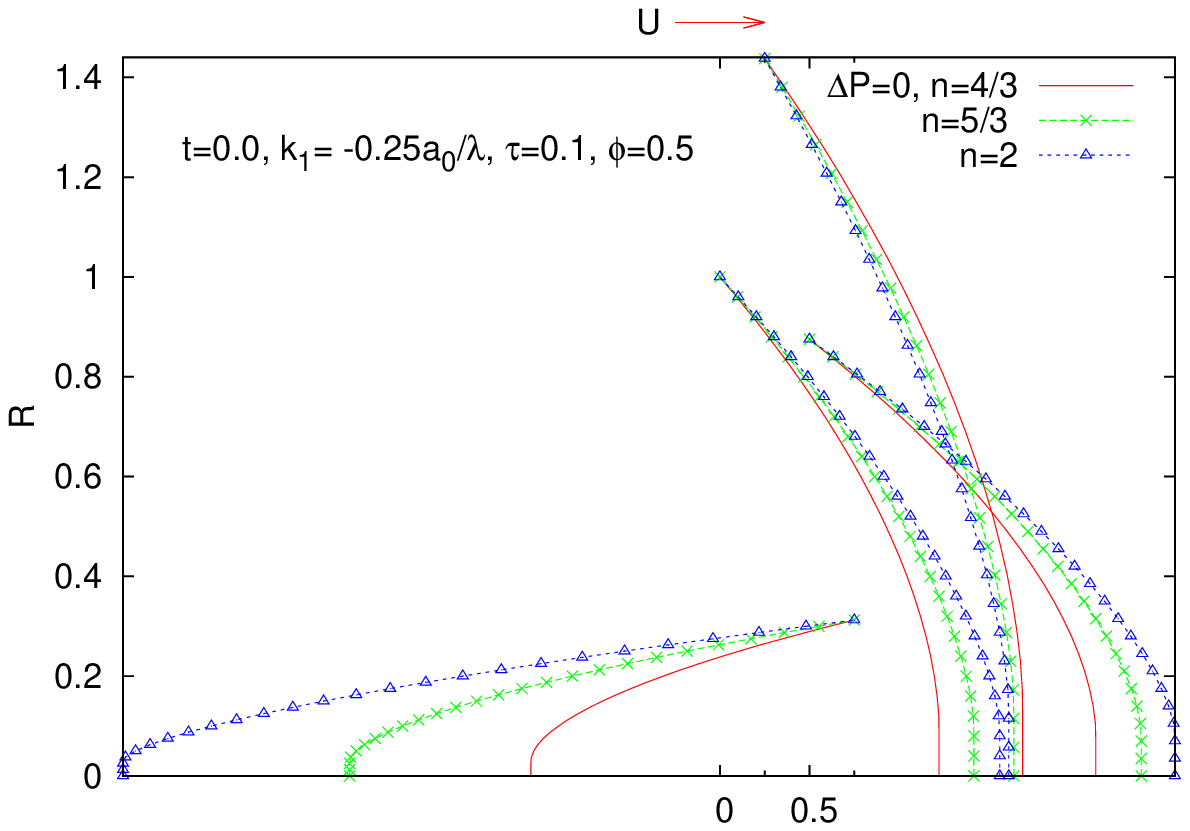}\includegraphics[width=3.3in,height=2.0in]{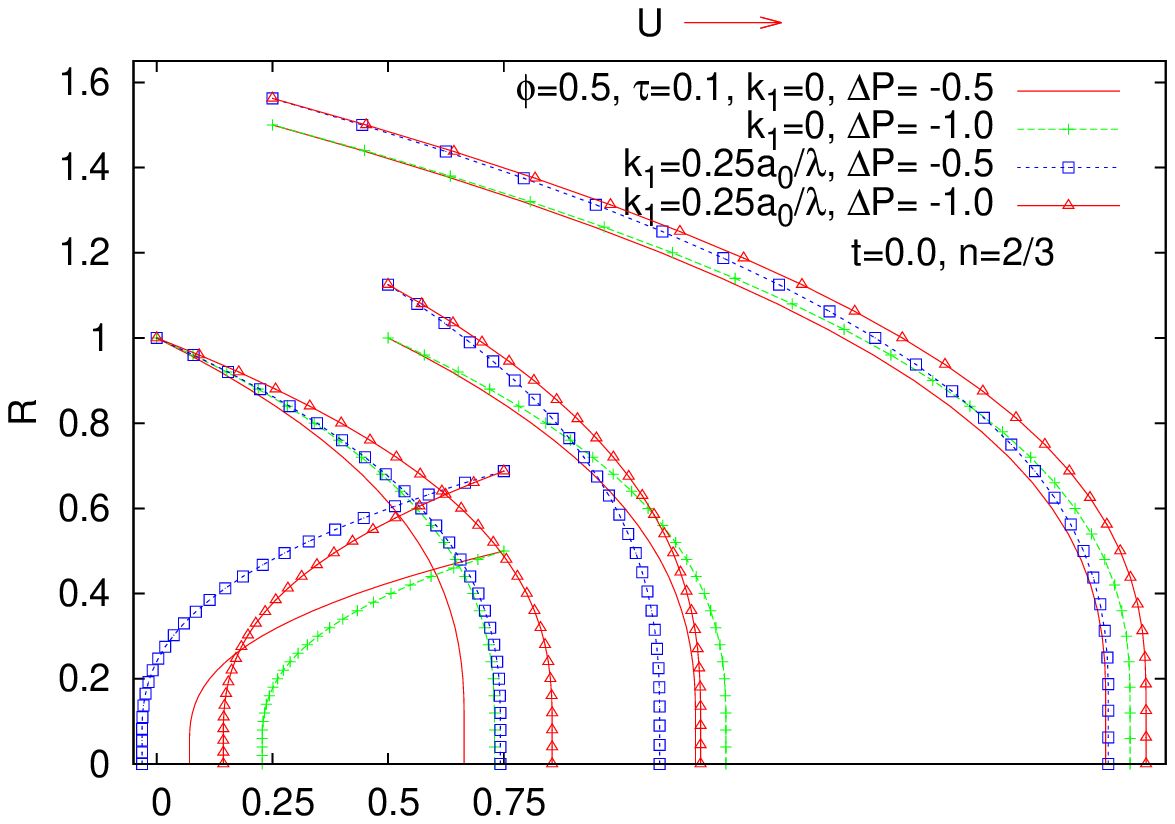}\\$~~~~~~~~~~~~~~~~~~~~~~~~~~~(k)~~~~~~~~~~~~~~~~~~~~~~~~~~~~~~~~~~~~~~~~~~~~~~~~~~~~~~~~~~~~~~~~~~~~~(l)~~~~~~~~~~~~~~~$
\end{figure}
\begin{figure}
\includegraphics[width=3.3in,height=2.0in]{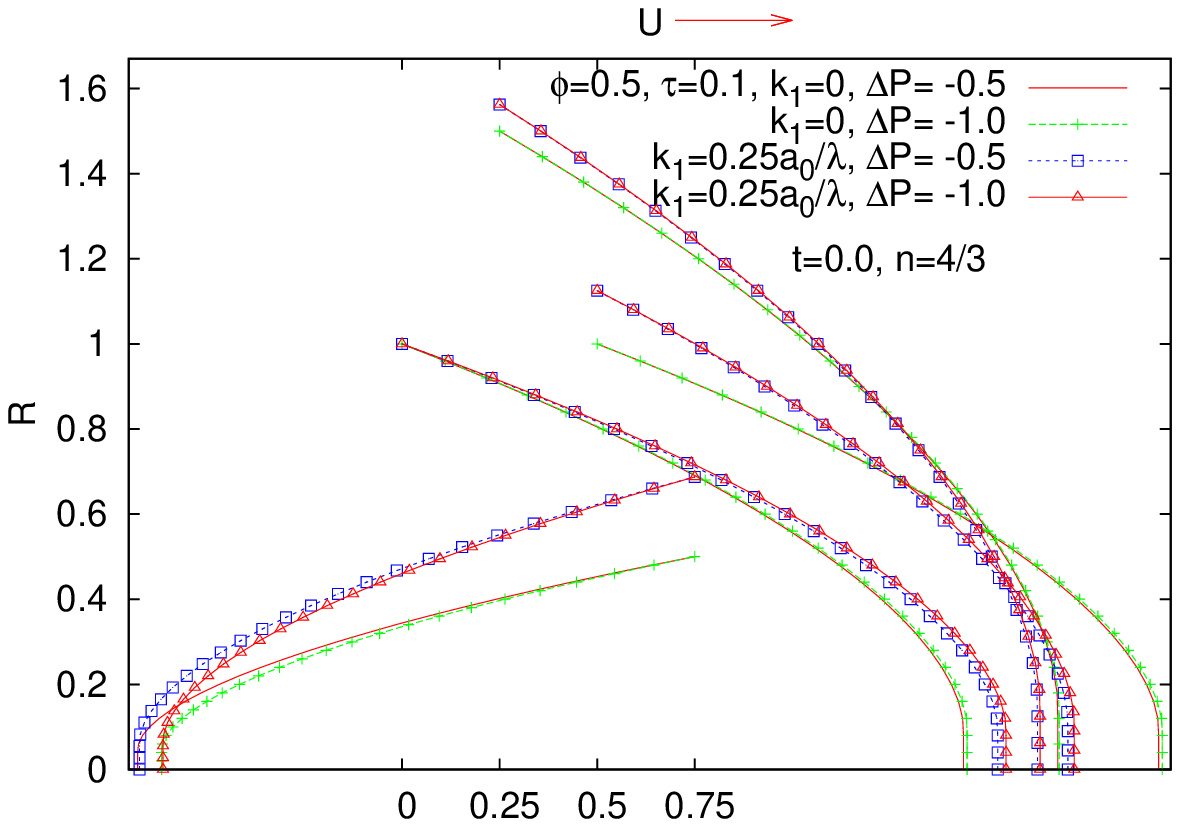}\includegraphics[width=3.3in,height=2.0in]{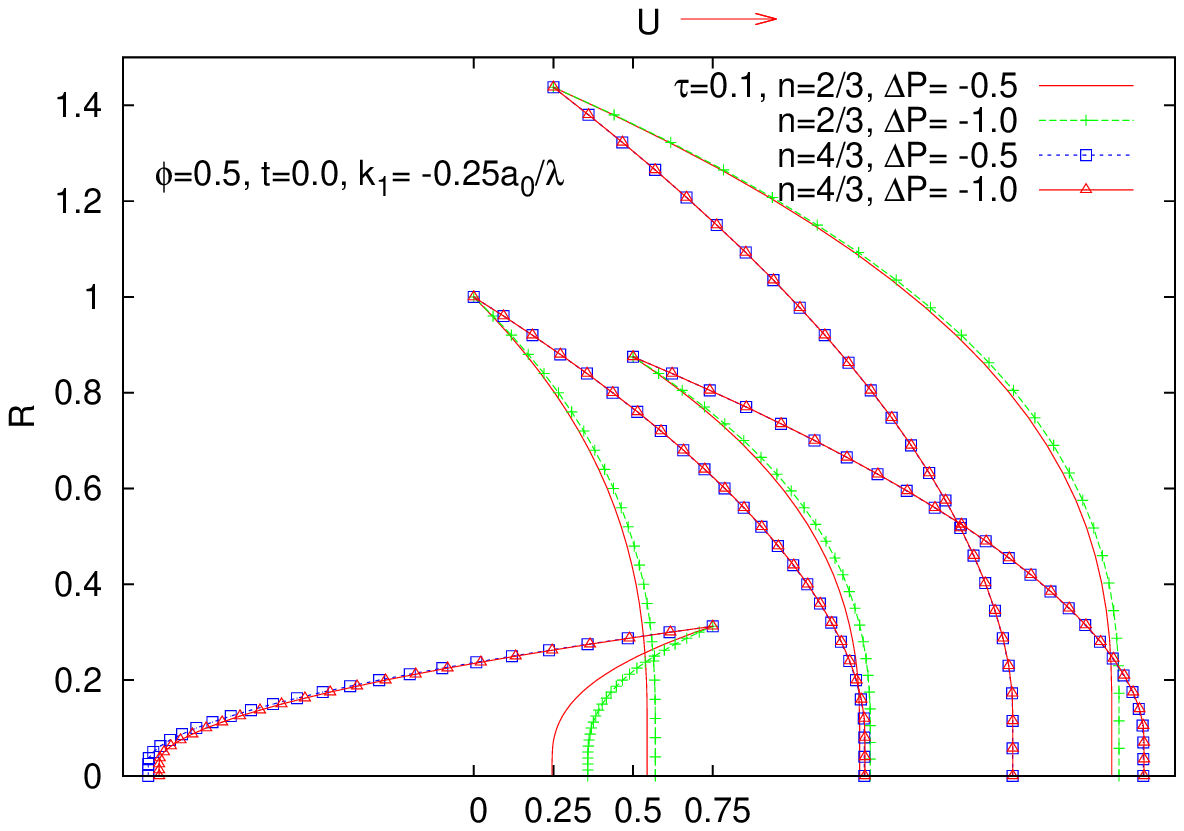}\\$~~~~~~~~~~~~~~~~~~~~~~~~~~~(m)~~~~~~~~~~~~~~~~~~~~~~~~~~~~~~~~~~~~~~~~~~~~~~~~~~~~~~~~~~~~~~~~~~~~~(n)~~~~~~~~~~~~~~~$
\includegraphics[width=3.3in,height=2.0in]{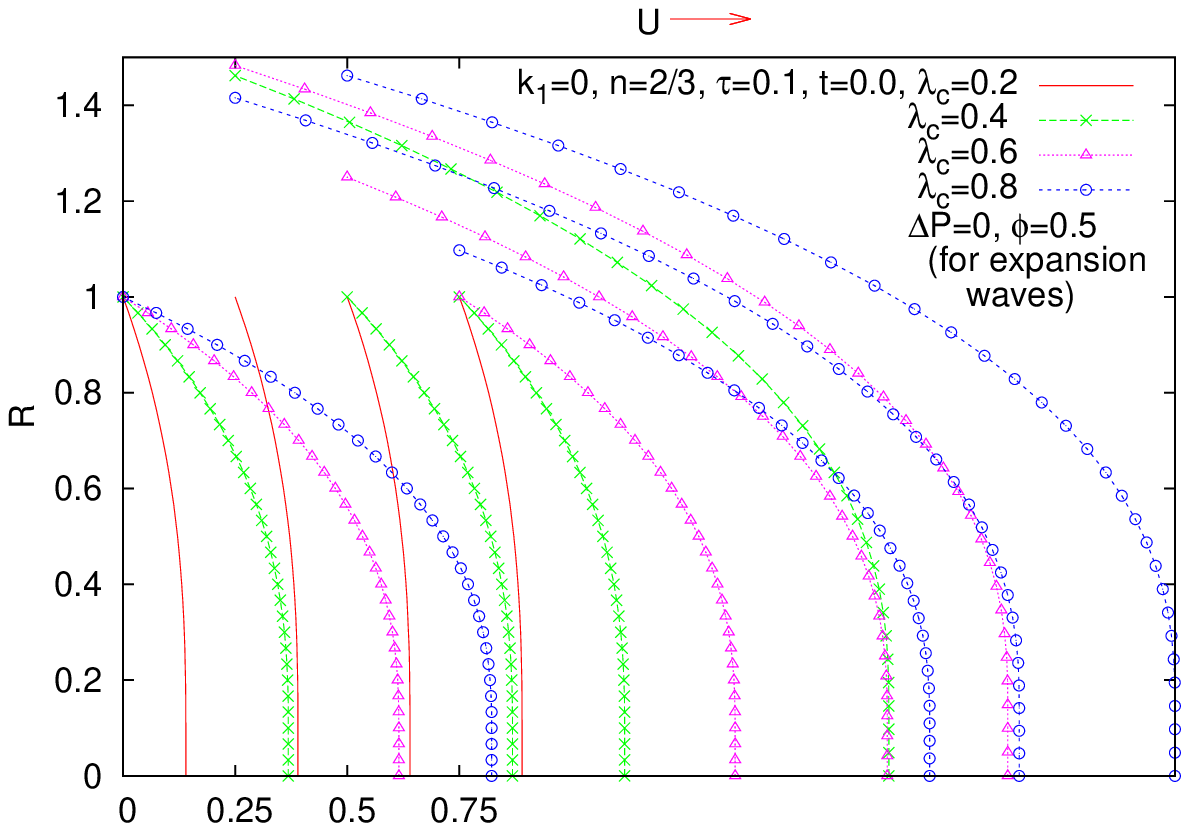}\includegraphics[width=3.3in,height=2.0in]{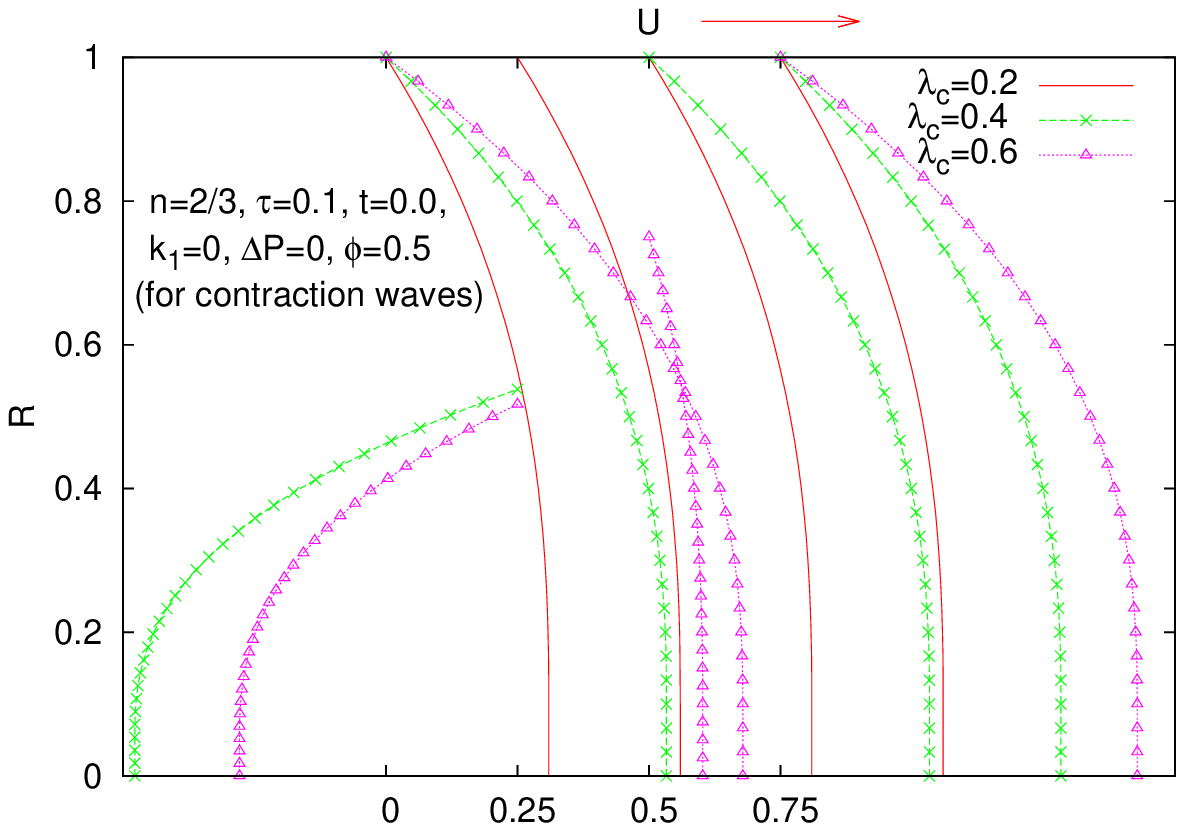}\\$~~~~~~~~~~~~~~~~~~~~~~~~~~~(o)~~~~~~~~~~~~~~~~~~~~~~~~~~~~~~~~~~~~~~~~~~~~~~~~~~~~~~~~~~~~~~~~~~~~~(p)~~~~~~~~~~~~~~~$
\caption{Distribution of velocity in various cases. The figures
reveal that flow reversal occur due to the contraction of the wave.}
\label{cnsns_manuscript_velo5.3.1-5.14.1}
\end{figure}

In Figs. \ref{cnsns_manuscript_velo5.3.1-5.14.1}(e)-(k), it is
worthwhile to observe the significant influence of the rheological
fluid index 'n' on the velocity distribution for flows in
uniform/non-uniform vessels. These figures reveal that the parabolic
nature of the velocity profiles is disturbed due to the non-Newtonian
effect. Magnitude of the velocity decreases at the maximum expansion
region; while in the remaining regions, a reverse trend is noticed,
when there is an increase in the value of 'n' (cf. Figs.
\ref{cnsns_manuscript_velo5.3.1-5.14.1}(e-g,j-k)). For a converging
vessel, the magnitude of the velocity is greater than that of a
uniform vessel; however, for a diverging vessel, our observation is
altogether different. Figs.
\ref{cnsns_manuscript_velo5.3.1-5.14.1}(l)-(n) illustrate the
influence of pressure on velocity distribution for shear thinning
/shear thickening fluids. The results presented for shear-thickening
fluid have been computed by taking $n=4/3$, while those for
shear-thinning fluid correspond to $n=2/3$. It may be noted that in
the case of a uniform/diverging vessel, if $\Delta P$ decreases, the
flow reversal tends to decrease; however, for a converging tube,
although there is reduction in the region of flow reversal, the change
is not too significant for shear thickening fluid. It is important to
mention that flow reversal takes place due to change in sign of the
vorticity or the shear stress along the wavy wall.

The results corresponding to SSD wave propagation are presented in
Figs. \ref{cnsns_manuscript_velo5.3.1-5.14.1}(o-p). It is seen that
for SSD expansion waves, there is no backward flow and that the
magnitude of the velocity is considerably less in this case as
compared to the case of sinusoidal wave propagation (cf. Fig.
\ref{cnsns_manuscript_velo5.1.5}). Moreover, as $\lambda_c$ increases,
velocity is seen to rise except around Z=0.25 where it decreases when
$\lambda_c$ exceeds 0.6. For SSD contraction waves, Fig.
\ref{cnsns_manuscript_velo5.3.1-5.14.1}(n) shows that if
$\lambda_c=0.2$, there is no backward flow within one wave length.  If
$\lambda_c>0.3$, backward flow is observed from Z=0.15 to Z=0.7.

 \begin{figure}
   \includegraphics[width=3.35in,height=1.8in]{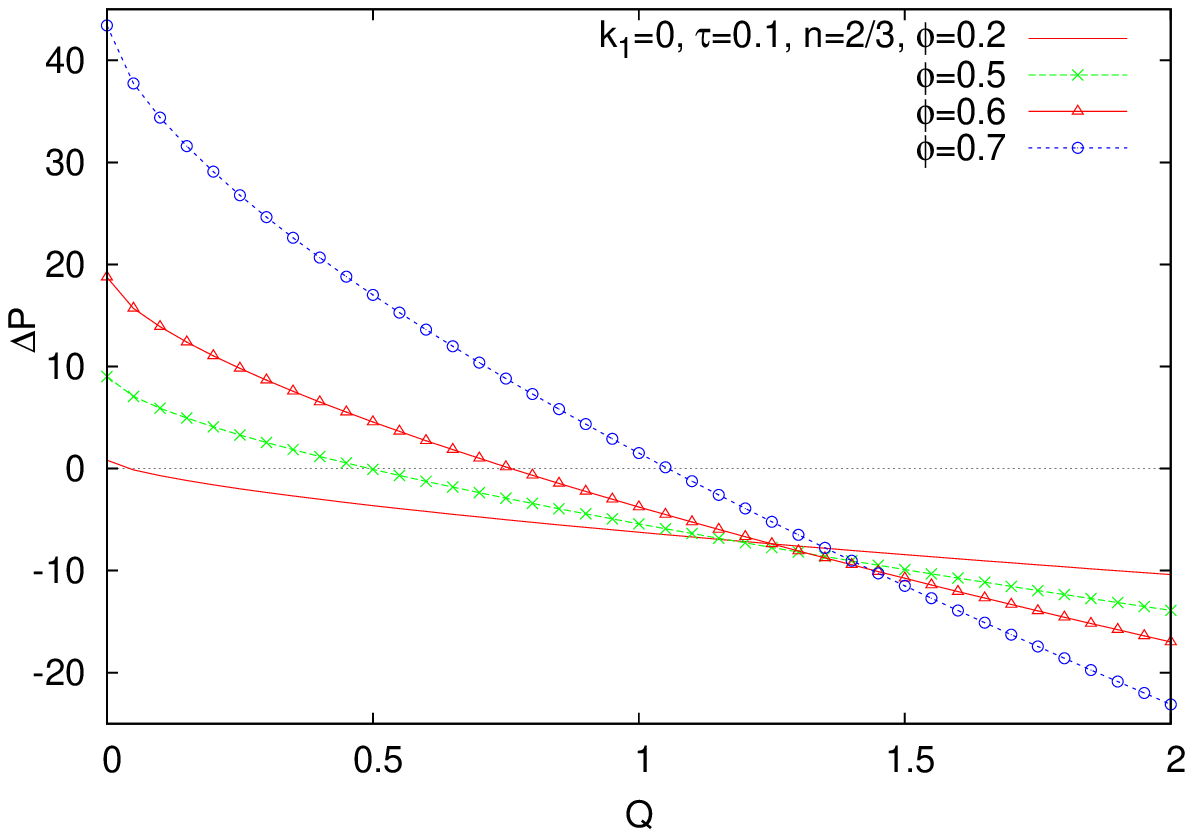}\includegraphics[width=3.35in,height=1.8in]{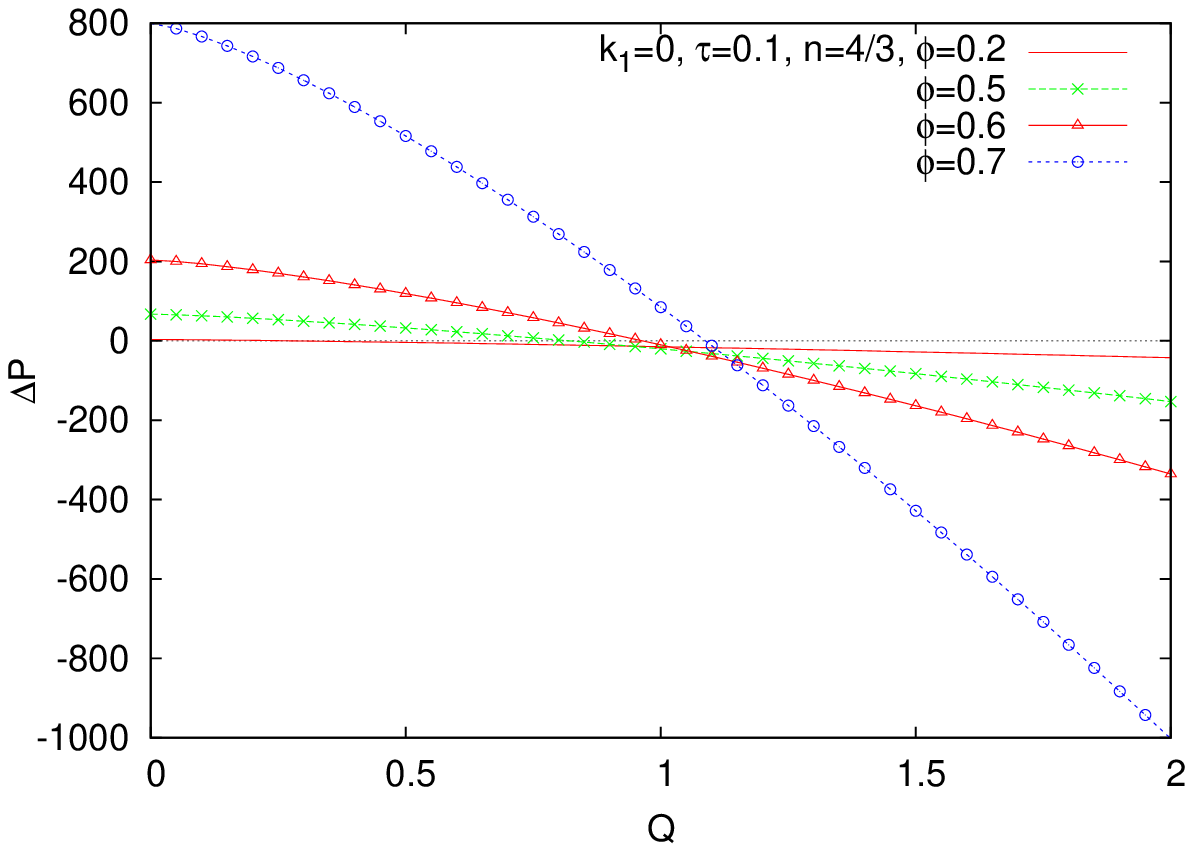}\\$~~~~~~~~~~~~~~~~~~~~~~~~~~~(a)~~~~~~~~~~~~~~~~~~~~~~~~~~~~~~~~~~~~~~~~~~~~~~~~~~~~~~~~~~~~~~~~~~~~~(b)~~~~~~~~~~~~~~~$
\includegraphics[width=3.35in,height=1.8in]{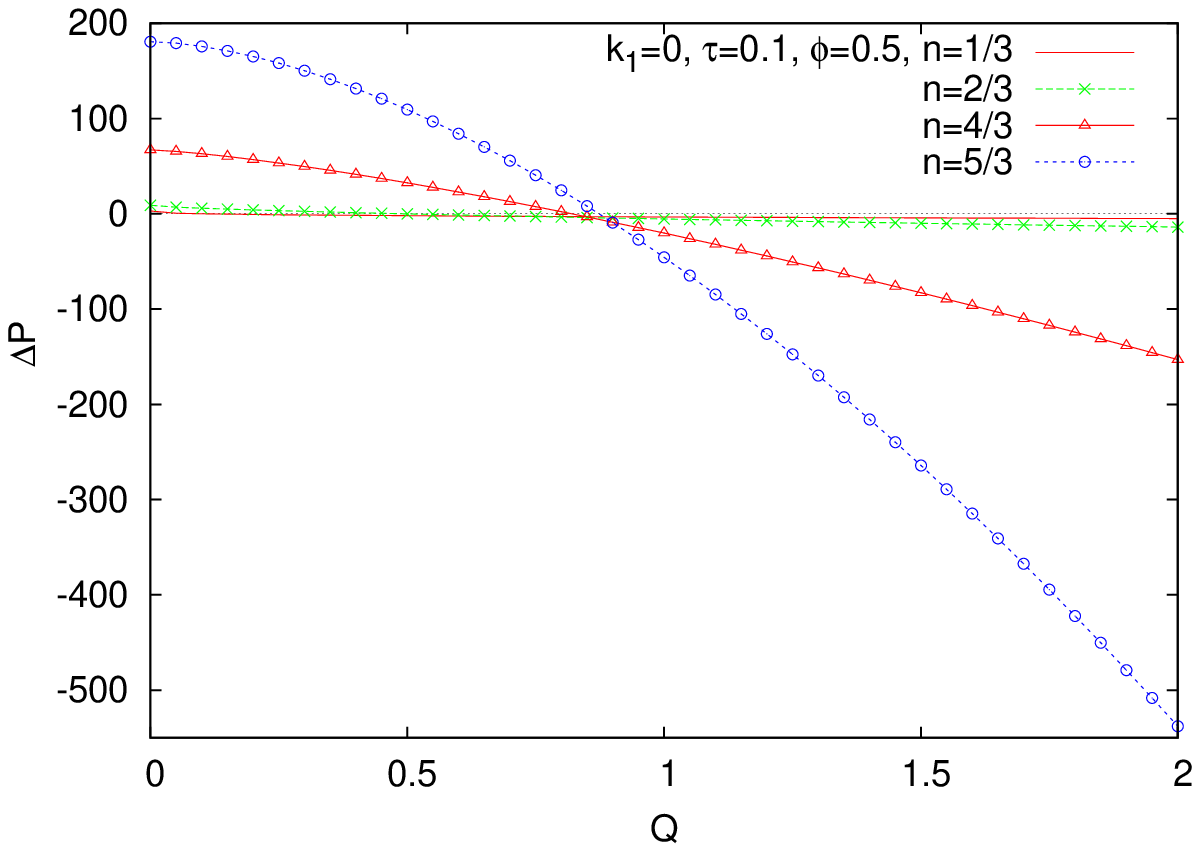}\includegraphics[width=3.35in,height=1.8in]{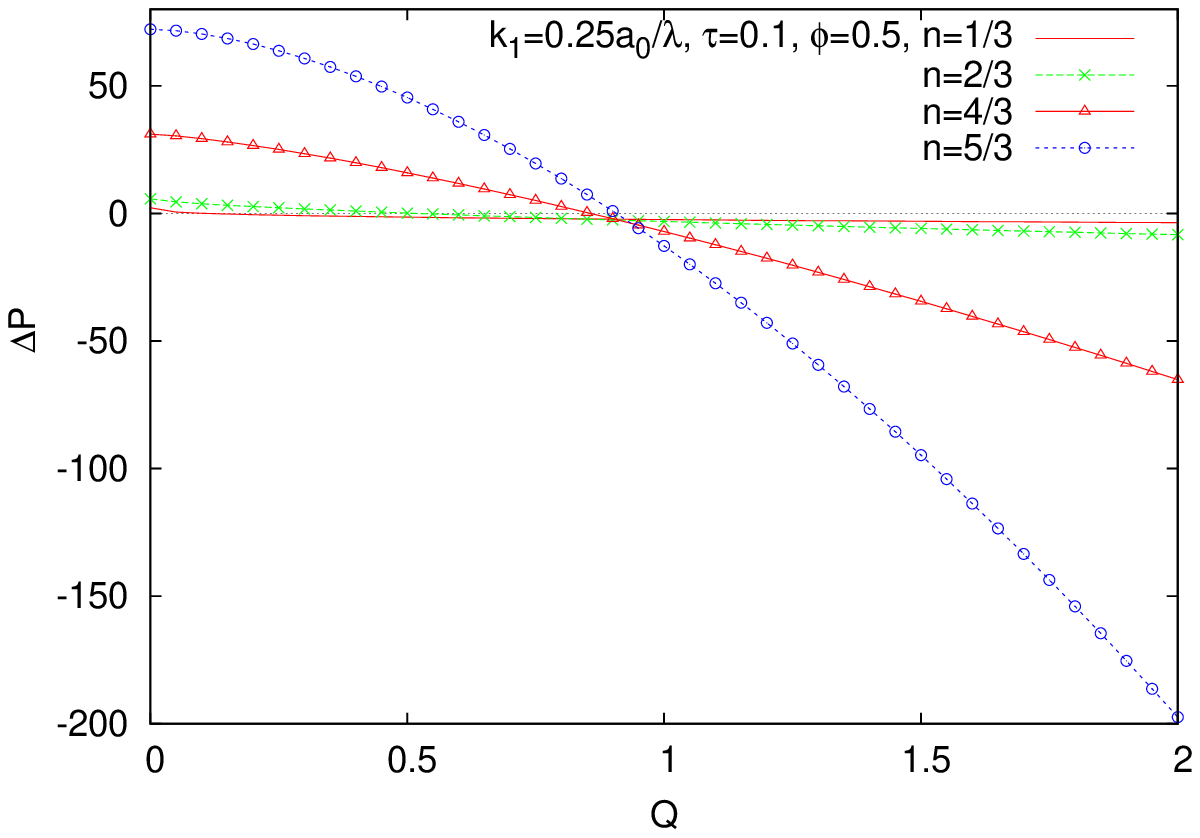}\\$~~~~~~~~~~~~~~~~~~~~~~~~~~~(c)~~~~~~~~~~~~~~~~~~~~~~~~~~~~~~~~~~~~~~~~~~~~~~~~~~~~~~~~~~~~~~~~~~~~~(d)~~~~~~~~~~~~~~~$
\includegraphics[width=3.35in,height=1.8in]{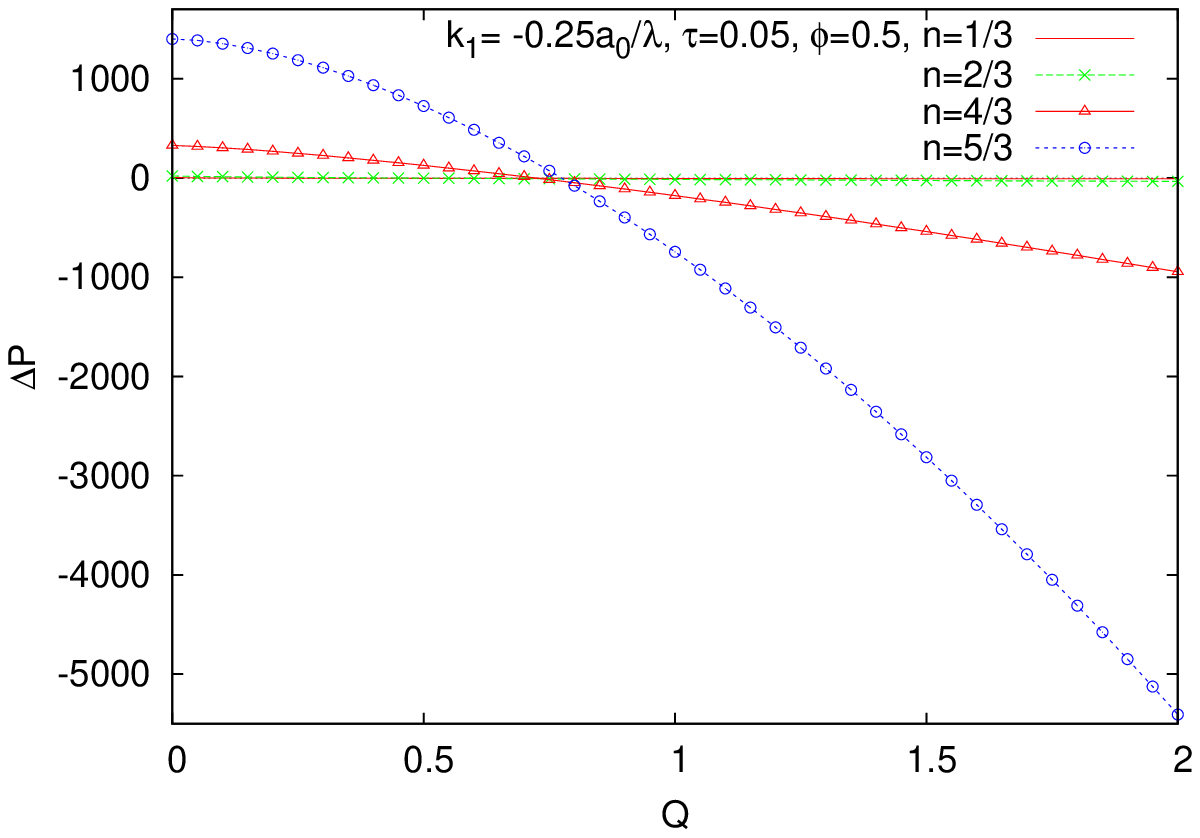}\includegraphics[width=3.35in,height=1.8in]{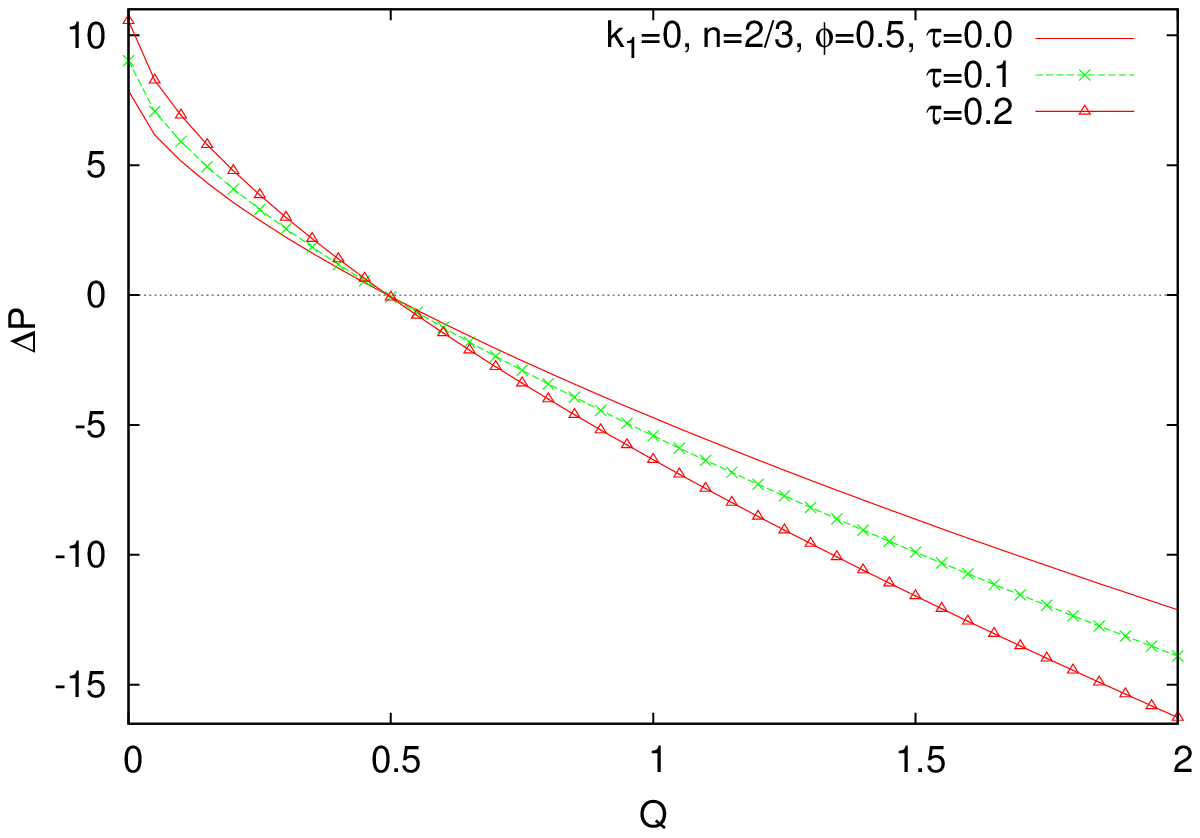}\\$~~~~~~~~~~~~~~~~~~~~~~~~~~~(e)~~~~~~~~~~~~~~~~~~~~~~~~~~~~~~~~~~~~~~~~~~~~~~~~~~~~~~~~~~~~~~~~~~~~~(f)~~~~~~~~~~~~~~~$
\includegraphics[width=3.35in,height=1.8in]{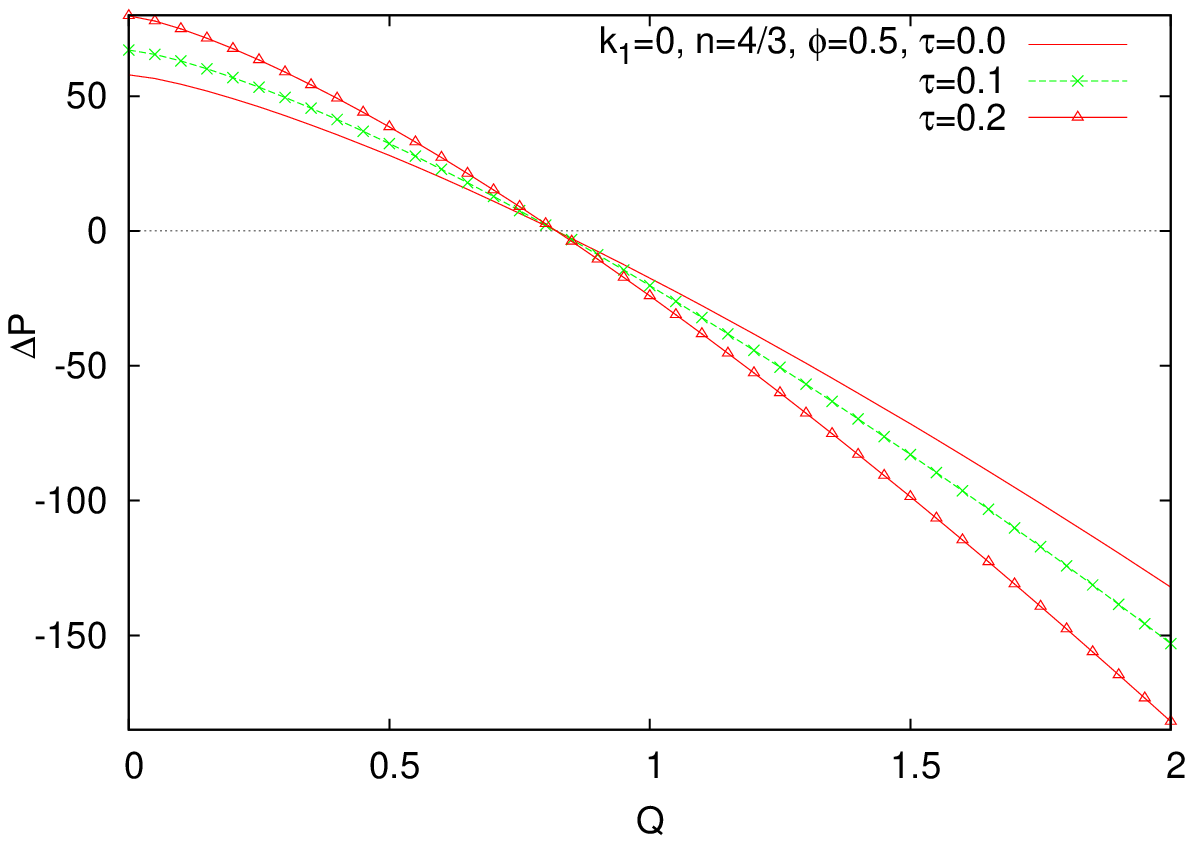}\includegraphics[width=3.35in,height=1.8in]{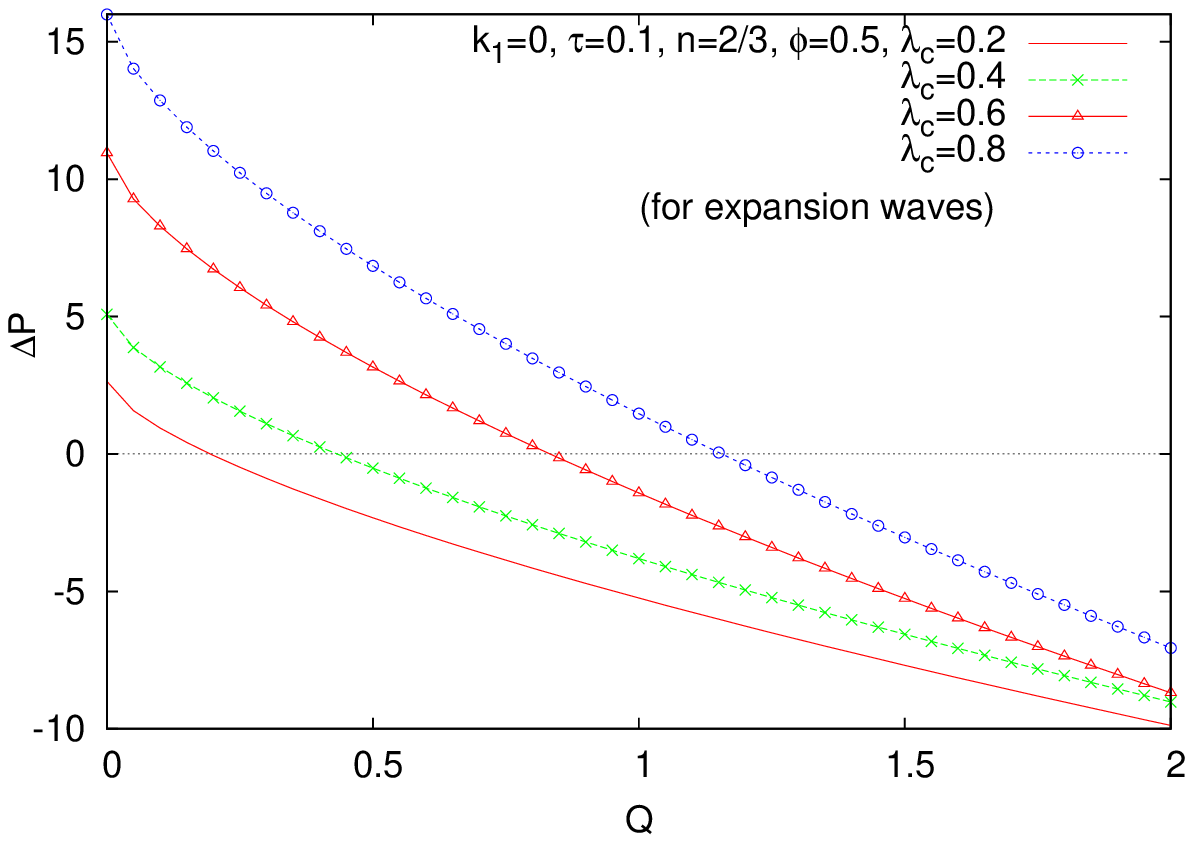}\\$~~~~~~~~~~~~~~~~~~~~~~~~~~~(g)~~~~~~~~~~~~~~~~~~~~~~~~~~~~~~~~~~~~~~~~~~~~~~~~~~~~~~~~~~~~~~~~~~~~~(h)~~~~~~~~~~~~~~~$
\end{figure}

\begin{figure}
\begin{center}
\includegraphics[width=3.5in,height=1.8in]{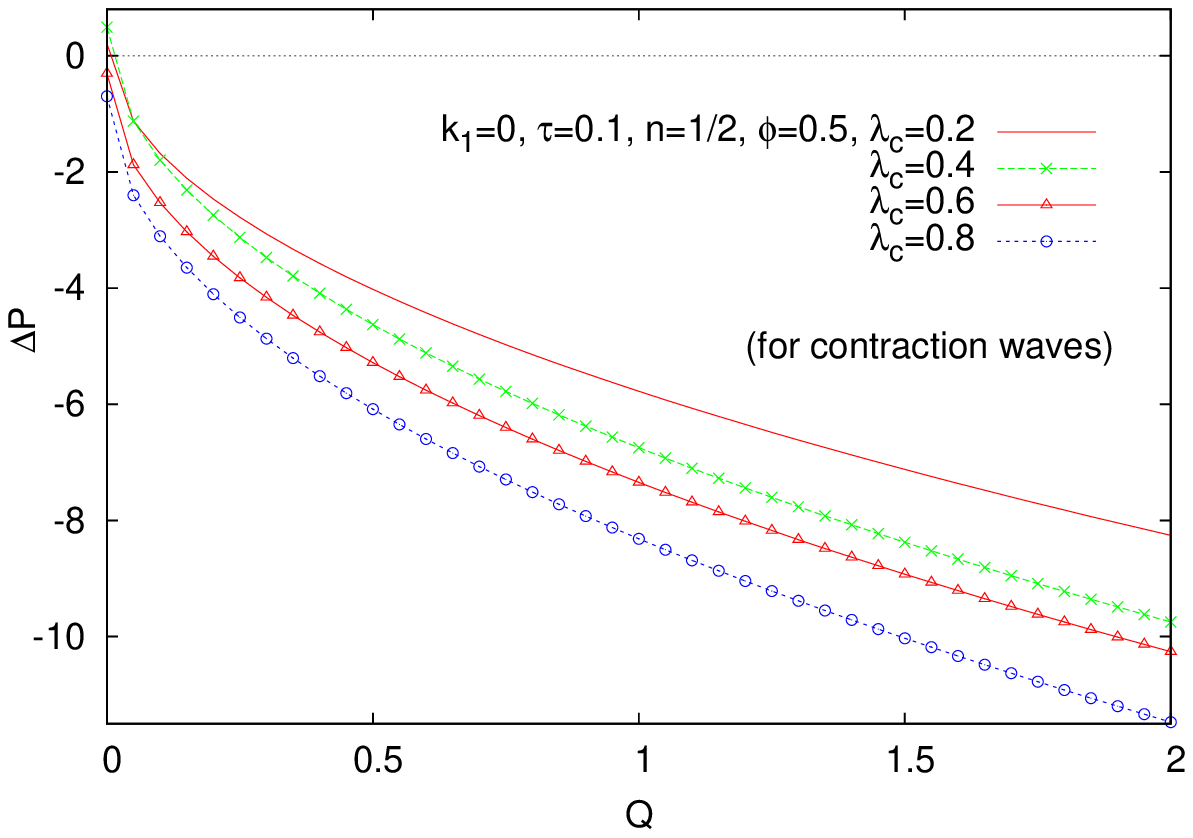}\\(i)
\caption{Pressure difference versus flow rate. It may be observed
that the area of the pumping region increases when the amplitude
ratio $\phi$ is enhanced. In addition, the variation of the vessel
radius in both converging and diverging cases as well as SSD
expansion/contraction wave modes have significant impact on the
pressure difference ($\Delta$P) as well as on the volumetric flow
rate (Q).} \label{cnsns_manuscript_pump5.1.1.1-5.12.1}
\end{center}
\end{figure}

\subsection{Pumping Behaviour}
The pumping characteristics can be determined through the variation of
time averaged flux with difference in pressure across one wave length
(cf. \cite{Shapiro}). It is known that if the flow is steady in the
wave frame, the instantaneous pressure difference between two stations
one wave length apart is a constant.  Since the pressure gradient is a
periodic function of $(Z-t)$, pressure rise per wave length $\Delta
P_{\lambda}$ is equal to $\lambda$ times the time-averaged pressure
gradient. In addition, the relation between the fixed frame and the
wave frame flux rates turns out to be linear, as in the case of
inertia-free peristaltic flow of a Newtonian fluid. From the said
relation, it is possible to calculate the amount of flow pumped by
peristaltic waves, even in the absence of mean pressure gradient. The
region in which $\Delta P=0$ is regarded to as the free pumping zone,
while the region where $\Delta P>0$ is said to be the pumping
zone. The situation when $\Delta P<0$ is favourable for the flow to
take place and the corresponding region is called the co-pumping
zone. Since the present study is concerned with peristaltic transport
of a non-Newtonian fluid, the aforesaid relationship between the
pressure difference and the mean flow rate is nonlinear.

Fig. \ref{cnsns_manuscript_pump5.1.1.1-5.12.1} illustrates the
variation of volumetric flow rate of the fluid by way of propagation
of peristaltic waves, for different values of the amplitude ratio
$\phi $, flow index number n, as well as the parameters $\tau$ and
$k_1$. Shapiro et al. \cite{Shapiro} used lubrication theory to show
that in the case of a Newtonian fluid, the flow rate averaged over one
wave length varies linearly with pressure difference. But for our study of
the peristaltic transport of a non-Newtonian fluid, the relationship
between the pressure difference and the mean flow rate is found to be
non-linear (cf. Figs \ref{cnsns_manuscript_pump5.1.1.1-5.12.1}). The
plots presented in this figure show that for a non-Newtonian fluid,
the mean flow rate Q increases as $\Delta P$ decreases. Figs.
\ref{cnsns_manuscript_pump5.1.1.1-5.12.1}(a-b) indicate that area of
the pumping region increases with an increase in the amplitude ratio
$\phi$ for both shear thinning and shear thickening fluids.
Figs. \ref{cnsns_manuscript_pump5.1.1.1-5.12.1}(c-e) illustrate the
influence of the rheological parameter `n' on the pumping performance
in uniform/diverging/converging vessels. It may be noted that the
pumping region ($\Delta P>0$) significantly increases as the value of
`n' increases, while in the co-pumping region ($\Delta P<0$) the
pressure-difference decreases when Q exceeds a certain
value. Figs. \ref{cnsns_manuscript_pump5.1.1.1-5.12.1}(f-g) show that
Q is not significantly affected by the value of the parameter $\tau$
in the case of free pumping. We further find that for both
shear-thinning and shear thickening fluids, pumping region increases
with $\tau$ increasing. Moreover, when Q exceeds a certain critical
limit, $\Delta P=0$, pressure difference decreases with an increase
in $\tau$.

The effect of SSD wave propagation on pumping is revealed in Figs.
\ref{cnsns_manuscript_pump5.1.1.1-5.12.1}(h-i). Unlike sinusoidal wave
form, Q increases significantly in pumping, free-pumping as well as
co-pumping regions for SSD expansion wave fronts as $\lambda_c$
increases. For SSD contraction wave fronts, our observation is
altogether different from SSD expansion wave fronts. However,
$Q\le0$ when $\Delta P<0$ for shear-thinning fluids in the case of
contraction waves. This observation is in contrast to the case of
sinusoidal wave propagation.
\begin{figure}
 \includegraphics[width=3.5in,height=2.0in]{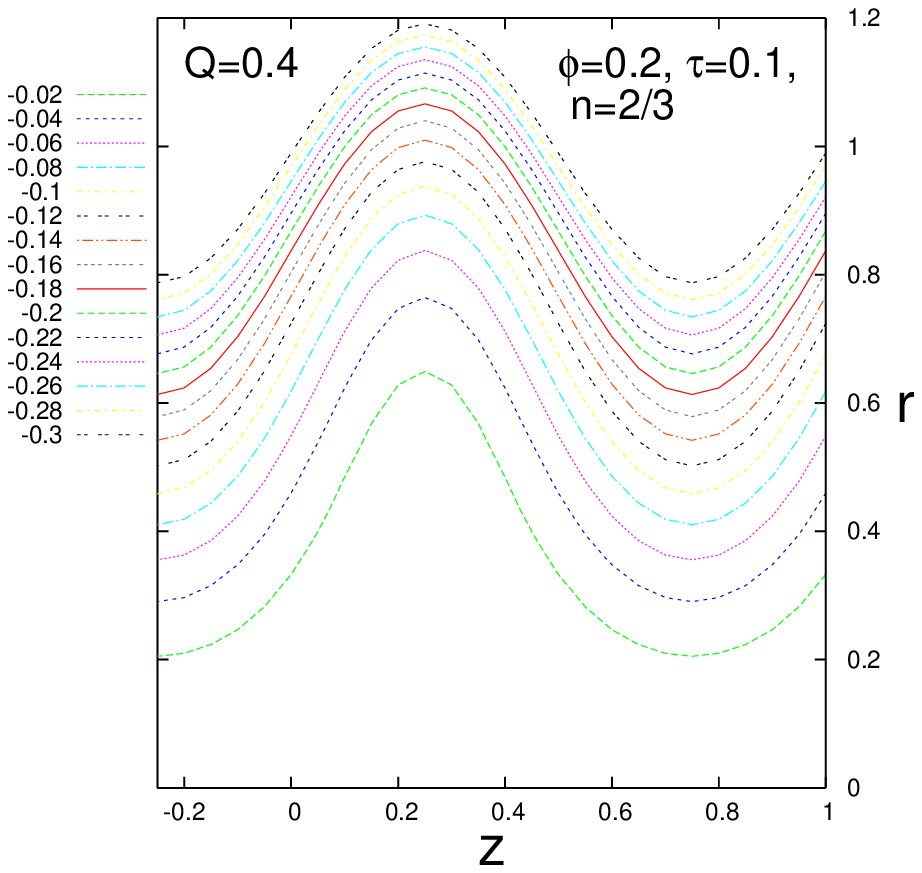}\includegraphics[width=3.5in,height=2.0in]{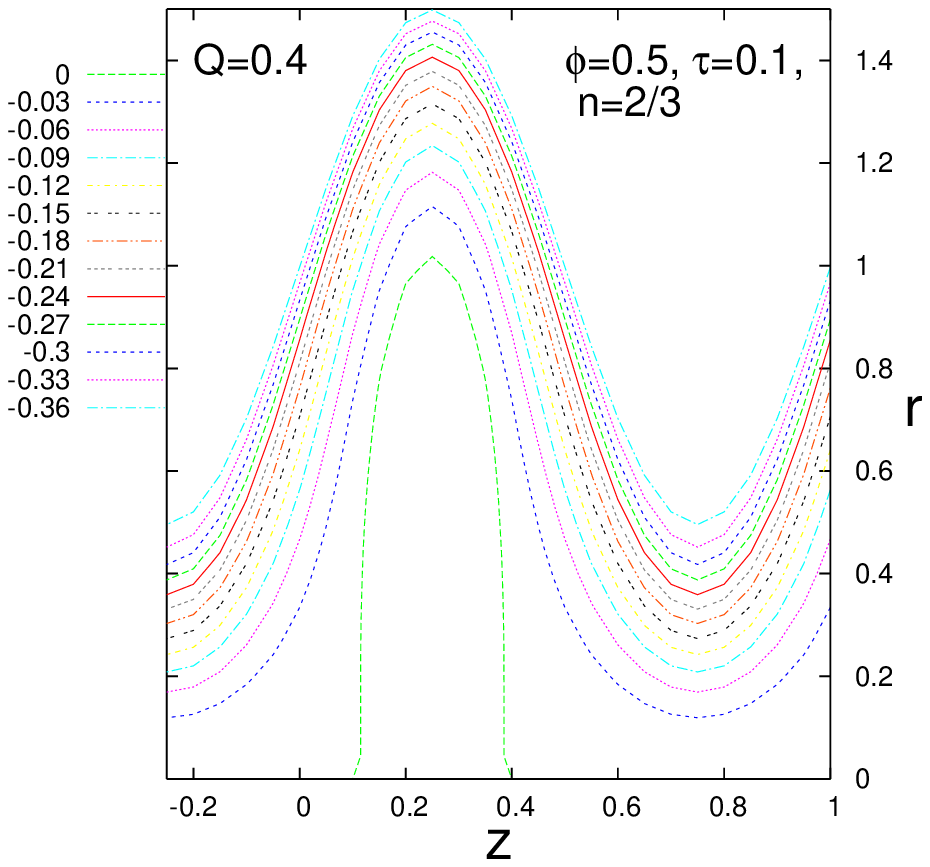}\\$~~~~~~~~~~~~~~~~~~~~~~~~~~~(a)~~~~~~~~~~~~~~~~~~~~~~~~~~~~~~~~~~~~~~~~~~~~~~~~~~~~~~~~~~(b)~~~~~~~~~~~~~~~~~~~~~~~~~$
\includegraphics[width=3.5in,height=2.0in]{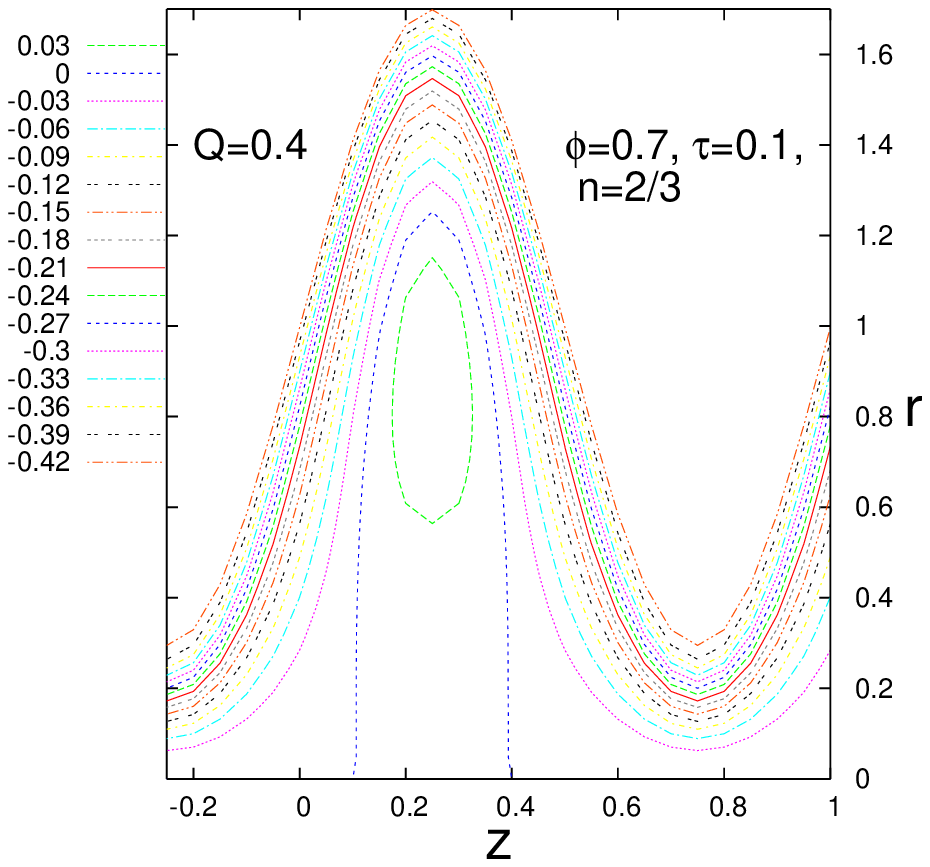}\includegraphics[width=3.5in,height=2.0in]{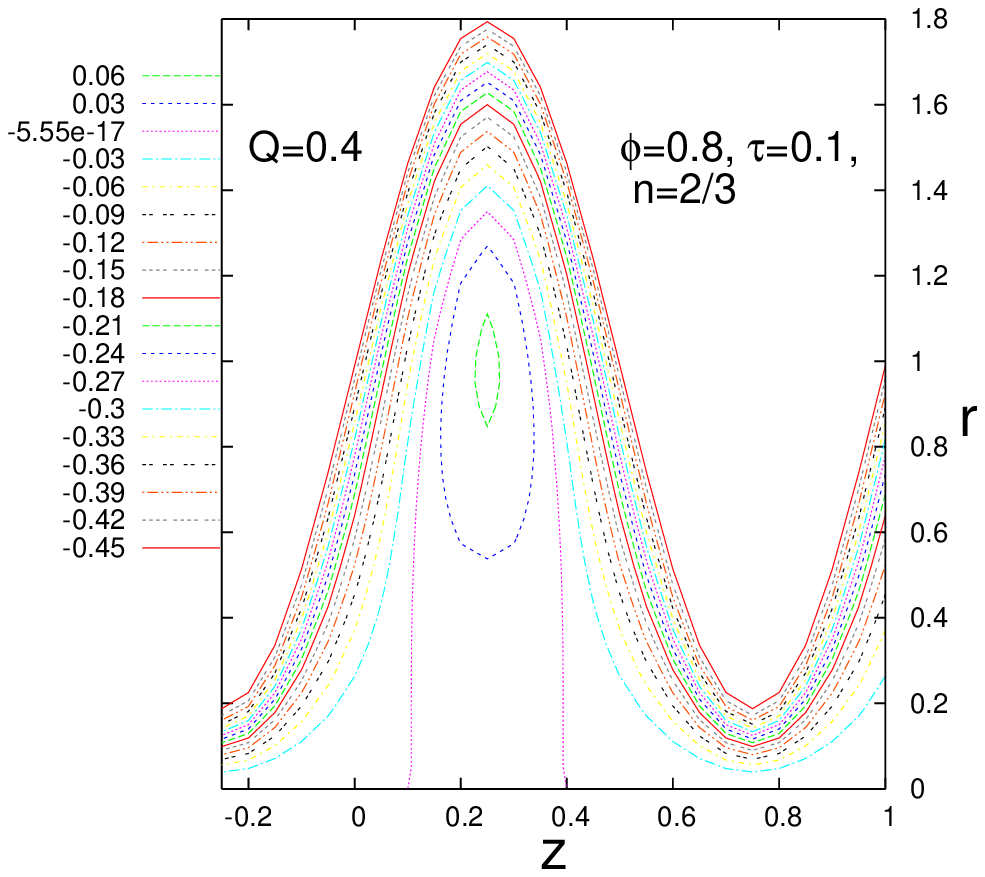}\\$~~~~~~~~~~~~~~~~~~~~~~~~~~~(c)~~~~~~~~~~~~~~~~~~~~~~~~~~~~~~~~~~~~~~~~~~~~~~~~~~~~~~~~~~(d)~~~~~~~~~~~~~~~~~~~~~~~~~$
\caption{Streamline patterns for peristaltic flow of a
shear-thinning fluid for different values of $\phi$ when n=2/3,
Q=0.4, $k_1=0$, $\tau=0$}
\label{cnsns_manuscript_stlp5.1.1.-5.1.6}
\end{figure}

\subsection{Streamlines and Trapping}
\begin{figure}
\includegraphics[width=3.5in,height=2.0in]{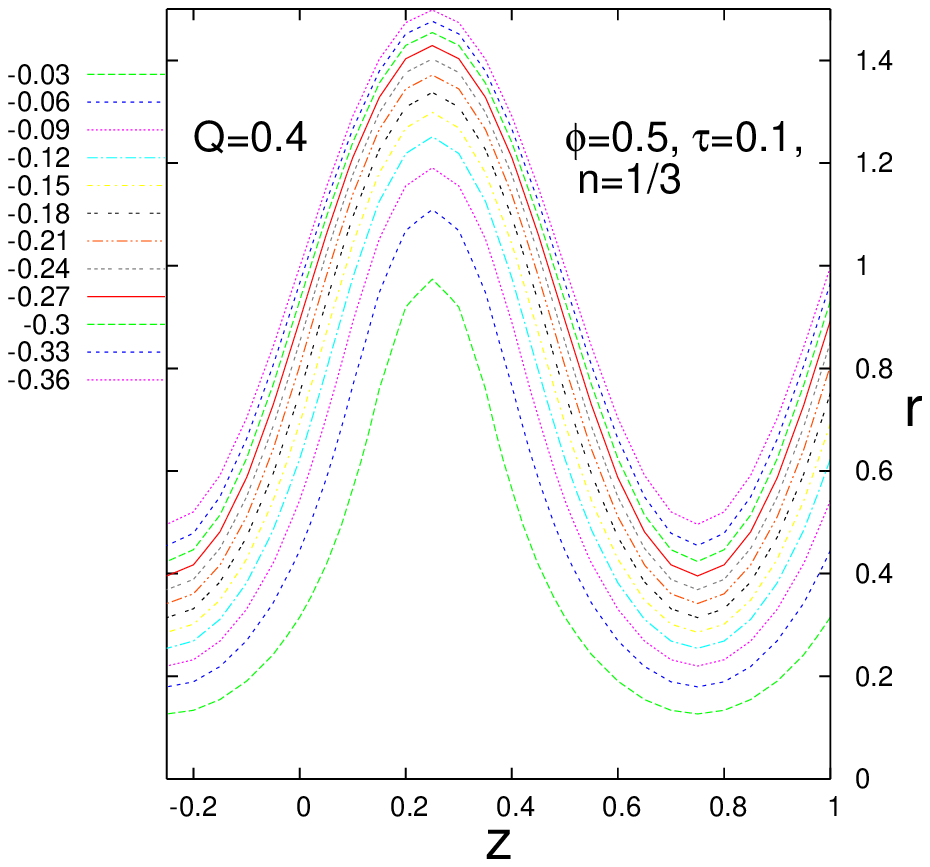}\includegraphics[width=3.5in,height=2.0in]{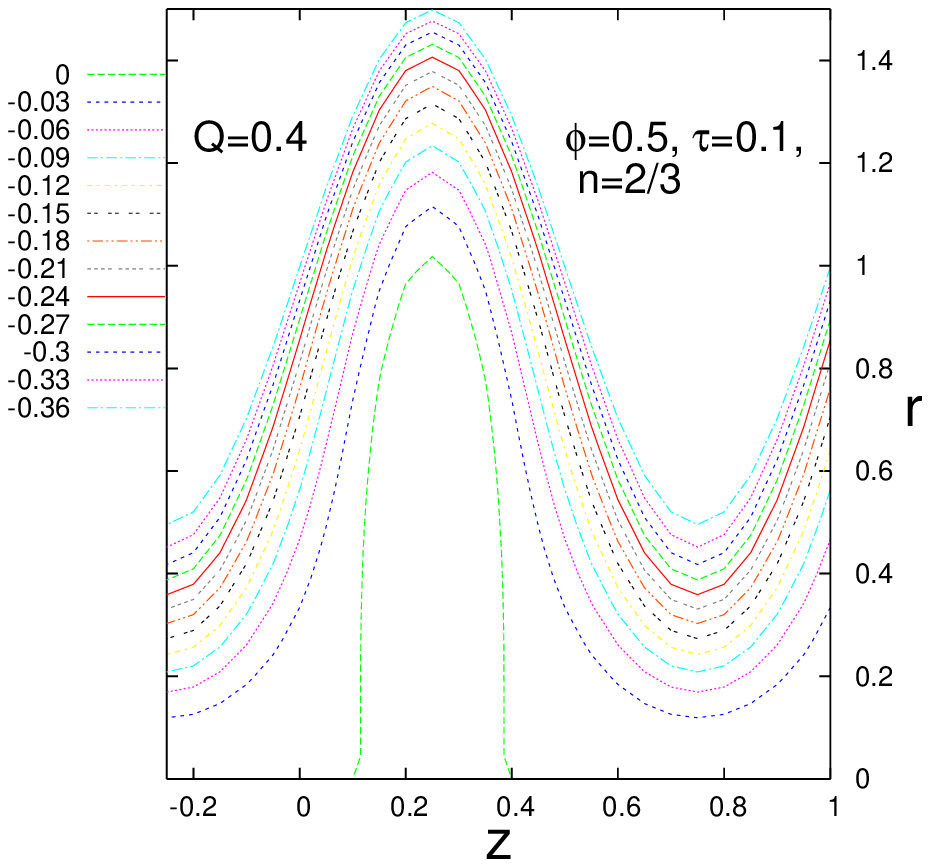}
\\$~~~~~~~~~~~~~~~~~~~~~~~~~~~(a)~~~~~~~~~~~~~~~~~~~~~~~~~~~~~~~~~~~~~~~~~~~~~~~~~~~~~~~~~~(b)~~~~~~~~~~~~~~~~~~~~~~~~~$
\includegraphics[width=3.5in,height=2.0in]{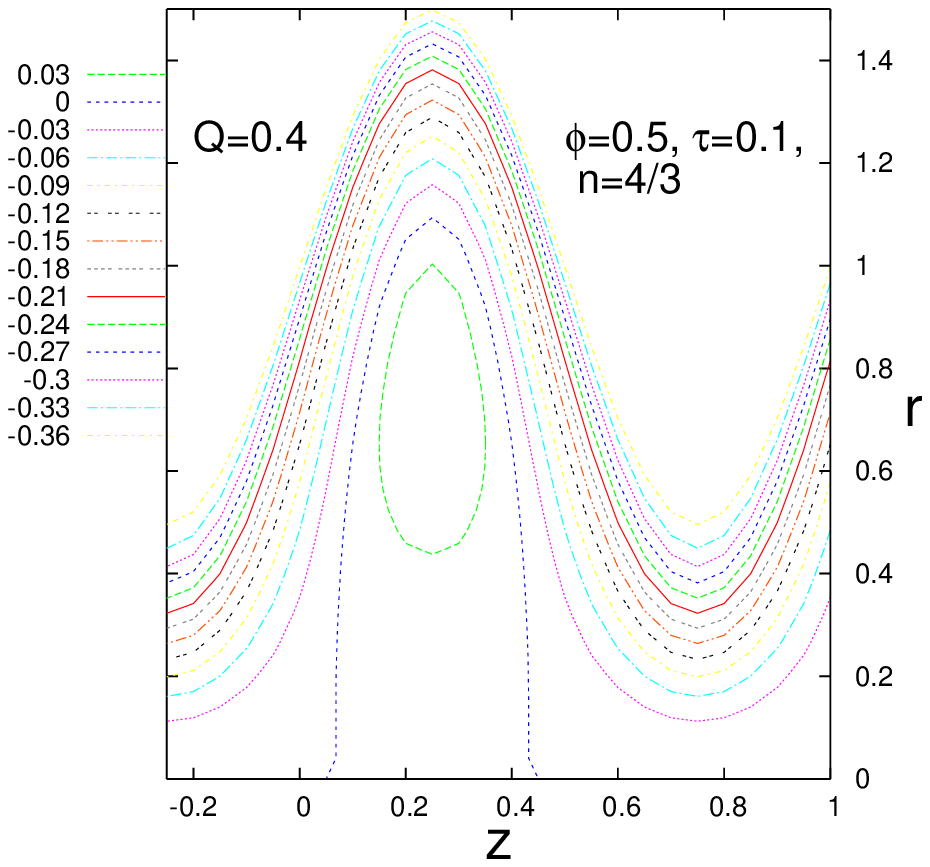}\includegraphics[width=3.5in,height=2.0in]{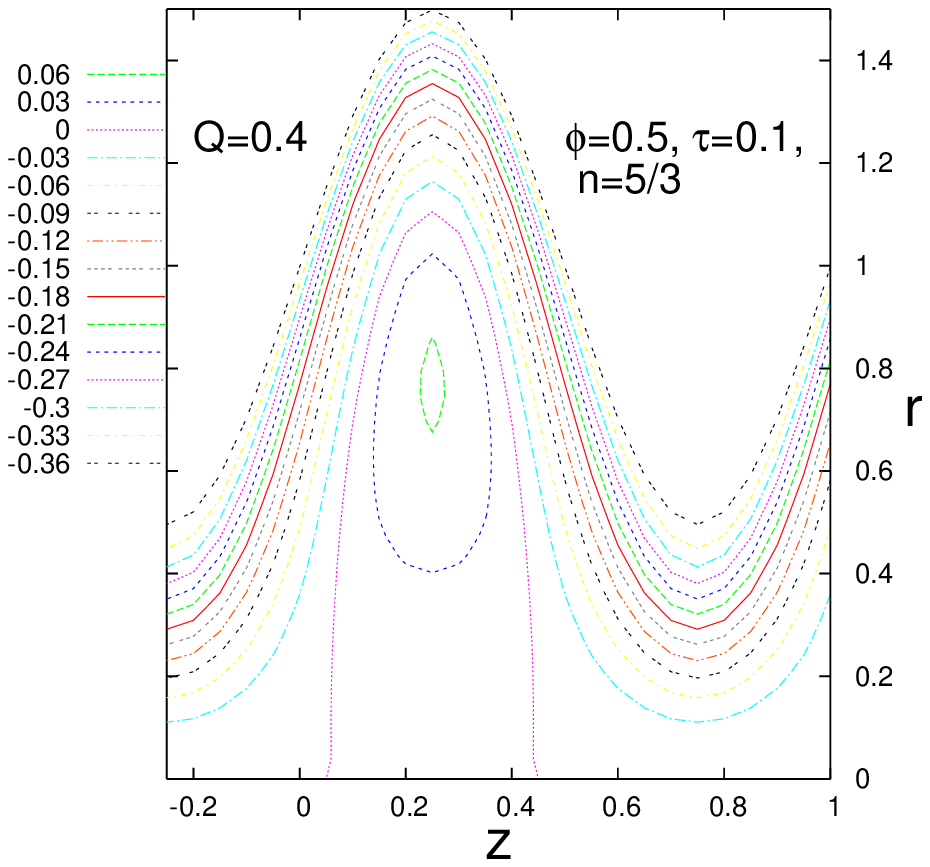}
\\$~~~~~~~~~~~~~~~~~~~~~~~~~~~(c)~~~~~~~~~~~~~~~~~~~~~~~~~~~~~~~~~~~~~~~~~~~~~~~~~~~~~~~~~~(d)~~~~~~~~~~~~~~~~~~~~~~~~~$
\caption{Streamline patterns and trapping in the case of peristaltic flow
 for different values of the physiological fluid index `n' when
$\phi=0.5,~\tau=0.1,~k_1=0,~Q=0.4$. The figures show that the
occurrence of trapping is highly dependent on the value of the fluid
index `n'.} \label{cnsns_manuscript_stlp5.2.1.-5.2.4}
\end{figure}

\begin{figure}
\includegraphics[width=3.5in,height=2.0in]{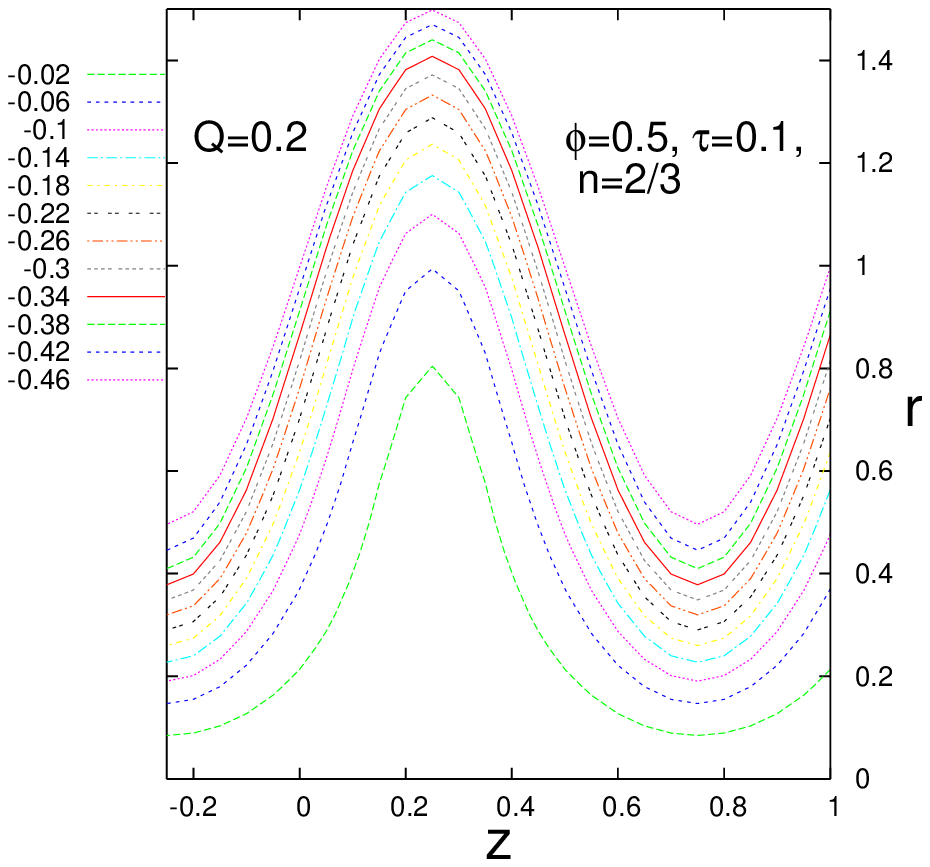}\includegraphics[width=3.5in,height=2.0in]{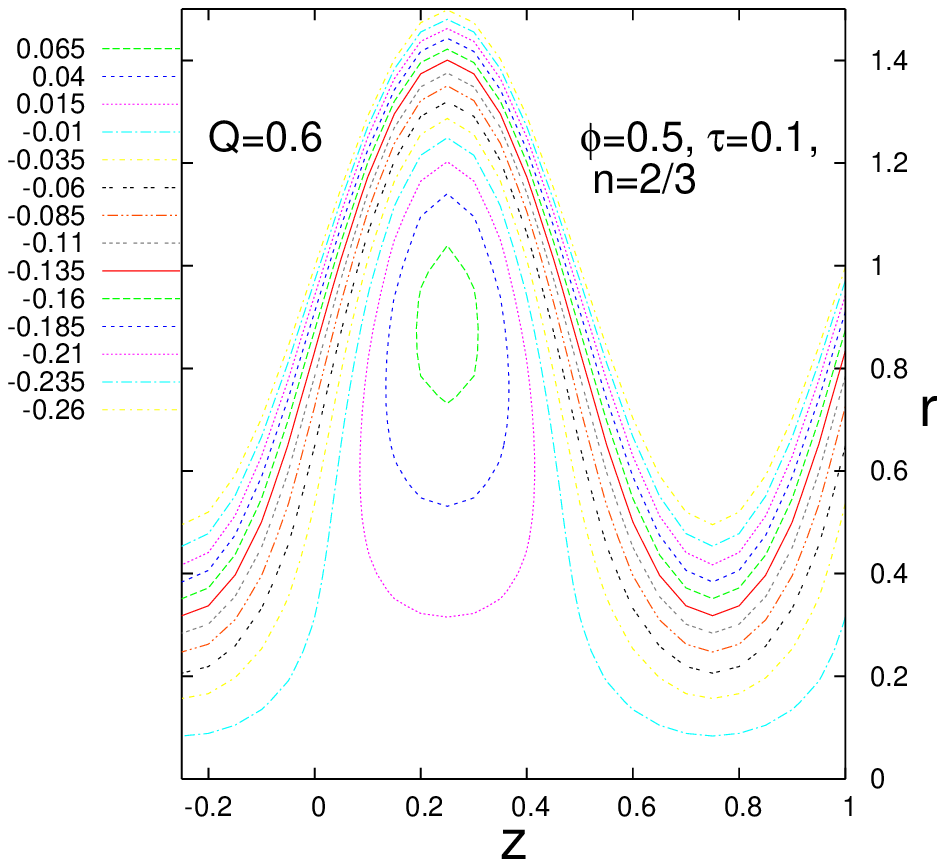}
\\$~~~~~~~~~~~~~~~~~~~~~~~~~~~(a)~~~~~~~~~~~~~~~~~~~~~~~~~~~~~~~~~~~~~~~~~~~~~~~~~~~~~~~~~~(b)~~~~~~~~~~~~~~~~~~~~~~~~~$
\includegraphics[width=3.5in,height=2.0in]{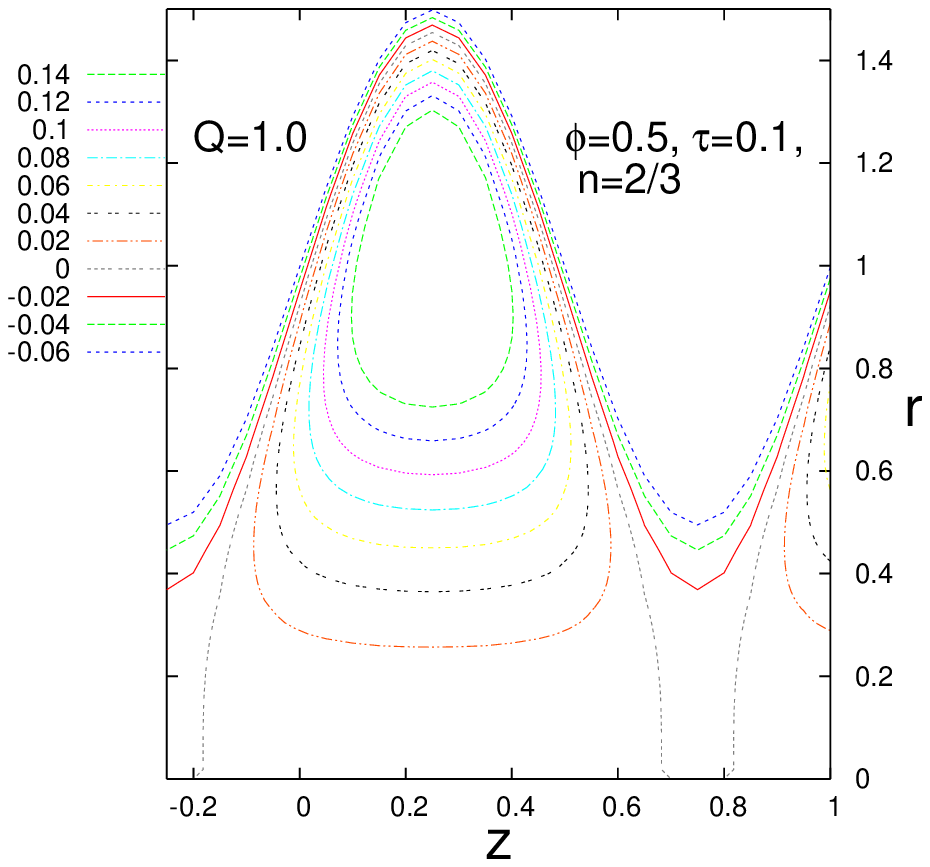}\includegraphics[width=3.5in,height=2.0in]{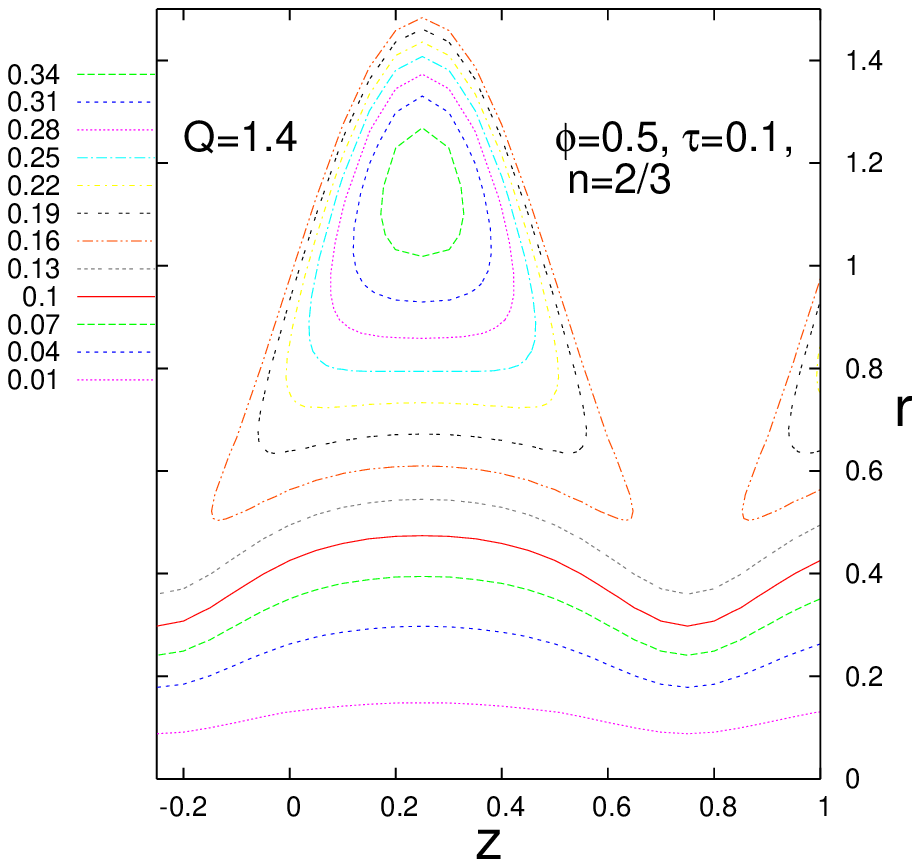}
\\$~~~~~~~~~~~~~~~~~~~~~~~~~~~(c)~~~~~~~~~~~~~~~~~~~~~~~~~~~~~~~~~~~~~~~~~~~~~~~~~~~~~~~~~~~~~~~~~~(d)~~~~~~~~~~~~~~~~~~~~~~~~~$
\caption{Streamline patterns and trapping for the effect of $Q$,
when $n=2/3,~\phi=0.5,~k_1=0, \tau=0.1$}
\label{cnsns_manuscript_stlp5.3.1.-5.3.4}
\end{figure}

\begin{figure}
\includegraphics[width=3.5in,height=2.0in]{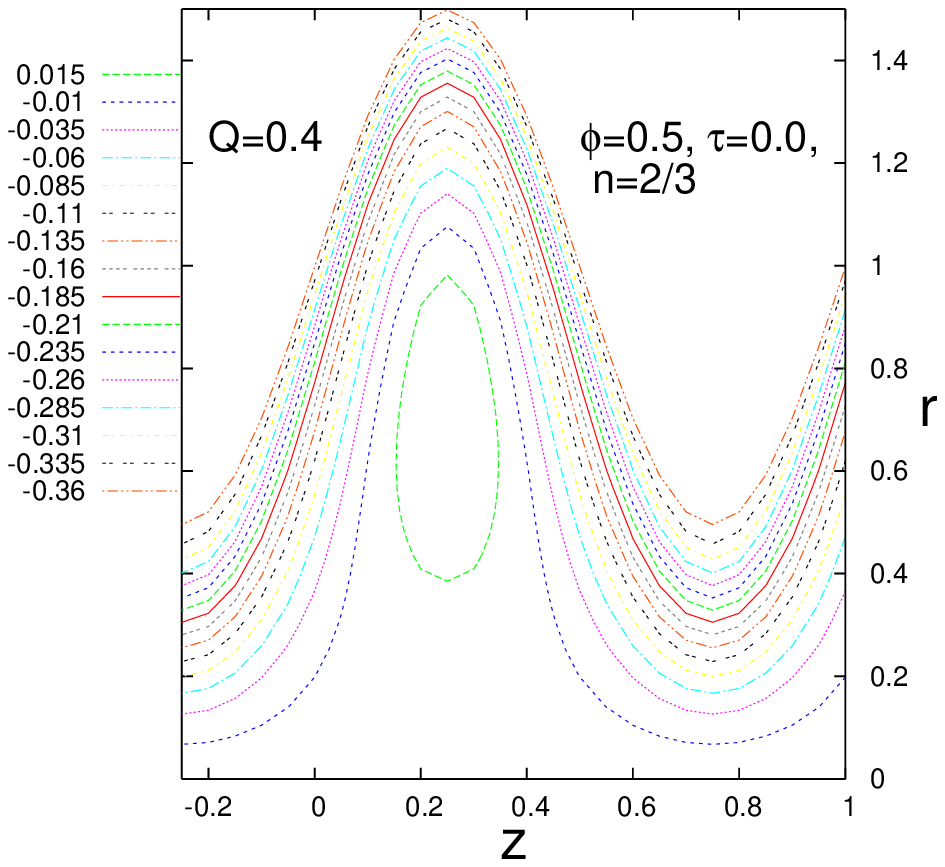}\includegraphics[width=3.5in,height=2.0in]{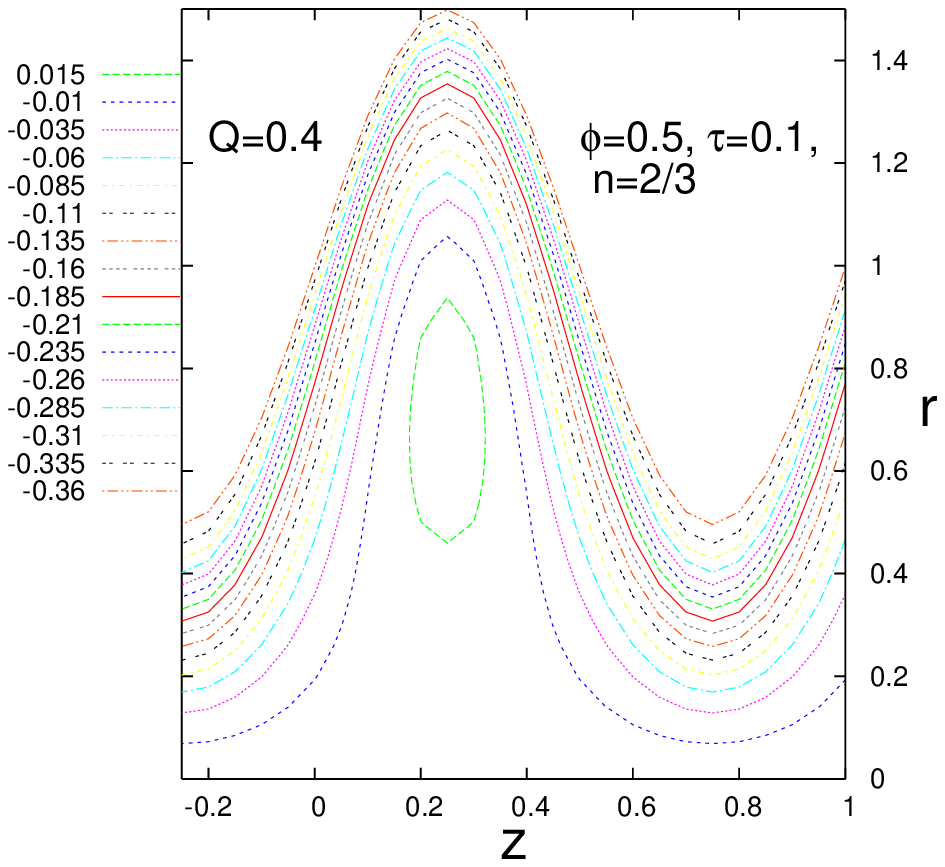}
\\$~~~~~~~~~~~~~~~~~~~~~~~~~~~(a)~~~~~~~~~~~~~~~~~~~~~~~~~~~~~~~~~~~~~~~~~~~~~~~~~~~~~~~~~~(b)~~~~~~~~~~~~~~~~~~~~~~~~~$
\includegraphics[width=3.5in,height=2.0in]{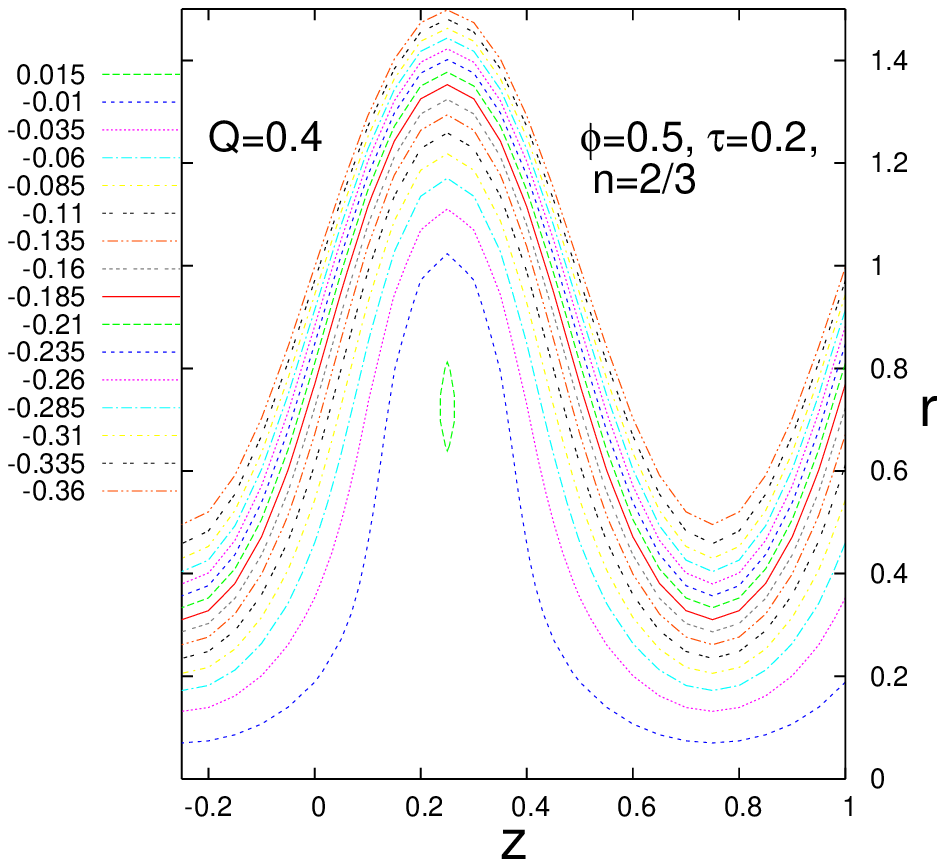}\includegraphics[width=3.5in,height=2.0in]{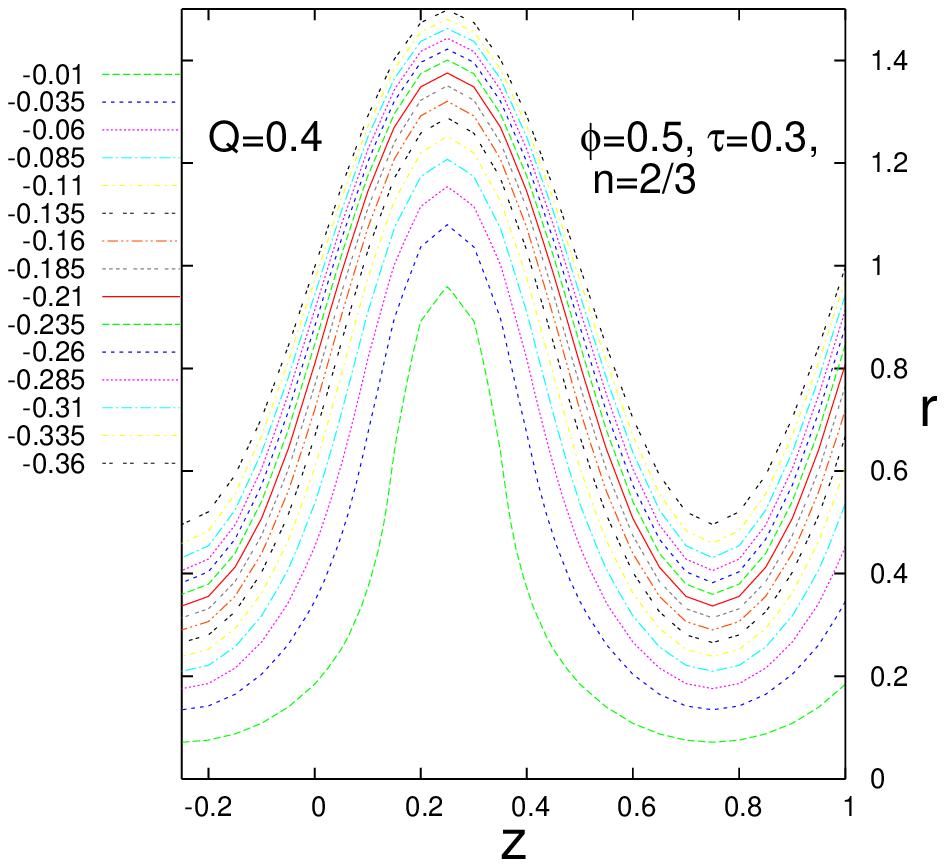}
\\$~~~~~~~~~~~~~~~~~~~~~~~~~~~(c)~~~~~~~~~~~~~~~~~~~~~~~~~~~~~~~~~~~~~~~~~~~~~~~~~~~~~~~~~~~~~~~(d)~~~~~~~~~~~~~~~~~~~~~~~~~~~~~$
\caption{Streamline patterns for different values of $\tau$
  in the case of peristaltic flow when
  $n=2/3,~\phi=0.5,~k_1=0,~Q=0.4$. It may be observed that the size of the bolus reduces with the increase in the value of
  $\tau$. The bolus totally disappears for $\tau\ge 0.3$.}
\label{cnsns_manuscript_stlp5.4.1.-5.4.4}
\end{figure}

It is known that one of the important characteristics of peristaltic
transport is the phenomenon of trapping. It occurs when streamlines on
the central line are split to enclose a bolus of fluid particles
circulating along closed streamlines in the wave frame of reference.
Then the trapped bolus moves with a speed equal to the wave
propagation velocity. This physical phenomenon may be responsible for
the formation of thrombus in blood. Let us consider the wave frame
transformations (x,y) moving with a velocity c away from a fixed frame
(X,Y) such that $x=X-ct,~y=Y,~u=U-c,~v=V,~p(x,y)=P(X,Y,t),$ in which
(u,v) and (U,V) are the velocity components, p and P stand for
pressure in wave frame and fixed frame of reference
respectively. Under the purview of the present study, Figs.
\ref{cnsns_manuscript_stlp5.1.1.-5.1.6}-\ref{cnsns_manuscript_stlp5.4.1.-5.4.4}
give an insight into the changes in the pattern of streamlines and
trapping that occurs due to changes in the values of different
parameters that govern the flow of blood in the wave frame of
reference.  Figs. \ref{cnsns_manuscript_stlp5.1.1.-5.1.6} provide the
streamline patterns and trapping in the case of a shear-thinning fluid
for different values of $\phi$. With an increase in $\phi$, the bolus
is found to appear in a distinct manner. Streamlines for different
values of the fluid index `n' are depicted in Fig.
\ref{cnsns_manuscript_stlp5.2.1.-5.2.4}. This figure indicates that
occurrence of trapping is strongly influenced by the value of the
fluid index. Fig. \ref{cnsns_manuscript_stlp5.3.1.-5.3.4} shows that
trapped bolus increases in size and also that it has a tendency to
move towards the boundary as the flow rate increases. Here it is
important to note that the bolus appearing for small values of $\tau$
decreases in size with an increase in $\tau$ (Figs.
\ref{cnsns_manuscript_stlp5.4.1.-5.4.4}).

\subsection{Distribution of Wall Shear Stress}
If the shear stress generated on the wall of a
blood vessel exceeds a certain limit, the constituents of blood are
likely to be damaged. The magnitude of the wall shear stress
has a vital role in the process of molecular convection at high
Prandtl or Schmidt number \cite{Higdon}. In view of these
observations, it is important to study the shear stress that is
developed during the hemodynamical flow of blood in small arteries. The
wall shear stress distributions for the present study are plotted in
Figs. \ref{cnsns_manuscript_shear5.1.1.-5.11.1} under varied
conditions. The distributions of wall shear stress at four different
time instants during one complete wave period have been presented in
Fig. \ref{cnsns_manuscript_shear5.1.1.-5.11.1}(a). It may be observed
from this figure that at each of these instants of time, there exist
two peaks in the wall shear stress distribution, with a gradual ramp
in between; however, negative peak of wall shear stress
$\tau_{min}$ is not as large as the maximum wall shear stress
$\tau_{max}$. The transition from $\tau_{min}$ to $\tau_{max}$ of
wall shear takes place in some zone between the minimum and
maximum radii of the vessel. At the location where the maximum
occlusion occurs, wall shear stress along with the pressure is
maximum. Since the pressure gradient to the left of this location
takes a positive value, the local instantaneous flow will take place
towards the left of $\tau_{max}$. This may lead to some serious
consequences. For example, if the shear rate at the crest exceeds
some limit, a dissolving wavy wall will have a tendency to level
out. Moreover, some bio-chemical reaction between the wall material
and the constituents of blood may set in. As a result of this, the
products of the chemical reaction may be deposited on the
endothelium and consequently wall amplitude may increase at a rapid
rate. This is likely to lead to clogging of the blood vessel.

 \begin{figure}
 \includegraphics[width=3.35in,height=1.95in]{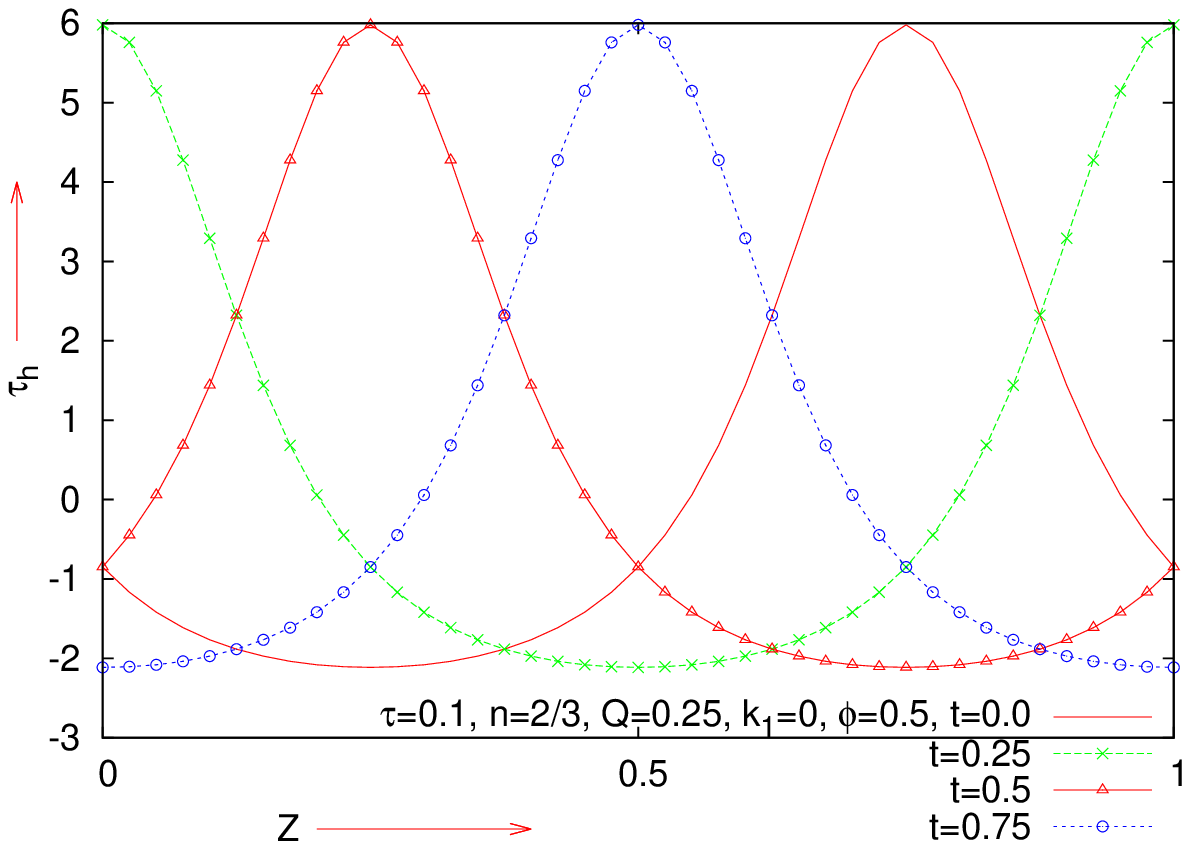}\includegraphics[width=3.35in,height=1.95in]{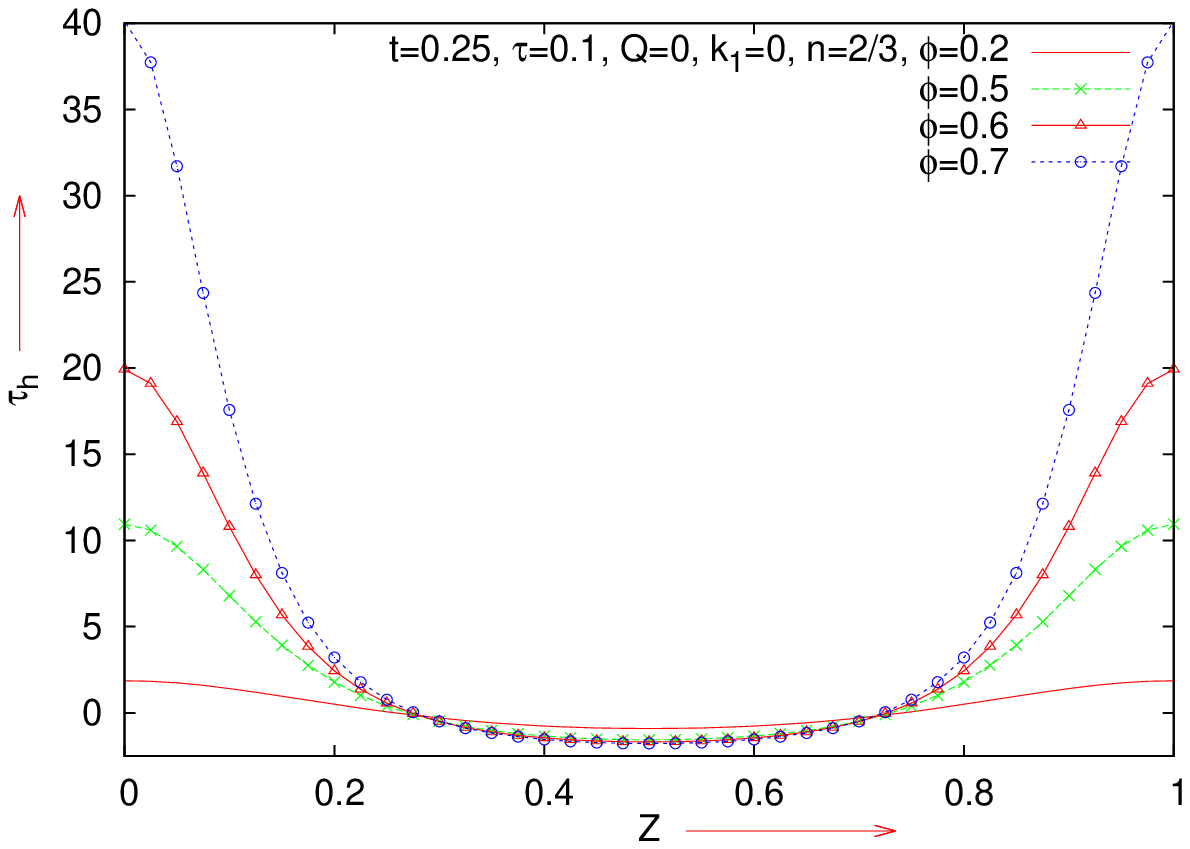}\\$~~~~~~~~~~~~~~~~~~~~~~~~~~~(a)~~~~~~~~~~~~~~~~~~~~~~~~~~~~~~~~~~~~~~~~~~~~~~~~~~~~~~~~~~~~~~~~~~~~~(b)~~~~~~~~~~~~~~~$
\includegraphics[width=3.35in,height=2.0in]{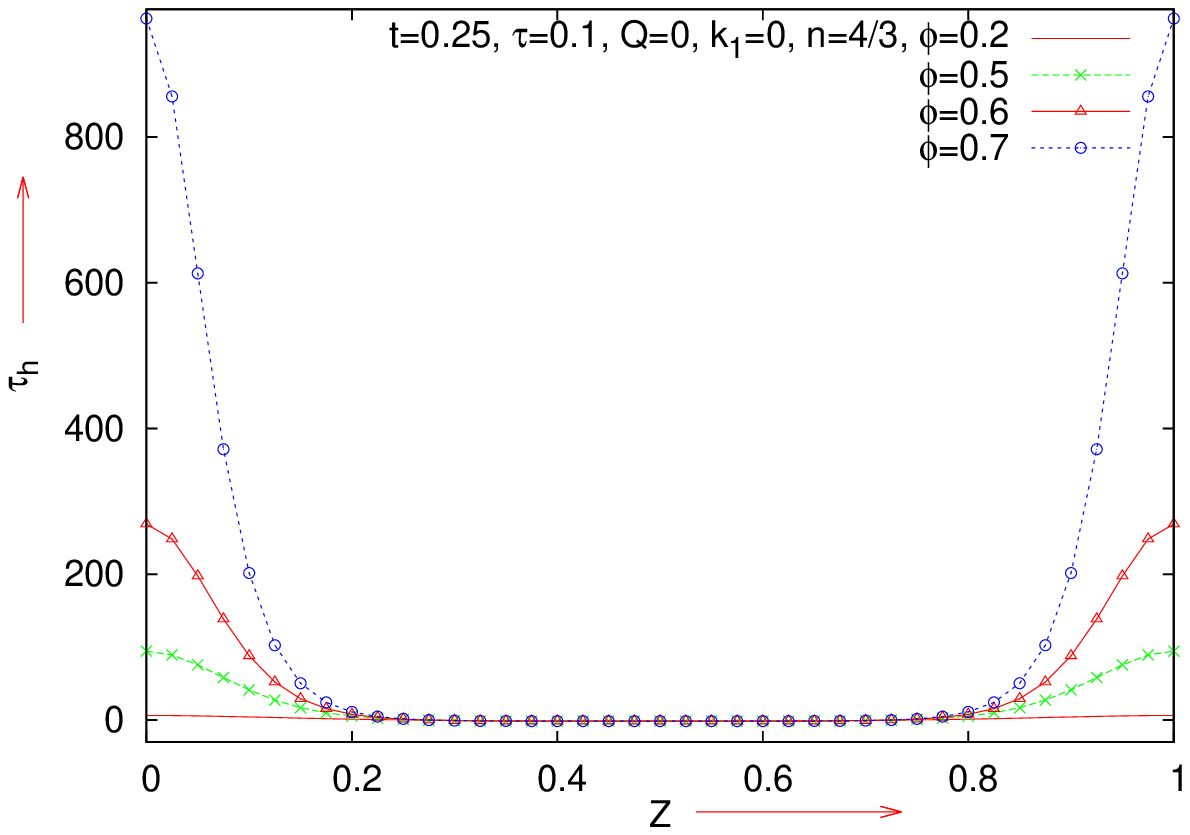}\includegraphics[width=3.35in,height=2.0in]{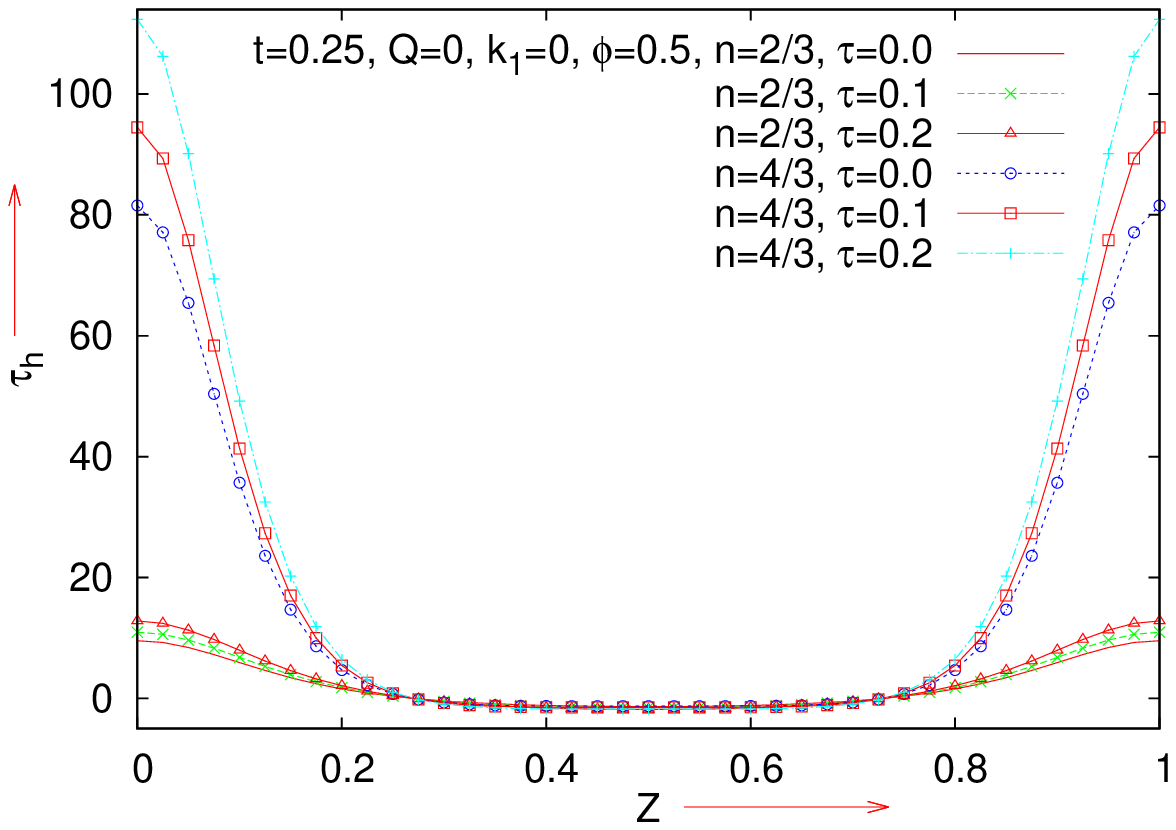}\\$~~~~~~~~~~~~~~~~~~~~~~~~~~~(c)~~~~~~~~~~~~~~~~~~~~~~~~~~~~~~~~~~~~~~~~~~~~~~~~~~~~~~~~~~~~~~~~~~~~~(d)~~~~~~~~~~~~~~~$
 \includegraphics[width=3.35in,height=2.0in]{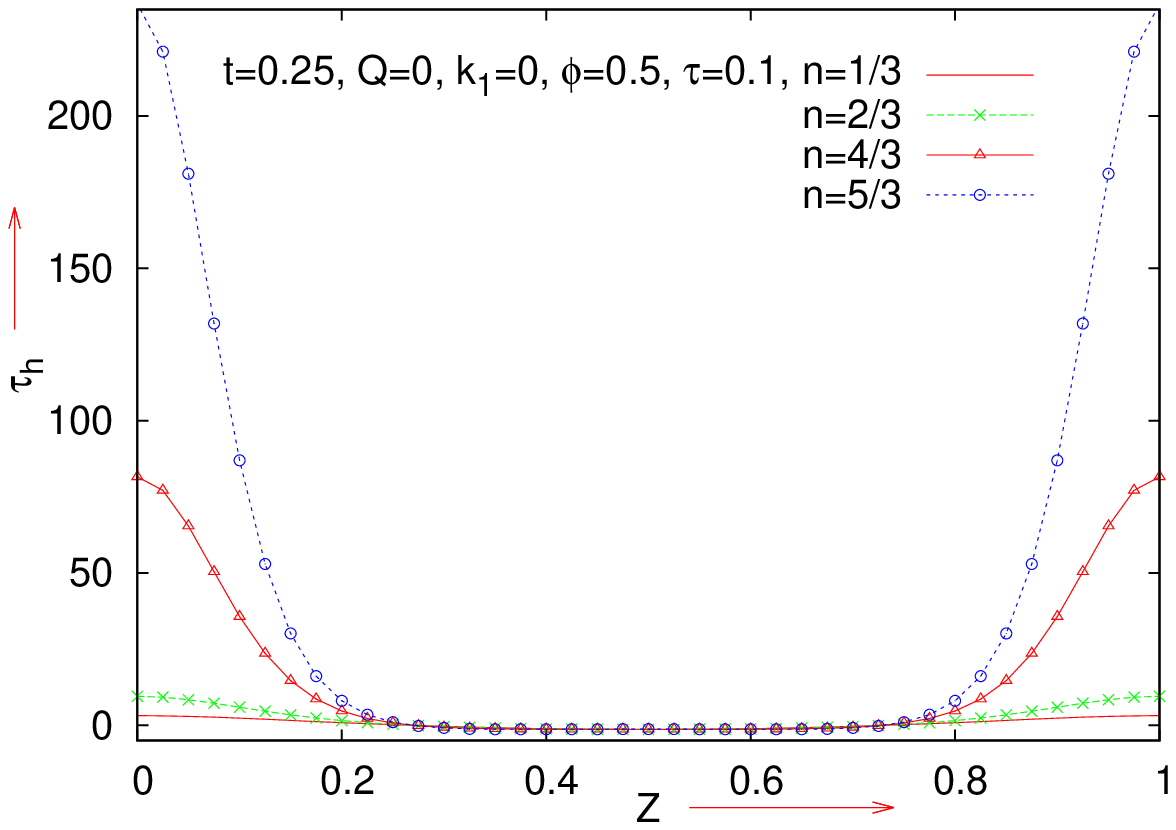}\includegraphics[width=3.35in,height=2.0in]{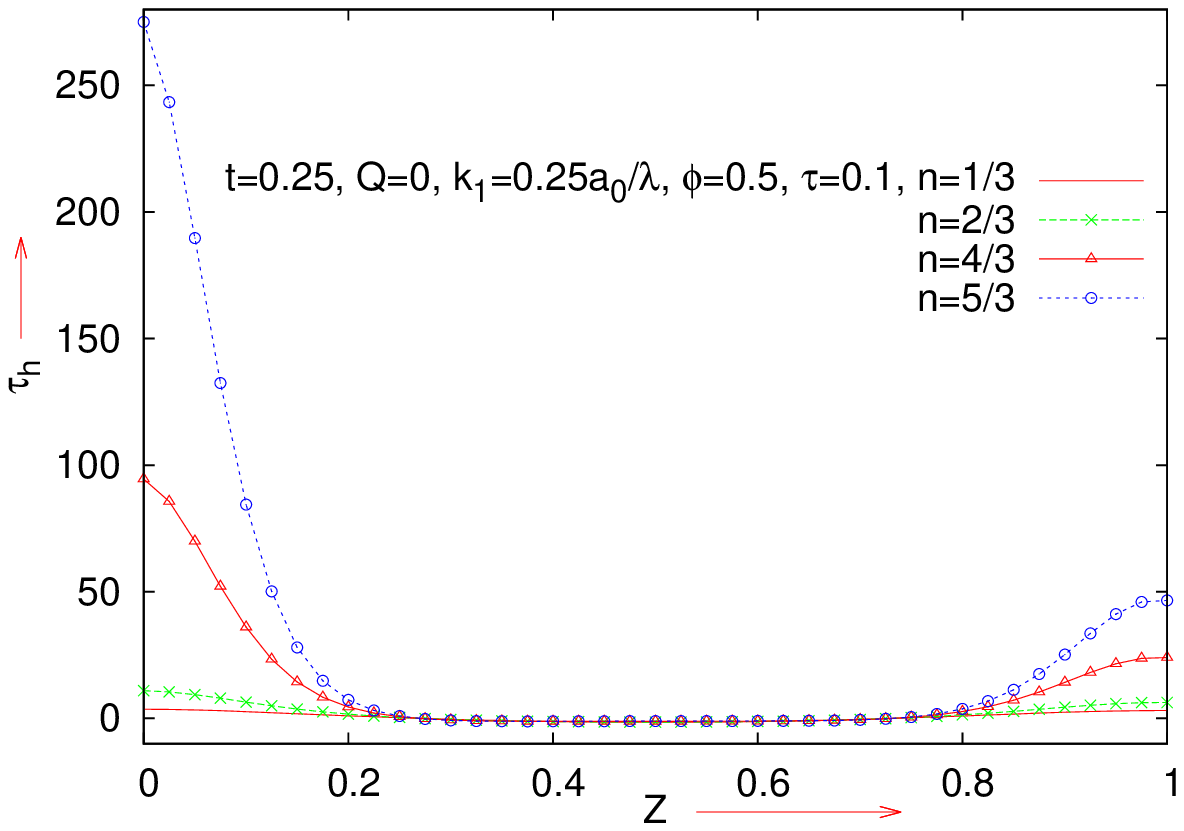}\\$~~~~~~~~~~~~~~~~~~~~~~~~~~~(e)~~~~~~~~~~~~~~~~~~~~~~~~~~~~~~~~~~~~~~~~~~~~~~~~~~~~~~~~~~~~~~~~~~~~~(f)~~~~~~~~~~~~~~~$
\end{figure}
\begin{figure}
\includegraphics[width=3.35in,height=2.0in]{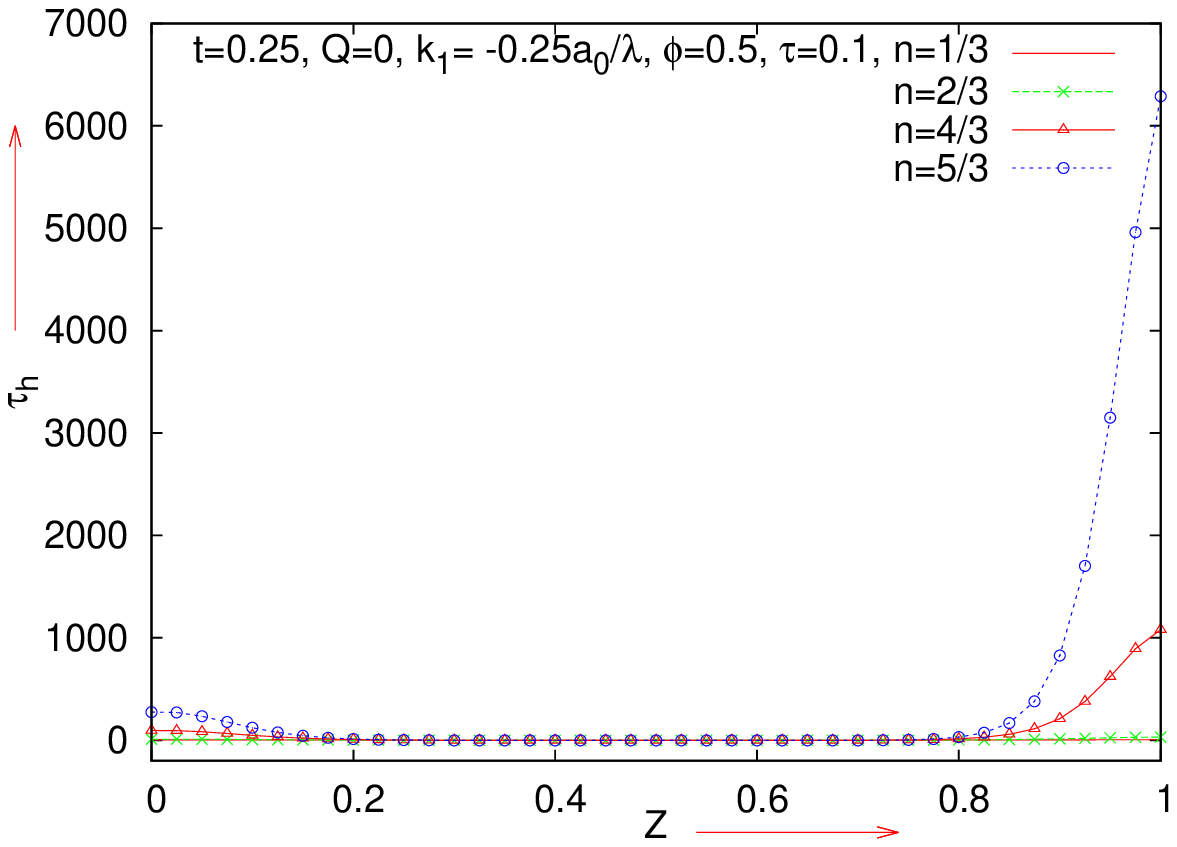}\includegraphics[width=3.35in,height=2.0in]{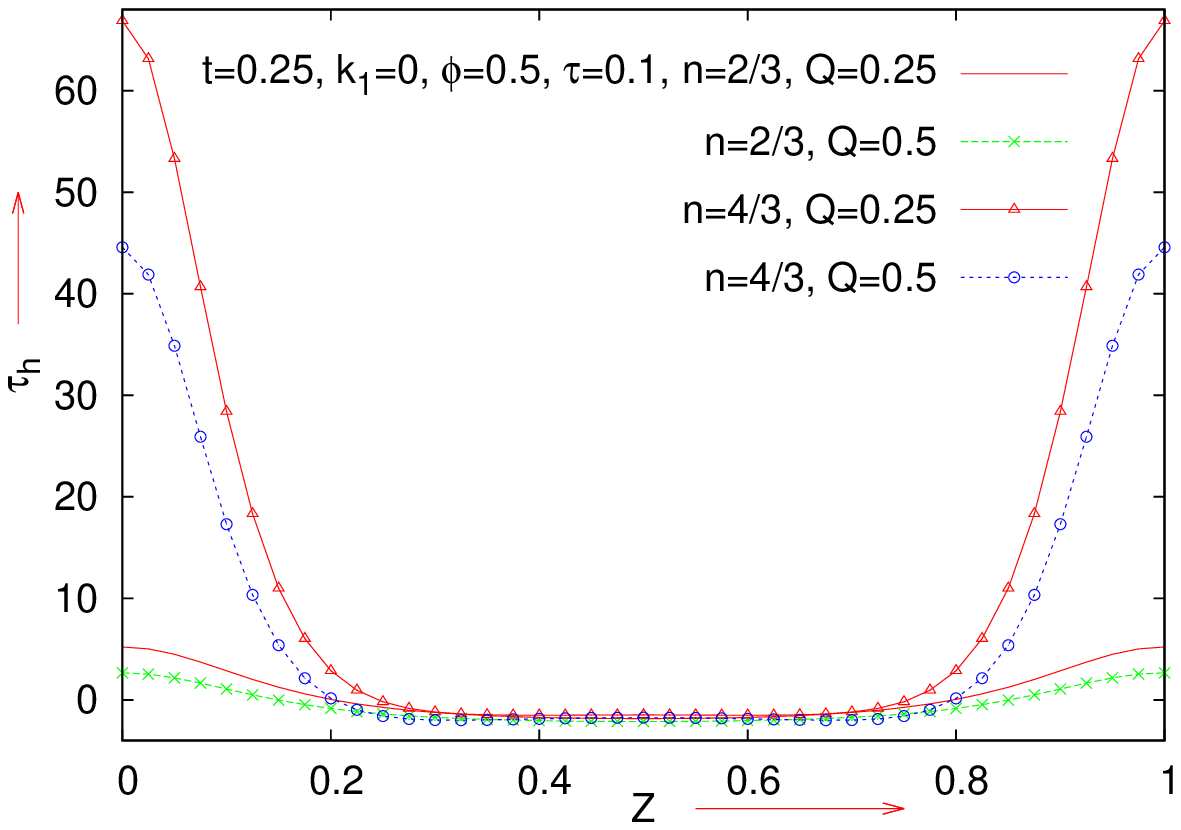}\\$~~~~~~~~~~~~~~~~~~~~~~~~~~~(g)~~~~~~~~~~~~~~~~~~~~~~~~~~~~~~~~~~~~~~~~~~~~~~~~~~~~~~~~~~~~~~~~~~~~~(h)~~~~~~~~~~~~~~~$
\includegraphics[width=3.35in,height=2.0in]{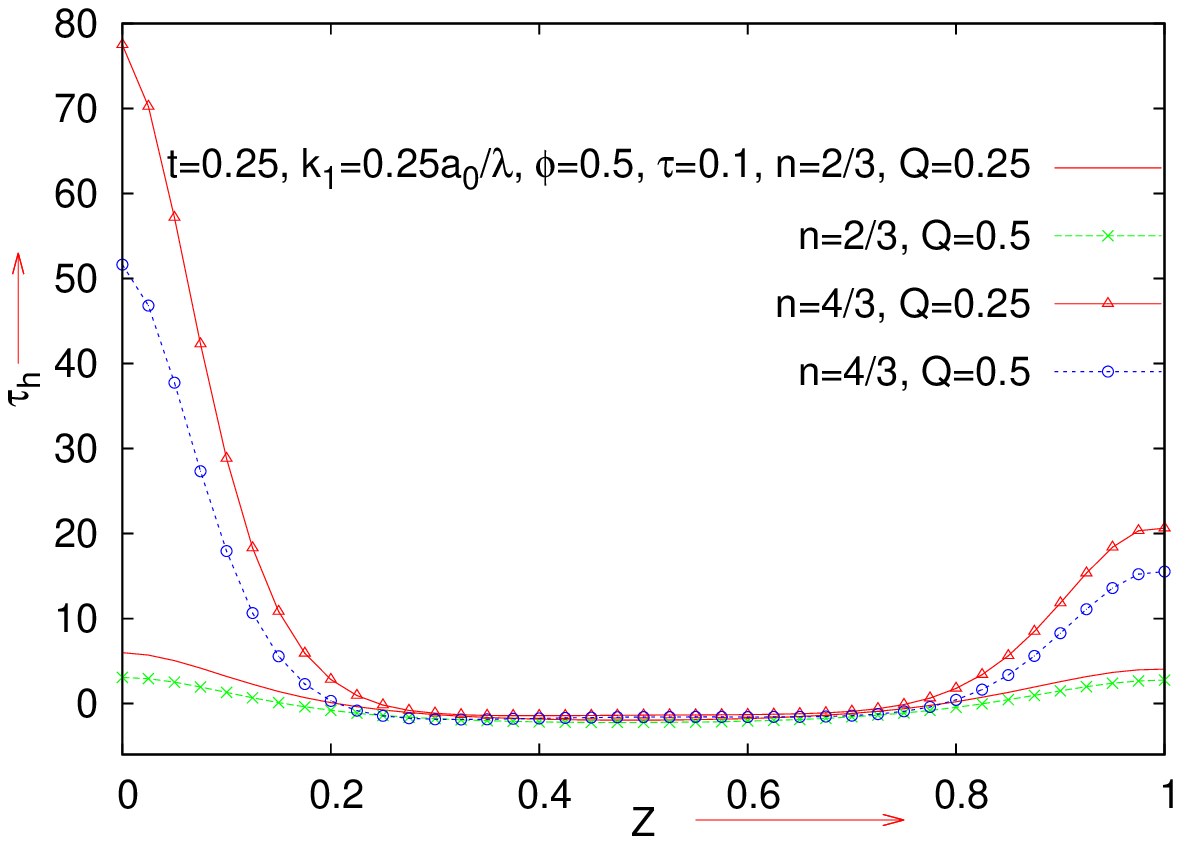}\includegraphics[width=3.35in,height=2.0in]{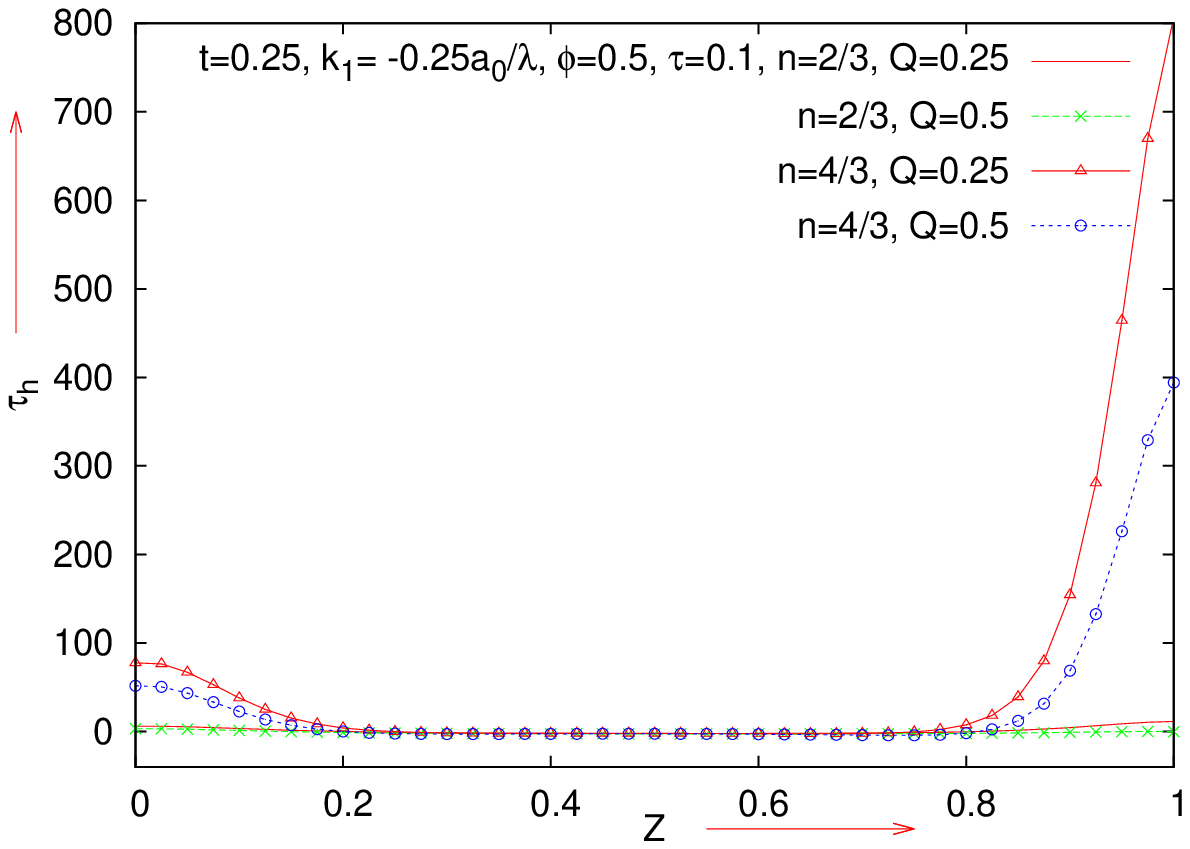}\\$~~~~~~~~~~~~~~~~~~~~~~~~~~~(i)~~~~~~~~~~~~~~~~~~~~~~~~~~~~~~~~~~~~~~~~~~~~~~~~~~~~~~~~~~~~~~~~~~~~~(j)~~~~~~~~~~~~~~~$
\includegraphics[width=3.35in,height=2.0in]{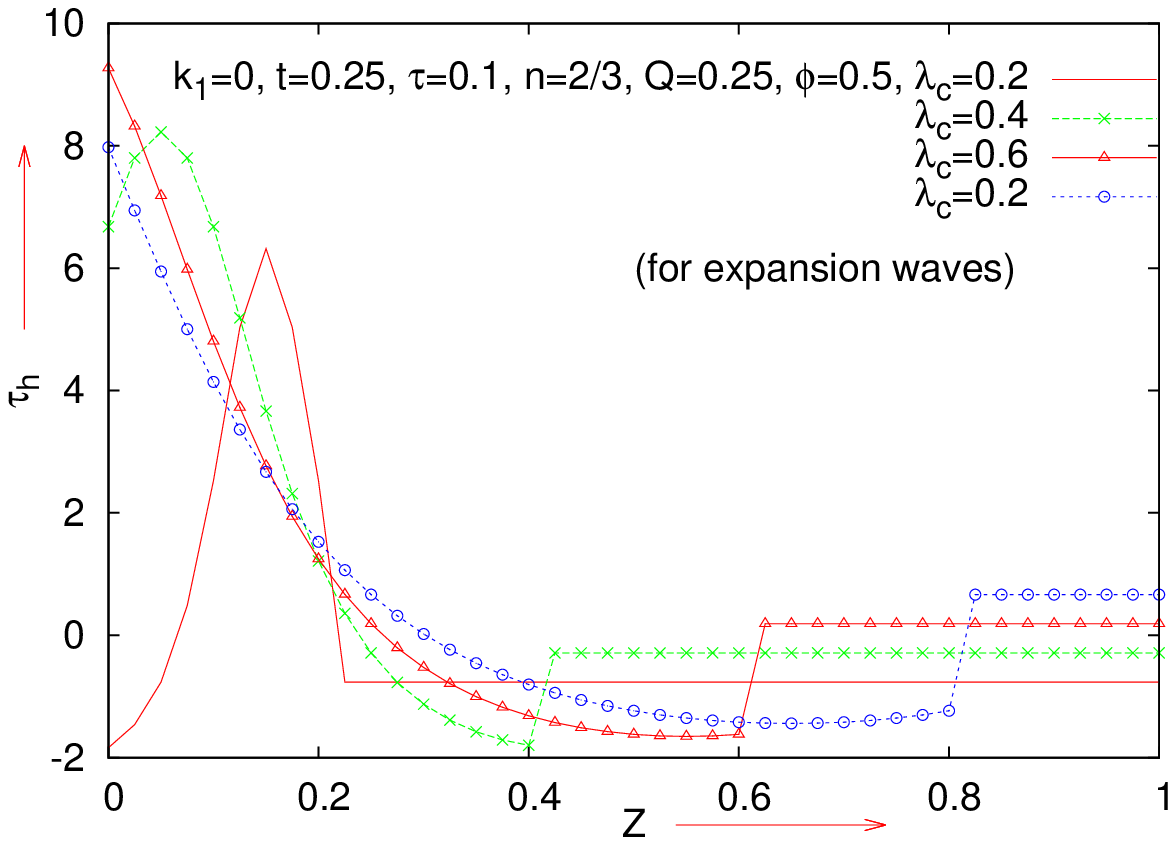}\includegraphics[width=3.35in,height=2.0in]{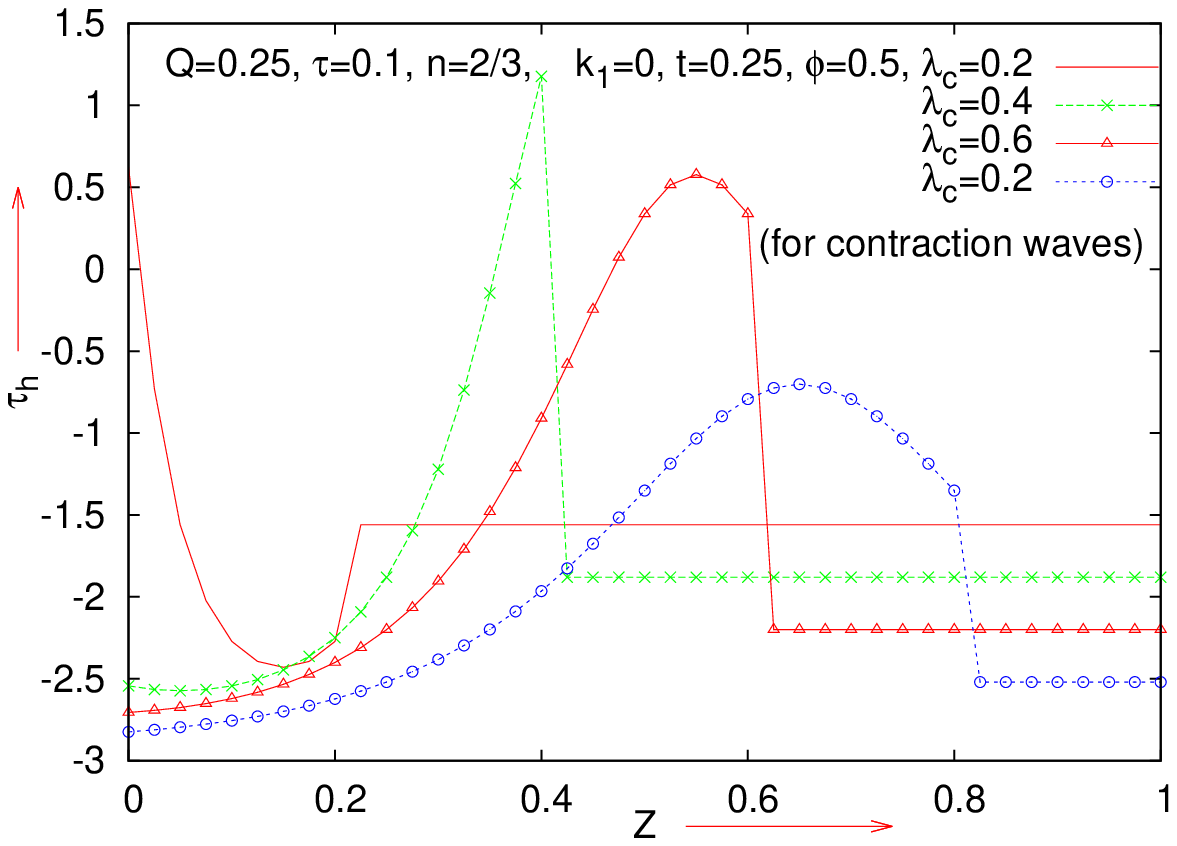}\\$~~~~~~~~~~~~~~~~~~~~~~~~~~~(k)~~~~~~~~~~~~~~~~~~~~~~~~~~~~~~~~~~~~~~~~~~~~~~~~~~~~~~~~~~~~~~~~~~~~~(l)~~~~~~~~~~~~~~~$
\caption{Wall shear stress distribution in different situations.
Wall shear stress attains its maximum/minimum at the location where
the radius of the vessel is smallest/largest.}
\label{cnsns_manuscript_shear5.1.1.-5.11.1}
\end{figure}

The peaks of the wall shear stress distribution on both sides of
$\tau_{max}$ are small and hence the local instantaneous flow will
occur in the direction of propagation of the peristaltic waves. For a
Herschel-Bulkley fluid, Figs.
\ref{cnsns_manuscript_shear5.1.1.-5.11.1}(b-c) show that in the
contracting region where occlusion takes place, there is a
remarkable increase in the wall shear stress due to an increase in
the value of $\phi$ for both shear-thinning and shear-thickening
fluids. With an increase in $\phi$, magnitude of $\tau_{min}$
increases in the expanding region for
shear-thinning/shear-thickening fluids, although the effect is not
very prominent. It may be observed from Fig.
\ref{cnsns_manuscript_shear5.1.1.-5.11.1}(d) that $\tau_{max}$
increases with increase in $\tau$, while $\tau$ has little effect on
$\tau_{min}$.

The quantum of influence of the rheological fluid index 'n' on the
distribution of wall shear stress is shown in Figs.
\ref{cnsns_manuscript_shear5.1.1.-5.11.1}(e-g) for uniform/non-uniform
blood vessels. In all types of vessels studied here, $\tau_{max}$
increases with increase in `n'. It is very important to mention
that the difference of shear stress between the outlet and the inlet
in the case of a converging vessel is exceedingly large in
comparison to the case of a diverging vessel. As the time averaged
flow rate increases, Figs.
\ref{cnsns_manuscript_shear5.1.1.-5.11.1}(h-j) indicate very clearly
that the wall shear stress tends to decrease for all types of
vessels examined here. One can observe from Figs.
\ref{cnsns_manuscript_shear5.1.1.-5.11.1}(k-l) that $\tau_h$ changes
its values within the region of SSD wave activation; beyond this, it
maintains a constant value.

\section{Summary and Conclusion}
The present paper deals with a study of the peristaltic motion of
blood in the micro-circulatory system, by taking into account the
non-Newtonian nature of blood and the non-uniform geometry of the
micro-vessels, e.g. arterioles and venules. The non-Newtonian
viscosity of blood is considered to be of Herschel-Bulkley type. The
effects of amplitude ratio, mean pressure gradient, yield stress and
the rheological fluid index n on the distribution of velocity and
wall shear stress as well as on the pumping phenomena, formation of
the streamline pattern and the occurrence of trapping are
investigated under the purview of the lubrication theory.
Experimental observations have revealed that in the case of roller
pumps, the fluid elements are prone to significant damage. Moreover,
during the process of transportation of fluids in living structures
executed by using arthro-pumps, the fluid particles are likely to be
appreciably damaged. Qualitative and quantitative studies of the
present problem for the wall shear stress have a significant
bearing on extracorporeal circulation. When the heart-lung machine
is used for patient care, the erythrocytes of blood are
likely to be damaged. Thus the present study bears the potential of
significant application in biomedical engineering and technology.

The study reveals that at any instant of time, there is a retrograde
flow region for Herschel-Bulkley type of non-Newtonian fluids like
blood when $\Delta P\le 0$. The regions of forward/retrograde flow
advance at a faster rate, if the values of n and $\phi$ are raised.
In the case of a uniform/diverging tube, the flow reversal decreases
when $\Delta P$ tends to be negative; however, for a converging tube
such an observation is not very prominent. In addition, this study
shows that non-uniform geometry of the vessel affects quite
significantly the distributions of velocity and the wall shear
stress as well as pumping and other flow characteristics. It is also
observed that the amplitude ratio $\phi$ and the rheological fluid index
`n' are very sensitive parameters that change the peristaltic
pumping characteristics and the distribution of velocity and wall
shear tress. The parabolic nature of the velocity profiles is also
significantly disturbed by the numerical  value of the rheological
fluid index `n'.

From the present study we may conclude that for the peristaltic flow
of blood through propagation of SSD expansion waves, the pumping
performance is better than that in the case of sinusoidal wave
propagation and that the backward flow region is totally absent in SSD
expansion wave propagation mode. On the basis of this study one may
also draw the conclusion that backward flow originates due to the
contraction of vessels.  \\

{\bf Acknowledgment:} {\it The authors wish to convey their thanks to
  all the reviewers (anonymous) for their comments and suggestions
  based upon which the present version of the manuscript has been
  prepared. One of the authors, S. Maiti, is thankful to the Council
  of Scientific and Industrial Research (CSIR), New Delhi and the
  University Grants Commission (UGC), New Delhi for awarding an SRF
  and the Dr. D. S. Kothari Post Doctoral Fellowship respectively
  during this investigation. The other author, Prof. J. C. Misra,
  wishes to express his deep sense of gratitude to Professor (Dr.)
  Manoj Ranjan Nayak, President of the Sikha O Anusandhan University,
  Bhubaneswar for providing a congenial environment and facilities for
  doing research.}

\end{document}